\documentclass[twocolumn]{aastex7}
\usepackage{float}
\usepackage{amsmath}
\usepackage{afterpage}
\usepackage{longtable}

\newcommand{\cloudy}{\texttt{CLOUDY}}
\newcommand{\pynbody}{\texttt{PYNBODY}}
\newcommand{\gasolineI}{{\sc GASOLINE}}
\newcommand{\gasolineII}{{\sc GASOLINE2}}
\newcommand{\changa}{{\sc ChaNGA}}

\definecolor{pink}{HTML}{de02ab}

\newcommand{\OVI}{O{\sc ~vi}}
\newcommand{\HI}{H{\sc ~i}}
\newcommand{\HII}{H{\sc ~ii}}

\newcommand{\HeI}{He{\sc ~i}}
\newcommand{\HeII}{He{\sc ~ii}}
\newcommand{\HeIII}{He{\sc ~iii}}

\newcommand{\CII}{C{\sc ~ii}}
\newcommand{\CIII}{C{\sc ~iii}}
\newcommand{\CIV}{C{\sc ~iv}}
\newcommand{\CV}{C{\sc ~v}}
\newcommand{\CVI}{C{\sc ~vi}}

\newcommand{\SiII}{Si{\sc ~ii}}
\newcommand{\SiIII}{Si{\sc ~iii}}
\newcommand{\SiIV}{Si{\sc ~iv}}

\newcommand{\rHI}{$r_{\,\Sigma_{\text{HI}}}$}

\newcommand{\NHI}{$N_{\text{HI}}$}
\newcommand{\NCII}{$N_{\text{CII}}$}
\newcommand{\NSiII}{$N_{\text{SiII}}$}
\newcommand{\NSiIII}{$N_{\text{SiIII}}$}
\newcommand{\NSiIV}{$N_{\text{SiIV}}$}
\newcommand{\NCIV}{$N_{\text{CIV}}$}
\newcommand{\NOVI}{$N_{\text{OVI}}$}
\newcommand{\OII}{O{\sc ~ii}}
\newcommand{\OIII}{O{\sc ~iii}}
\newcommand{\OIV}{O{\sc ~iv}}
\newcommand{\OV}{O{\sc ~v}} 
\newcommand{\OVII}{O{\sc ~vii}}

\def\MM{\textsf{M\texttt{+}M}}
\begin{document}

\title{Marvelous Metals: Surveying the Circumgalactic Medium of Simulated Dwarf Galaxies}

\author[orcid=0000-0003-0200-6986]{Daniel R. Piacitelli}
\affiliation{Rutgers University, Department of Physics and Astronomy, Piscataway, NJ 08854, USA}\email[show]{piacitelli.danielr@gmail.com}

\author{Alyson M. Brooks}
\affiliation{Rutgers University, Department of Physics and Astronomy, Piscataway, NJ 08854, USA}\affiliation{Center for Computational Astrophysics, Flatiron Institute, 162 Fifth Ave, New York, NY 10010, USA}\email{abrooks@physics.rutgers.edu}

\author[0000-0001-6779-3429]{Charlotte Christensen}\affiliation{Physics Department, Grinnell College, 1116 Eighth Avenue, Grinnell, IA 50112, USA }\email{christenc@grinnell.edu}

\author[0000-0001-7589-6188]{N. Nicole Sanchez}\affiliation{Carnegie Observatories, 813 Santa Barbara Street, Pasadena, CA 91101 USA}\affiliation{California Institute of Technology, TAPIR 350-17, 1200 E. California Boulevard, Pasadena, CA 91125-0001 USA}\email{nnicolesanchez@gmail.com}

\author[0000-0003-3520-6503]{Yakov Faerman}
\affiliation{Department of Astronomy, University of Washington, Seattle, WA 98195, USA}
\affiliation{School of Physics and Astronomy, Tel Aviv University, Tel Aviv 69978, Israel}\email{yakov.faerman@gmail.com}

\author{Sijing Shen}\affiliation{Institute of Theoretical Astrophysics, University of Oslo, PO Box 1029, Blindern 0315, Oslo, Norway}\email{sijing.shen@astro.uio.no}

\author[0000-0001-7831-4892]{Akaxia Cruz}
\affiliation{Center for Computational Astrophysics, Flatiron Institute, 162 Fifth Ave, New York, NY 10010, USA}\email{acruz@flatironinstitute.org}

\author[0000-0002-9642-7193]{Ben Keller}
\affiliation{Department of Physics and Materials Science, University of Memphis, 3720 Alumni Avenue, Memphis, TN 38152, USA}\email{bkeller1@memphis.edu}

\author[0000-0001-5510-2803]{Thomas R. Quinn}
\affiliation{Department of Astronomy, University of Washington, Seattle, WA 98195, USA}\email{trq@astro.washington.edu}

\author{James Wadsley}
\affiliation{Department of Physics \& Astronomy, McMaster University, ABB-241, 1280 Main Street West, Hamilton, Ontario, L8S 4M1, Canada}\email{wadsley@mcmaster.ca}

\begin{abstract}
Dwarf galaxies are uniquely sensitive to feedback processes and known to experience substantial mass and metal loss from their disk. Here, we investigate the circumgalactic medium (CGM) of 64 isolated dwarf galaxies ($6.0<\rm{log(M}_*/M_{\odot})<9.5$) at $z=0$ from the Marvel-ous Dwarfs and Marvelous Massive Dwarfs simulations. Our galaxies produce column densities broadly consistent with current observations. We investigate these column densities in the context of mass and metal retention rates and CGM physical properties. We find $48\pm11\%$ of all baryons within $R_{200c}$ reside in the CGM, with $\sim70\%$ of CGM mass existing in a warm gas phase, $10^{4.5}<T<10^{5.5}$K that dominates beyond $r/R_{200c}\sim0.5$. The warm and cool ($10^{4.0}<T<10^{4.5}$K) gas phases each retain $5-10\%$ of metals formed by the dwarf galaxy. The significant fraction of mass and metals residing in the warm CGM phase provides an interpretation for the lack of $z\sim0$ low ion detections beyond $b/R_{200c}\sim0.5$, as the majority of mass in this region exists in higher ions. We find a weak correlation between galaxy mass and total CGM metal retention despite the fraction of metals lost from the halo increasing from $\sim10\%$ to $>40\%$ towards lower masses. Our findings highlight the CGM (particularly its warm phase) as a key reservoir of mass and metals for dwarf galaxies across stellar masses, underscoring its importance in understanding the baryon cycle in the low-mass regime. Finally, we provide individual simulated galaxy properties and quantify the fraction of ultraviolet observable mass to support future observational programs aimed at performing a metal budget around dwarf galaxies.
\end{abstract}

\section{Introduction}\label{sec:Introduction}
Across galaxy masses, feedback processes are expected to play a key role in shaping galaxy evolution. For low-mass galaxies, the shallower gravitational wells result in stellar feedback processes being highly efficient at driving metal-enriched mass loss from the disk of the galaxy, thereby regulating star formation \citep[e.g.][]{Keller_2016}. Some observational estimates indicate that up to 95\% of the metals produced by a dwarf galaxy’s stellar population may no longer reside in its disk \citep[e.g.,][]{2015ApJ...815L..17M}. However, the final fate of the ejected mass and metals remains an open question. Specifically, to answer these questions, it is essential to constrain the fraction of ejected mass and metals that escape into the intergalactic medium (IGM) and how much remains within the halo. In this direction, the circumgalactic medium of dwarf galaxies (CGM-DG) and its mass and metal content are key in understanding the galactic baryon cycle in the low-mass regime.

A growing number of observational studies have sought to map the distribution of the material in the CGM-DG. Out to $z\sim 0.7$, the CGM of isolated dwarf galaxies has been commonly detected in \HI\, absorption \citep{LiangChen_14, Johnson_17, Bordoloi_18, Zheng_24, Mishra_24}. In fact, \HI\, detections have reached well beyond the virial radius of the galaxy, and equivalent width or column densities show a declining profile with increasing impact parameter from the galaxy \citep[e.g.][]{LiangChen_14, Johnson_17, Mishra_24}.

Despite the common detections of \HI, studies targeting metal absorption in the CGM-DG have yielded mixed results. Low-$z$ studies have only successfully detected metal ions (\CII, \CIV, \SiII, \SiIII, and \SiIV) within the inner CGM ($\sim 0.5 R_{vir}$), but only non-detections (upper limits) have been found in the outer CGM and beyond \citep{Bordoloi_14, LiangChen_14, Burchett_16, Johnson_17, QuBregman_22, Zheng_24}. Similarly, at higher redshifts, detections of lower metal ions like \CII, \SiII, and \SiIII\, have been constrained within $r < 0.5 R_{vir}$ \citep[e.g.][]{Mishra_24}. These low detection rates may suggest that the mass of metal-enriched gas declines significantly beyond $r \sim 0.5 R_{vir}$. However, at higher redshifts ($0.077<z<0.73$), \OVI\, has been detected around dwarf galaxies out to twice their virial radius \citep{Johnson_17, Tchernyshyov_22, Qu_24, Mishra_24}.  Since \OVI\, is expected to exist in warmer conditions and lower ions in cooler conditions, these detections and non-detections imply that there is a cool component in the CGM-DG that is largely concentrated to the inner regions of the halo and a warm, extended component in the outer regions of the halo.

Additionally, observational studies have sought to quantify the mass and metals residing in the cool and warm phases of the CGM-DG. \citet{Zheng_24} introduces empirical models for the CGM mass distribution that adopt gas density and volume filling factor values based on observed \HI\, column densities. Utilizing these models,  \citet{Zheng_24} estimates the cool CGM phase ($T\sim10^4$ K) around $M_*=10^{8.3} M_{\odot}$ galaxies contains a mass of $M_{CGM,cool} \sim 10^{8.4} M_{\odot}$---which only accounts for $\sim 2\%$ of the galaxy's baryonic budget relative to the cosmic abundance ($f_{CGM,cool}=M_{CGM,cool}/M_{200m}\Omega_b/\Omega_m$). Recently, \citet{2025ApJ...982L..30F} used existing \HI\ column densities to derive a power-law density profile for the CGM-DG and provide observationally motivated constraints on the cool gas mass in the CGM-DG. They find that measured HI columns suggest that $<10\%$ of the halo baryon budget resides in the cool, photoionized phase.

Regarding metal content, \citet{Johnson_17} estimates that the cool phase ($T\sim10^4$ K) harbors only $2-6\%$ of the silicon budget, based on expected supernovae yields. Similarly, \citet{Zheng_24} predicts that the cool phase contains $\sim10\%$ of the metals ever produced by the dwarf galaxy's star formation history. For the warmer, \OVI-traced phase of the CGM-DG, studies have predicted a \OVI\, mass of $M_{CGM, \text{\OVI}} \sim 10^{5.0-6.0}M_{\odot}$ \citep{Johnson_17, Tchernyshyov_22, Zheng_24} which corresponds to $\sim8\%$ of the oxygen budget \citep{Johnson_17,Zheng_24}. 

Ultimately, these observations and predictions indicate an observable CGM in the halos of dwarf galaxies, much of which is unaccounted for. Further work investigating the CGM-DG's different temperature phases holds promise in mapping the CGM beyond  $r \sim 0.5 R_{vir}$ and constraining the physical state of the CGM-DG. Modern hydrodynamic simulations can support these goals by providing predictions for ion masses and column densities, the physical conditions of the CGM gas, and halo mass and metal retention.

On the theoretical side, there exists a handful of works studying the state of the CGM-DG within various simulations. In an increased resolution run of 6 IllustrisTNG dwarf galaxies, \citet{2024arXiv241216440T} finds the majority of CGM gas for $M_{200c} \sim 10^{10.5}-10^{11.0} M_{\odot}$ galaxies exists at temperatures between $10^3 < \text{T} < 10^4$K which spans a range of densities. This differs from the Auriga simulations for the same halo masses, where the median mass-weighted temperature of CGM gas lies within $10^{4.5} < \text{T} < 10^{5}$K  \citep{Cook_24}. Dwarf galaxy studies from FIRE, like those of IllustrisTNG, find that the CGM-DG roughly follows pressure equilibrium (FIRE: Figure 11 of \citet{DwarfCGM_in_FIRE}, IllustrisTNG: Figure 3 of \citep{2024arXiv241216440T}).

In terms of metal content, simulations appear to show that dwarf galaxies are more efficient at ejecting metals from their disk than Milky Way-mass galaxies \citep[e.g.,][]{Christensen_2016, Muratov2017, Hafen2019, Mina2021}. However, different simulations do not necessarily agree on the fraction of metals retained within the halo and CGM of dwarf galaxies. FIRE \citep{Hafen2019} finds that dwarf galaxies can retain between $30-100\%$ of their metals within all gas inside $R_{200c}$ and $10-60\%$ in their CGM, while \citet{Christensen_2018} finds lower retention when considering all gas within $R_{200c}$ ($30-60\%$) and CGM ($15-25\%$). These differences likely arise from different feedback modeling; however, although both studies span a wide galaxy mass range, they ultimately utilize a relatively small sample of galaxies in the dwarf galaxy regime.

Ultimately, the baryon cycle of dwarf galaxies; the state and metal content of the CGM-DG; the relationship between the CGM-DG and the evolution of the dwarf galaxy; and the bulk properties of the galaxy and halo remain open questions. Furthermore, current theoretical literature investigating these questions is limited by small sample sizes of individual galaxies, so that the galaxy-galaxy scatter in CGM-DG properties remains underconstrained. We address these issues and questions in this work by using a larger sample of isolated, high-resolution, fully cosmological dwarf galaxies to study the metal content and physical structure of the CGM-DG and provide theoretical predictions and estimates for future observations studying the low-mass CGM.


The paper is organized as follows. Section \ref{sec:Simulations} includes details on the simulation suites we use in this work.  Section \ref{subsec:GalObservables} compares our galaxies with observed scaling relations, while Section \ref{subsec:CGMObservables} describes the synthetic column densities generation and agreement with CGM observations. Section \ref{sec:CGMBudgets} presents mass and metal retention rates of the halo ($r<R_{200c}$) and CGM ($0.15R_{200c}<r<R_{200c}$). Section \ref{sec:CGMPhase} details the physical properties (density, temperature, metallicity) and ionization fractions of the CGM and the total mass accessible via UV observations. Section \ref{sec:Discussion} discusses the implications of our results for CGM-DG observations, how this work relates to previous work, and its limitations. We conclude and summarize in Section \ref{sec:summary}.

\section{Simulations}\label{sec:Simulations}

This work uses two suites of simulations, the Marvel-ous Dwarfs and the Marvelous Massive Dwarfs. These are cosmological zoom simulations \citep[e.g.,][]{Onorbe_2013} and are run with the N-Body, smoothed particle hydrodynamics code \changa\, \citep{CHANGA}. Both suites track the evolution of hydrogen, helium, oxygen, iron, and total metallicity. These simulations and the subgrid physics implemented have been shown to reproduce observed scaling relations for stellar-halo mass, metallicity, luminosity, gas content, and size and produce galaxies with dark matter cores \citep{2021ApJ...923...35M,Azartash-Namin24, Ruan2025}. Given their well-tested success, these state-of-the-art simulations are the ideal tools for this study. The subgrid physics that is common between suites is summarized below. The differing specifications of the two suites are discussed in Section \ref{subsec:Marvel} and \ref{subsec:Marvel_Mass}, and the final sample selection and pertinent definitions are presented in Section \ref{subsec:MM}.

\begin{figure}
    \centering
    \includegraphics[width=1.0\linewidth]{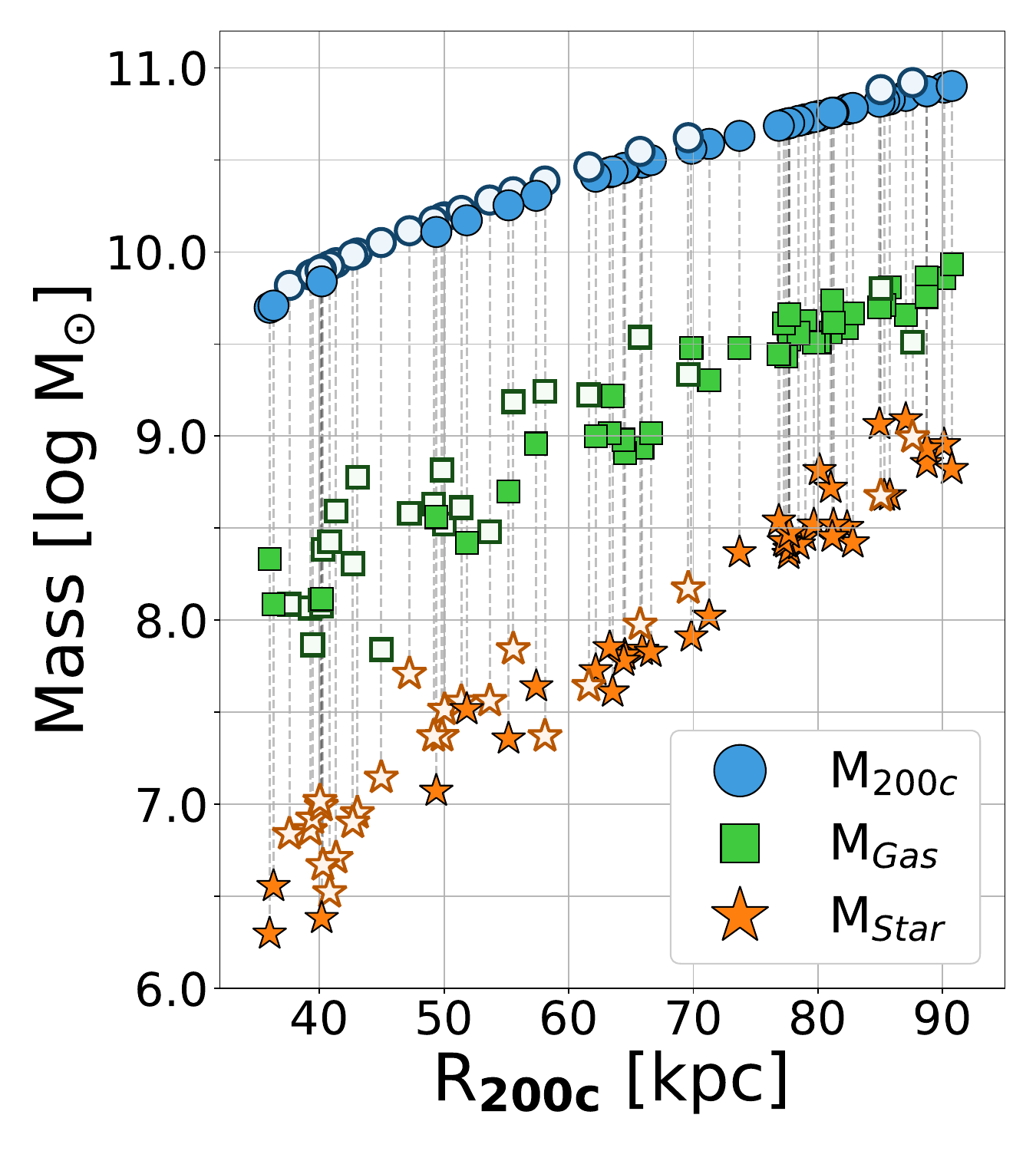}
    \includegraphics[width=1.0\linewidth]{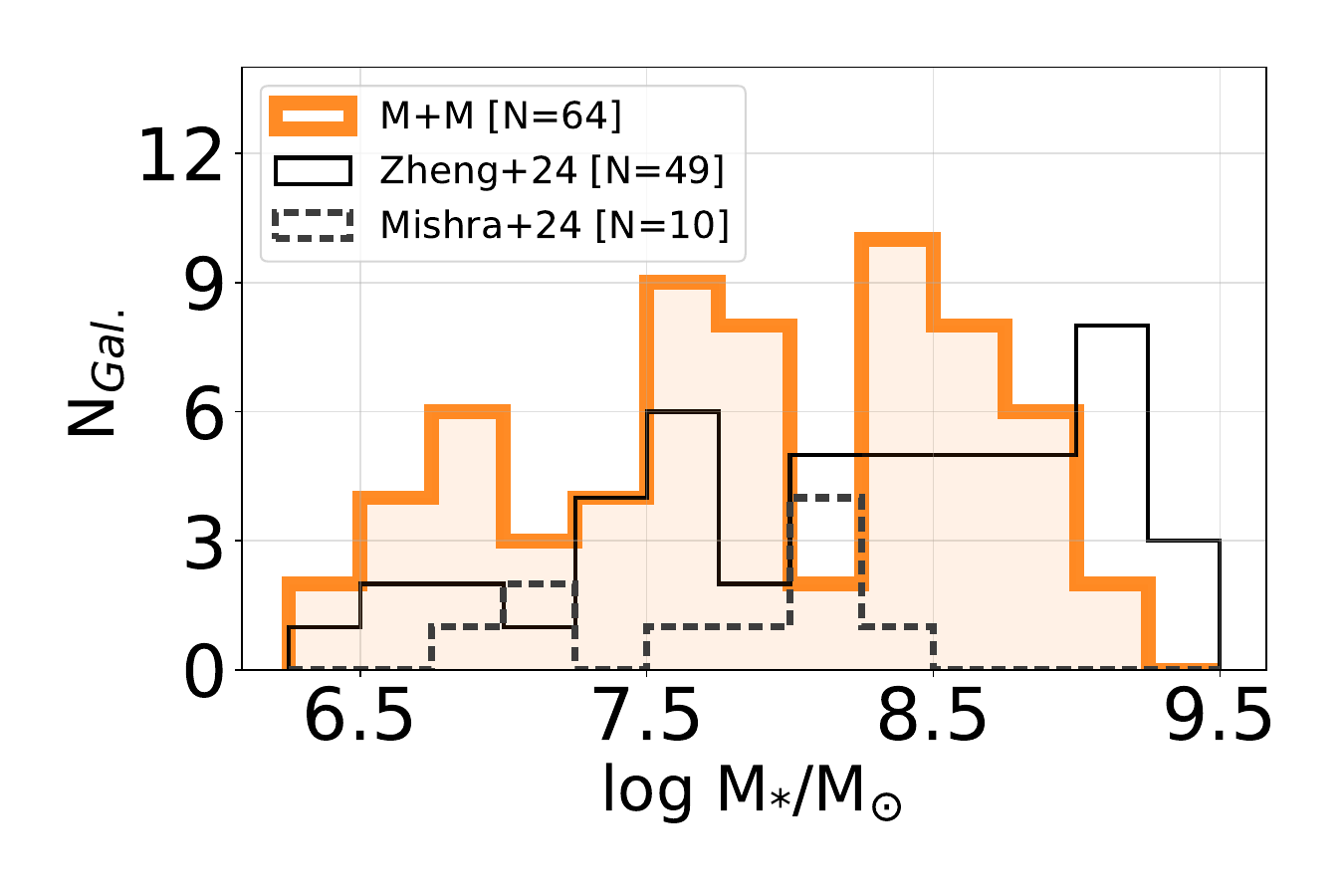}
    \caption{\textbf{Top panel:} Summary of the virial mass (blue circle), total gas mass in the halo (green square), and stellar mass (yellow star) for each galaxy in the Marvel simulation (open-faced markers) and the Marvelous Massive Dwarfs simulations (filled markers). The galaxies in the Marvelous Massive Dwarfs have slightly greater virial masses for a given virial radius than the Marvel galaxies due to the difference in cosmology (see Section \ref{sec:Simulations}).  \textbf{Bottom panel:} Histogram of stellar masses for observed galaxies (\citet{Zheng_24}: solid black line, \citet{Mishra_24} dashed grey line) and the histogram of stellar masses for the \MM\, sample (orange). All simulated stellar masses are 0.6 of the total sum of star particle masses within a halo, as this is a better match to photometric stellar mass determinations \citep{2013ApJ...766...56M}. All galaxies included in \citet{Zheng_24} and the $z<0.3$ galaxies included in \citet{Mishra_24} are shown in this plot.}
    \label{fig:MM-Summary}
\end{figure}

Metal line cooling is implemented using cooling and heating rates calculated before runtime \citep{2010MNRAS.407.1581S}. These rates are determined using the nebular photoionization code \cloudy\, and assume the redshift-dependent UV photoionizing background provided in \citet{HM12}. This UV background is chosen to match the same applied during runtime when modeling photoionization.\footnote{See \citet{Obreja2019} and \citet{Taira2025} for discussions of how the choice of UV background impacts the CGM.} Cooling and heating rates are interpolated during runtime based on the density, temperature, metallicity, and redshift. In addition to metal cooling that follows \cloudy\, tables, the total radiative cooling in our simulations also incorporates cooling/heating from Compton processes and cooling from primordial species (\HI, H$_2$, \HII, \HeI, \HeII, \HeIII). Additionally, both suites include prescriptions for dust, H$_2$ self-shielding, and dust shielding of \HI.

Star formation is only permitted in gas particles that are sufficiently cold ($T<1000$K) and dense ($n
>0.1\, \rm{m}_{\rm{H}}\, \rm{cm}^{-3}$) \citep{2012MNRAS.425.3058C}. Once these criteria are met, star formation then occurs probabilistically. Equation \ref{eqn:SF_prescription} describes the probability, $p$, of gas particles forming stars in a time step $\Delta t$:

\begin{equation}
    p=\frac{m_{gas}}{m_{star}} (1- \text{exp}\biggl[ \frac{-c_{0}^{*} X_{\text{H}_2}\Delta t}{t_{form}} \biggr]), \label{eqn:SF_prescription}
\end{equation}

where $m_{gas}$ and $m_{star}$ are the particle gas mass and initial mass of the potential star particle, respectively. $c_{0}^{*}$ is the star formation efficiency parameter, for which we adopt a value of $c_{0}^{*}=0.1$. $X_{\text{H}_2}$ is the fraction of molecular hydrogen. Finally, $t_{form}$ is the local free-fall time. Once star formation has been triggered, a star particle is created with an initial stellar mass distribution according to the \citet{Kroupa2001_IMF} Initial Mass Function. We note that despite the density threshold for star formation being relatively low/diffuse, star formation usually occurs only in gas particles of $n
>100\, \rm{m}_{\rm{H}}\, \rm{cm}^{-3}$, due to the dependence on $X_{\text{H}_2}$.

When a SN event occurs, thermal energy, mass, and metals are deposited into the surrounding environment. The subsequent evolution of the SN remnant is modeled differently by the two suites, and both implement different total values for the injected thermal energy (see \ref{subsec:Marvel} \& \ref{subsec:Marvel_Mass} for more details). Despite this, we note that throughout this work, we do not find strong differences between these suites and their CGM properties.  We include figures in Appendix \ref{appendix:SuitebySuite} explicitly showing the comparison of column densities between the two suites and differentiate the simulations in plots in the main text where feasible.

Across both suites, all SN Ia events eject $1.4 M_{\odot}$ of mass, including $0.63 M_{\odot}$ of iron, and $0.13 M_{\odot}$ of oxygen \citep[following yields presented in ][]{Thielemann86}, which is deposited into the surrounding gas particles. Stellar winds are also identically modeled in both suites. Mass and metal return via winds is modeled for stars within the stellar mass range of $1-8 M_{\odot}$. The fraction of mass returned to the ISM takes the functional form presented in \citet{Weidemann_87} and is then imparted, along with its fractional metal content, to gas particles in the vicinity of the star particle. The scheme for metal diffusion, which drives the distribution of metals after SN-driven outflows, is also presented in \citet{2010MNRAS.407.1581S}. 

Black hole (BH) formation and feedback are largely common between the two suites, but the numerical values of certain parameters differ and are provided in the following subsections. Generally, BHs are allowed to form from gas particles that must have low metallicity, high density, cool temperature, and low molecular gas content. Additionally, the gas particle must exceed the Jeans Mass criterion. These requirements restrict BH formation to gas that would collapse quickly and cool slowly \citep{Sharma2020, Bellovary21}. The initial mass of the BH differs slightly between suites and is summarized below. Upon formation, the BH particle accretes mass from surrounding gas particles until it depletes the neighboring mass within one softening length or reaches the seed mass. During accretion, thermal feedback from the BH is injected into the nearest neighbor particles at a rate equal to the mass accretion rate. The thermal evolution of the feedback-heated gas particles then follows the SN feedback prescription. 
The BH model has been found to reproduce galaxies that align with the central supermassive black hole mass - bulge stellar dispersion ($M_{BH} - \sigma$), stellar mass - halo mass ($M_* - M_{halo}$), and mass - metallicity scaling relations \citep{Tremmel2017}.  However, initial results that included the Marvel-ous Dwarfs simulations showed that BHs in dwarfs grow little over a Hubble time and do not strongly impact the evolution of the dwarfs \citep{Bellovary21}.  The BH evolution in the Marvelous Massive Dwarfs will be studied in future work, but we assume there is little impact on the CGM from the BH prescription, similar to the Marvel-ous Dwarfs.  


\subsection{Marvel-ous Dwarfs}\label{subsec:Marvel}
The Marvel-ous Dwarfs (Marvel hereafter) simulations are cosmological-zoom simulations presented in \citet{2021ApJ...923...35M} and utilized in \citet{Bellovary21, Christensen24, Azartash-Namin24, Riggs24}.  In total, there are four $25^3\text{ Mpc}^3$ regions (``cptmarvel'', ``rogue'', ``storm'', ``elektra'') encompassed by the Marvel suite, which were simulated with WMAP3 cosmology \citep{2007ApJS..170..377S}. These simulations are run by choosing a region of the universe that contains several dwarf galaxies and rerunning this zoomed region with higher resolution.  For each region, the most massive galaxies are $M_{*} \sim 10^{9.3} M_{\odot}$ while the smallest galaxies are $M_{*} \sim 3000 M_{\odot}$ ultra-faint dwarf galaxies. All galaxies are 1.5-7 Mpc away from a Milky Way-mass galaxy \citep{Christensen24}. Marvel has a force resolution of 60 pc, gas particle mass resolution of 1410 $M_{\odot}$, initial star particle mass resolution of 420 $M_{\odot}$, and dark matter particle mass resolution of 6650 $M_{\odot}$.

Stellar feedback from SNII follows the ``blastwave'' (BW) supernova feedback model \citep{Stinson06-BW}. Only massive stars of $8-40 M_{\odot}$ produce SNII. When this occurs, thermal energy ($E_{BW}=1.5\times10^{51}$erg) is deposited into the gas particles within the radius of a SNe blastwave according to \citet{Chevalier1974}, as modeled by Equation 9 in \citet{Stinson06-BW}. In addition to thermal energy, mass and metals (oxygen and iron) are also deposited, with total mass values according to yields derived in \citet{Raiteri96}. After the initial thermal dump, the adiabatic or Sedov phase then begins, which is the period where the blastwave expands but is not able to cool efficiently due to long cooling timescales. To simulate this inefficiency of cooling and prevent numerical overcooling, the BW model disables cooling in the SNe-heated gas particles during the Sedov phase. Subsequently, the blastwave enters the snowplow phase, where momentum is conserved, and cooling is re-enabled.

BH formation proceeds in gas particles that satisfy the following conditions: low metallicity (log$Z<{-4}$), high density ($1.5\times10^4$ cm$^{-3}$), low temperature ($T< 2\times 10^4$ K), and low molecular gas content ($f_{\text{H}_2}<10^{-4}$). 
The resulting BH has an initial mass of $5\times10^4 M_{\odot}$, and BH feedback follows the BW SN feedback model.


\subsection{Marvelous Massive Dwarfs}\label{subsec:Marvel_Mass}

The Marvelous Massive Dwarfs are cosmological-zoom simulations of dwarfs selected from the Romulus25 cosmological volume simulation \citep{Tremmel2017} and utilized in \citet{Keith2025} and \citet{Ruan2025}.  The Massive Dwarfs are run within a Planck cosmology \citep{2016A&A...594A..13P} and initial conditions were generated using \textsf{GenetIC} \citep{GenetIC2021}. The simulations have a force resolution of 87 pc, gas particle mass resolution of 3300 $M_{\odot}$, initial star particle mass resolution of 994 $M_{\odot}$, and dark matter particle mass resolution of 17,900 $M_{\odot}$.

Stellar feedback follows the ``superbubble'' (SB) supernova feedback model \citep{Keller14-SB}. This model is motivated by the spatial and temporal clustering of star formation and, consequently, stellar feedback, which leads to ``superbubbles'' of feedback rather than isolated events. The SB model is similar to the BW model in that it deposits thermal energy ($E_{SB}=1.0\times10^{51}$erg) 
and the same total mass and metals into the surrounding medium. However, the SB model deposits this energy into 1 neighboring gas particle, and the SNe energy then diffuses into the surrounding gas particles according to thermal conduction and evaporation subgrid models \citep{CowieMcKee_77, Keller14-SB}. To prevent numerical overcooling, SB models SNe-heated particles as multiphase fluid elements or ``two-phase particles,'' if the mass of the particles is not fully heated. These two-phase particles model two temperature phases, a hot and cold phase, which remain in pressure equilibrium with one another and have individual masses, densities, and temperatures. This method aims to simulate the SNe heated gas (hot phase) that sweeps up ISM material (cold phase) seen in ``superbubbles.'' Mass exchange from the cold to hot phase is calculated via the subgrid thermal evaporation model. Once the cold mass has been fully evaporated or if the hot phase cools below $10^5\text{K}$, the two-phase particle returns to the single-phase state. For analysis in this work, we average the temperature and density of any two-phase particles to treat them as single-phase particles. We find these two-phase particles account for $\sim5\%$ of the CGM by mass and $\sim10\%$ of the CGM by metal mass, yet these particles tend to have little effect on column densities and temperature budgets.

BH formation proceeds in gas particles that satisfy the following conditions: low metallicity (log$Z<{-5}$), high density ($1.5\times10^4$ cm$^{-3}$), cool temperature ($T< 5\times 10^3$ K), and low molecular gas content ($f_{\text{H}_2}<2\times10^{-3}$).
The resulting BH has an initial mass of $1.2\times10^{5} M_{\odot}$ and BH feedback follows the SB feedback model.

\subsection{The \MM\, Sample}\label{subsec:MM}
Using the Marvel and Marvelous Massive Dwarfs simulations, star-forming (specific star formation rates $\geq 10^{-11}$ in the 100 Myr prior to $z=0$) dwarf galaxies were selected to match the stellar mass range of $6.0\leq \text{log}(M_*/M_{\odot}) \leq 9.5$ used in \citet{Zheng_24}. The $z = 0$ snapshots are used to correspond to the low-redshift sample presented in \citet{Zheng_24}. Additionally, only dwarf galaxies that are at least 200 kpc away from the center of mass of a $M_* > 10^6$ M$_{\odot}$ galaxy were chosen to avoid contamination from other halos in synthetic column densities and global CGM properties. With these criteria, 24 dwarf galaxies were selected from the Marvel suite, and 40 dwarf galaxies were selected from the Marvelous Massive Dwarfs suite to comprise the Marvel+Marvelous Massive Dwarfs (``\MM'') sample. 

A summary of stellar mass ($M_*$), virial radius ($R_{200c}$), virial mass ($M_{vir}$), and total halo gas mass ($M_{gas}$) is shown in the top panel of Figure \ref{fig:MM-Summary} and a comparison of stellar masses between the \MM\, sample and \citet{Zheng_24} is shown in the bottom panel. A sample of $z<0.3$ galaxies observed in \citet{Mishra_24} is also included and used for comparison throughout this work. We find adequate agreement between the stellar mass distributions of the full simulated sample and the total observed sample. In this work, the virial radius of a galaxy is taken as the radius within which the average density is 200 times the critical density of the Universe ($R_{200c}$), which is lower by a factor of $0.6$ than the matter density definition ($R_{200m}$).

In the \MM\, sample, we define the CGM as all material within $(0.15 - 1)R_{200c}$. This boundary was chosen to select particles residing in the extended CGM and remove those within the disk of the galaxy. This method of defining the CGM as a fixed fraction of the virial radius is similar to many other studies in the field \citep[e.g.][]{ Hafen2019, DwarfCGM_in_FIRE, Cook_24}. We note, however, that with decreasing halo mass, the size of the disk relative to the halo shrinks. Consequently, at the low-mass end, we exclude mass that an observer would likely consider the CGM. We opt to adopt a standardized definition across masses for simplicity and clarity, but discuss the repercussions of this in Appendix \ref{appendix:CGMDef}.

\begin{figure*}[t]
    \centering
    \includegraphics[width=0.9\linewidth]{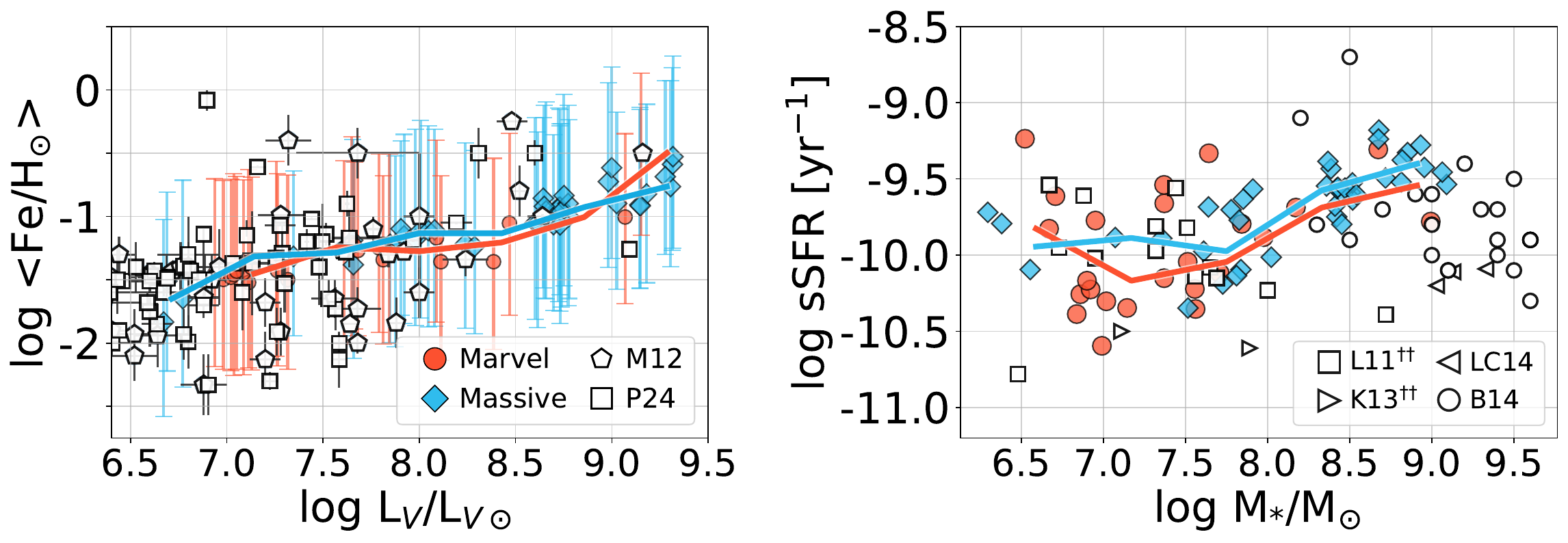}
    \caption{Luminosity - Metallicity relation (Fe/H vs $L_V$, \textbf{left}) and galaxy specific star formation rates (sSFR vs $M_*$, \textbf{right}). Red circles represent galaxies from the Marvel suite, and blue diamond markers represent the Marvelous Massive suite; median values of both suites are shown as solid lines. \textbf{Left panel:} Shown is the logarithm of the mean stellar metallicity relative to solar abundance. Error bars on simulated points show the 16th-84th percentile of the simulated stellar metallicity distribution. Open-face pentagon markers are observational data from \citet{McConnachie12}, and open-face square markers are observational data from the Local Volume Database \citep{Pace2024LVDB}. \textbf{Right panel:} sSFR values for the \MM\, sample are calculated over a timescale of 100 Myr. Open-face markers are observations that derive SFRs as follows: \citet{Bordoloi_14} uses detected nebular emission lines, and \citet{LiangChen_14} uses rest-frame UV absolute magnitude. \citet{Karachentsev2013} and \citet{Lee2011} values utilize far UV magnitudes and are corrected for Galactic extinction in \citet{Zheng_24} following the methods presented in their Appendix. We find adequate agreement between the \MM\, sample and observations, as well as similar behavior across suites. }
    \label{fig:ScalingRelations}
\end{figure*}

\section{Comparison to Observations}\label{sec:CGMObservables}
In this section, we verify that our galaxies match key empirical scaling relations, as well as observed CGM column densities, before using them to interpret the CGM of observed dwarf galaxies.

\subsection{Galaxy Observables}\label{subsec:GalObservables}

Scaling relations between physical properties of galaxies have been used to empirically describe galaxy evolution \citep[e.g.][]{2020ApJ...891..181M,2023ApJ...951..138L}. For example, the luminosity-metallicity relation provides insight into the chemical evolution of a galaxy since the present-day metal content of a galaxy arises from the combined effects of star formation, feedback-driven metal loss, and accretion of pristine gas \citep{2007ApJ...658..941D, 2020ApJ...891..181M}.

The left panel of Figure \ref{fig:ScalingRelations} shows the mean luminosity-metallicity relation for the \MM\, sample (red and blue markers) compared to Local Group dwarf galaxy observations \citep[open grey circles;][]{McConnachie12} and the Local Volume Database \citep[open grey diamonds;][]{Pace2024LVDB}\footnote{In the case of duplicate measurements between \citet{McConnachie12} and \citet{Pace2024LVDB}, we choose to show the measurements from \citet{McConnachie12}.}. For each simulated galaxy, we represent the 16th-84th percentile of stellar metallicities as error bars, and plot the median values of both suites as solid lines. We find the stellar metallicities of the \MM\, sample reproduces the majority of observations within the simulated errorbars. Further we observe no systematic difference between suites, despite limited observations on the higher mass dwarf end. The agreement between simulations and observations confirms that realistic masses of metals are retained within the stellar populations. 

The galaxy specific star formation rate-stellar mass relation \cite[e.g.][]{2007ApJS..173..293W} provides observational constraints on specific star formation rates (sSFR) for a given galaxy stellar mass ($M_*$). The right panel of Figure \ref{fig:ScalingRelations} presents sSFR values for the \MM\, sample and observations (open-faced markers). We find a substantial degree of scatter ($\sim1.5$ dex) for $\text{log}(M_*/M_{\odot}) < 8.0$ in the simulations, which agrees with the scatter in observations in this mass range. While there is a similar degree of scatter for $\text{log}(M_*/M_{\odot}) > 8.0$, the simulations tend to produce galaxies biased towards greater sSFR. However, across stellar masses, the \MM\, sample exhibits sSFRs that agree with observational values. Across suites, we find both produce galaxy populations ranging in similar values of sSFR. We also note that the bias towards greater sSFR for $\text{log}(M_*/M_{\odot}) > 8.0$ appears to be present in both suites. Given the agreement between simulations and observations in Figure \ref{fig:ScalingRelations}, we find that the \MM\, sample contains galaxies with realistic star formation rates for their stellar masses. 

The agreement shown in Figure \ref{fig:ScalingRelations} indicates our simulations produce galaxies with realistic star formation rates and metal contents, thereby motivating the choice to use these simulations to study the CGM-DG.  

\subsection{CGM Observables -- Column Densities}\label{subsec:CGMObservables}

\begin{figure*}
    \centering
    \includegraphics[width=0.335\textwidth]{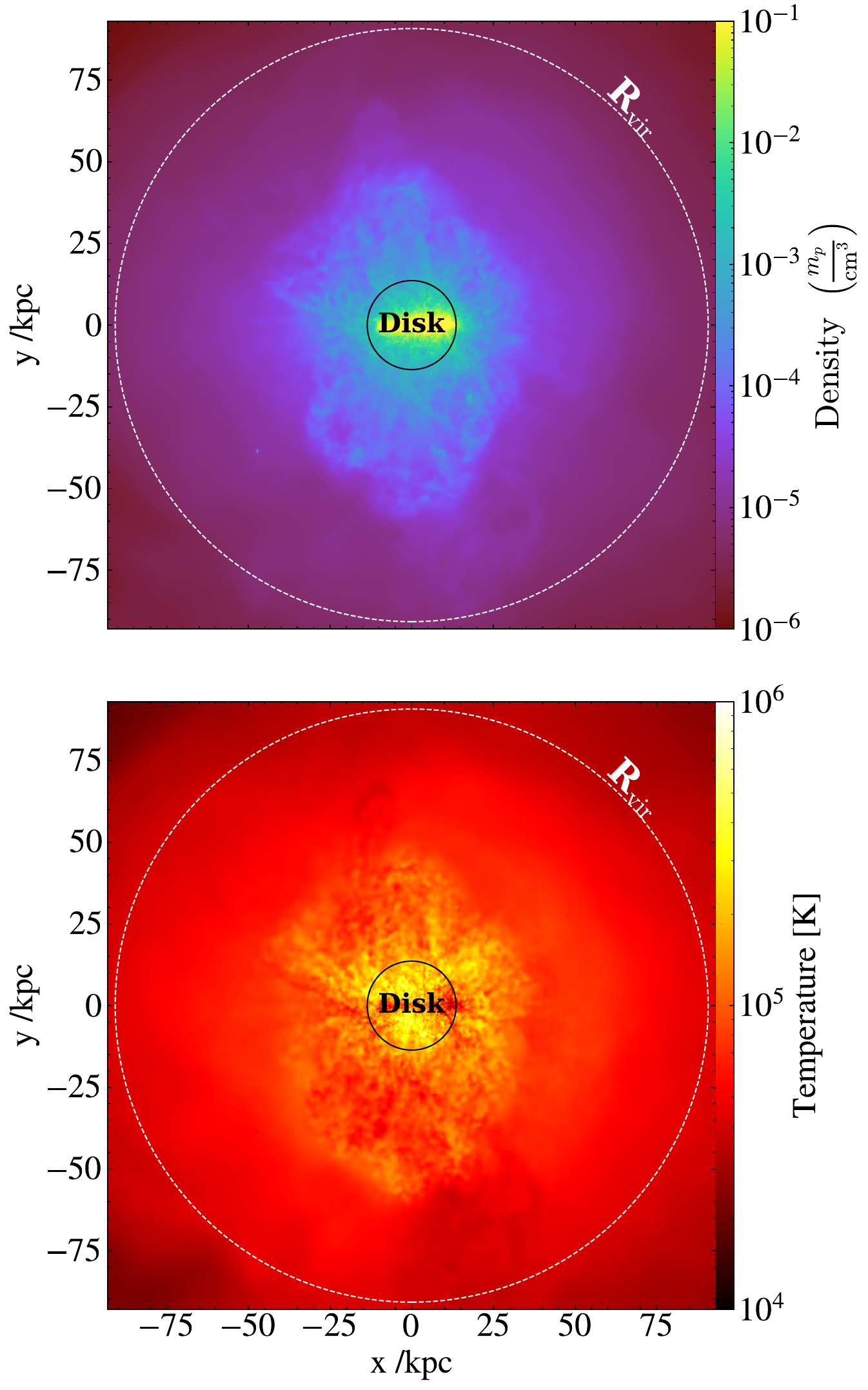}
    \includegraphics[width=0.6\textwidth]{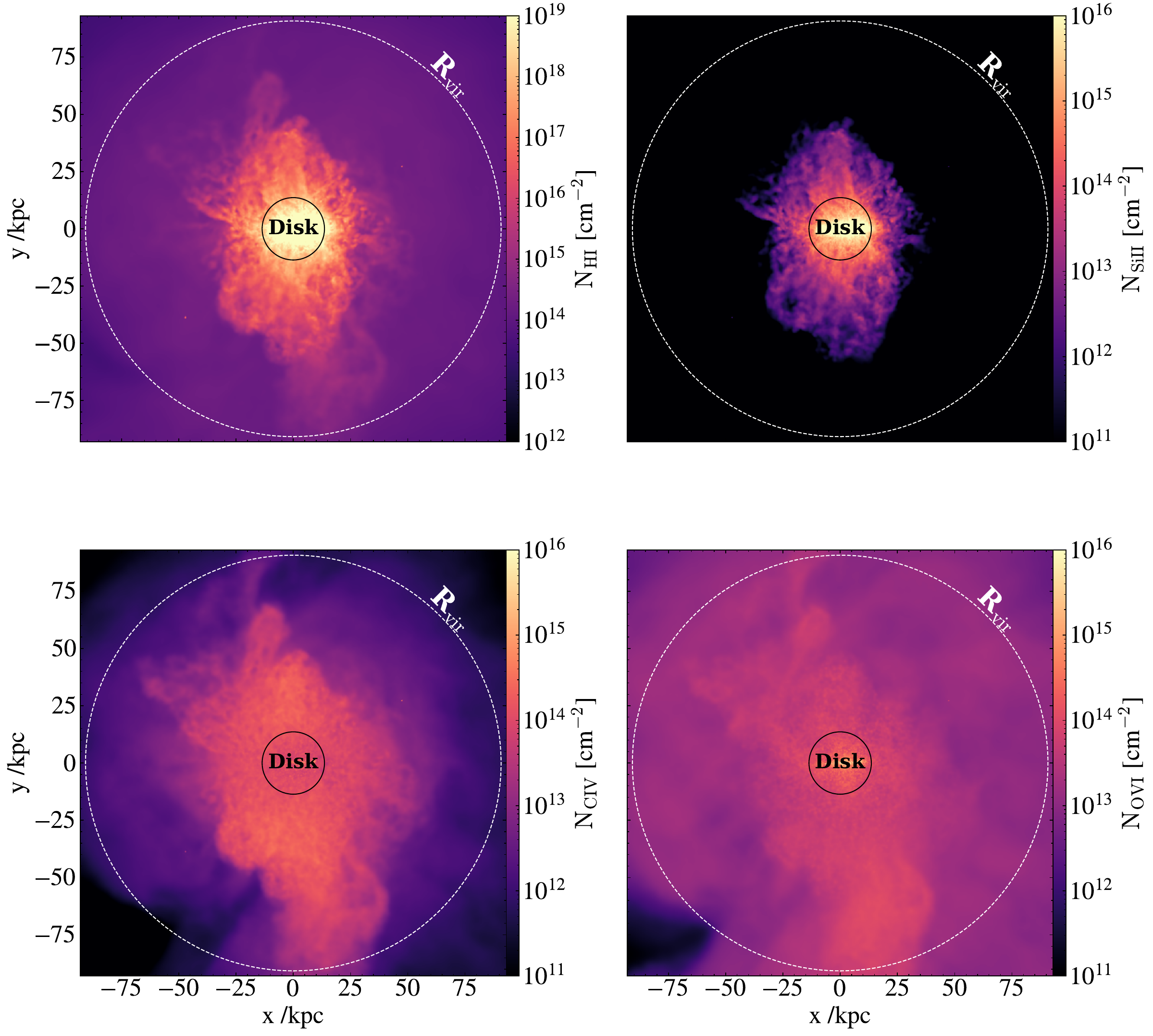}
    \caption{Projected maps of CGM gas density (upper left), gas temperature (lower left), \HI\, column density (upper center), \SiII\, column density (upper right), \CIV\, column density (lower center), and \OVI\, column density (lower right) using \textsf{TRIDENT} \citep{Trident} and \textsf{yt} \citep{yt}. Maps are made for a typical galaxy in the \MM\, sample (a log$ M_*/M_{\odot}=8.82\,$ galaxy from the Marvelous Massive suite). In each panel, we show the $R_{200c}$ as the dashed white line and $0.15R_{200c}$ as the solid black line. The galaxy disk is oriented in the panel such that the disk is edge-on.}
    \label{fig:N_maps}
\end{figure*}

\begin{figure*}
    \centering
    \includegraphics[width=0.95\linewidth]{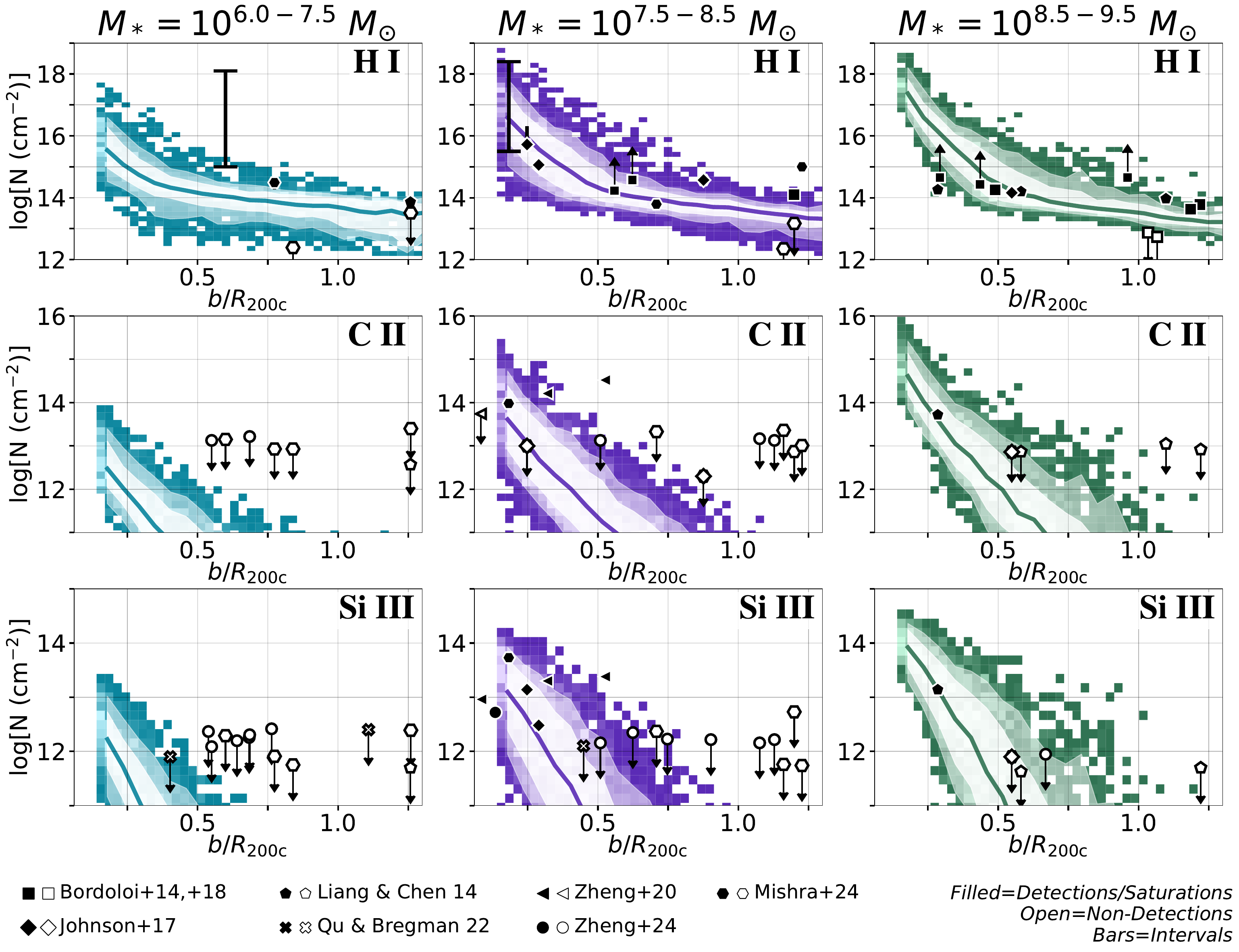}
    \caption{Adapted Figure 4 from \citet{Zheng_24} showing column densities (N) for low ions (\HI, \CII, and \SiIII) as a function of normalized impact parameter ($b/R_{200c}$). The column density results from the \MM\, sample are shown as 2-D probability density histograms with the brightest cells showing a greater probability at a given $b/R_{200c}$. The median of all simulated columns as a function of $b/R_{200c}$ is shown as a solid line with two shaded regions representing the 16th-84th ($1\sigma$) and 5th-95th percentiles ($2\sigma$). Observed column densities are shown as points where each marker is attributed to a certain observational paper (legend shown at the bottom of the Figure). Open markers are non-detections and are treated as upper limits, while filled markers are detections; filled markers with upward arrows are saturations and are treated as lower limits. Black bars are intervals provided in \citet{Mishra_24} that show the observationally constrained range (upper and lower limits) of column densities for saturated systems. Each column shows all simulated and observational data for the CGM around galaxies within a given stellar mass bin (denoted at the top). From left to right, the three stellar mass bins (and the number of simulated galaxies in each bin) are as follows: $M_* = 10^{6.0-7.5} ~M_{{\odot}}$ (N=19), $M_* = 10^{7.5-8.5} ~M_{{\odot}}$ (N=29), and $M_* = 10^{8.5-9.5} ~M_{{\odot}}$ (N=16). We find the \MM\, sample reproduces the majority of existing column density observations for the low ions included in this work (see also Table \ref{tab:N_stats}).}
    \label{fig:LowIonColumnDensitiesResults}
\end{figure*}
Figure \ref{fig:N_maps} shows projected column density maps in \HI, \SiII, \CIV, and \OVI\, of a representative galaxy in the \MM\, sample (log $ M_*/M_{\odot}=8.82\,$ galaxy from the Marvelous Massive suite). We find \HI, \SiII, and \CIV\, column densities are greatest closest to the galaxy and decrease with increasing distance from the galaxy. \SiII\, columns tend to be largest only in the inner CGM and very low in the outer CGM. \HI\, columns are also largest in the inner CGM; however, in the outer regions \HI\, tends to have a uniform \NHI$\sim 10^{14}\text{cm}^{-2}$ coverage. \OVI\, column densities tend not to vary strongly with distance from the galaxy. Furthermore, we find that column densities in species of low ionization potential (\HI\, and \SiII) tend to trace denser and cooler structures in the inner CGM, while species of intermediate and high ionization potential (\CIV\, and \OVI) exhibit column densities that trace a volume-filling medium. This qualitative behavior is generally in agreement with other simulations \citep[e.g.,][]{Cook_24}.

Column densities for the \MM\, sample were calculated by generating 325 lines of sight around each galaxy. Each sightline is 0.25 Mpc in length to avoid intersecting another galaxy's CGM while still probing the surrounding environment of the dwarf galaxy. Additionally, each sightline is randomly oriented with respect to the disk and centered at a random impact parameter, $b$, from the disk. This methodology allows us to randomly sample each galaxy in the \MM\, sample at a range of projected distances and inclination angles. To fully sample the CGM and avoid intersecting the disk of the galaxy, the sightlines are generated within $b/R_{200c} = [0.15,1.3]$. The physical length scales for the virial radii in the \MM\, sample are shown in Figure \ref{fig:MM-Summary}. 

Sightline positions and path length elements are generated using \textsf{TRIDENT} \citep{Trident}, which relies on the simulation analysis tool \textsf{YT} \citep{yt}. As aforementioned, the \MM\ sample simulations track hydrogen, oxygen, and iron throughout runtime by tracking the mass fraction of each element for each gas particle (the same method as used in \citealp{Mina2021}). The abundances of other elements are added in during the analysis phase by using the relative abundances of a tracked element and a non-tracked element from \citet{Asplund2009} (i.e., $f_C = f_O \times (A_C/A_O)$). Ion fractions for each element are generated using \cloudy\, \citep{Ferland17} and added to each particle in the simulation during post-processing based on the particle's density, temperature\footnote{\changa\, converts particle internal energy into temperature via the ideal gas law. The mean molecular weight is computed from non-equilibrium abundances of \HI\, and \HII, and \HeI, \HeII, and \HeIII, with metals contributing an additional term that scales with metallicity ($\mu_Z=17.6003$). Metal cooling is also tabulated, see further details in \citet{2010MNRAS.407.1581S} and Section \ref{sec:Simulations}} The same redshift-dependent photoionizing UV background from \citet{HM12} that is implemented in the simulations is used to generate ion fractions as well. We assume that the ionizing radiation from the central galaxy is negligible based on the low star formation rates seen in the \MM\ sample (Figure \ref{fig:ScalingRelations}); however, see Section \ref{subsec:Discussion_Limitations} for a discussion on how this may affect our results. Column densities are then calculated by summing the product of the ion number density and the path length element across the line of sight.  

To compare to observations, we present results for the ions \HI, \CII, \CIV, \SiIII, \SiIV, and \OVI. We choose to omit \SiII\, in this section, as it shows similar trends to \CII; however, we include it in the Appendix \ref{appendix:SuitebySuite} (along with figures showing the column density results differentiated by simulation suite).

\begin{figure*}
    \centering
    \includegraphics[width=0.95\linewidth]{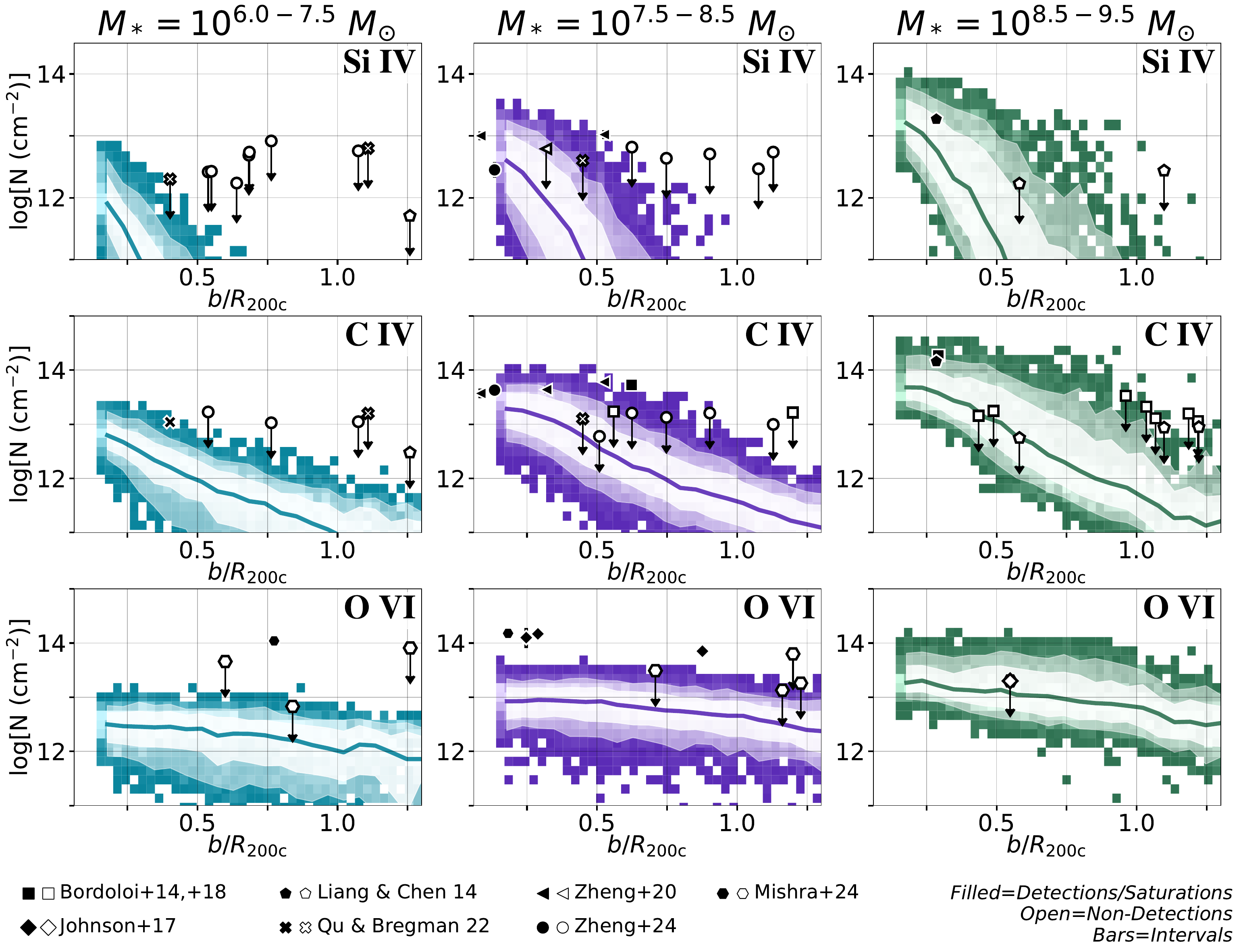}
    \caption{Same as Figure \ref{fig:LowIonColumnDensitiesResults} (adapted Figure 4 from \citet{Zheng_24}) now showing column densities for higher ions (\SiIV, \CIV, and \OVI). We find the \MM\, sample reproduces the majority of existing \SiIV\, column density observations, while the simulations tend to underpredict \CIV\, and \OVI\, column densities (see Section \ref{subsec:Discussion_OVI} for further discussion).} 
    \label{fig:HighIonColumnDensitiesResults}
\end{figure*}

Figure \ref{fig:LowIonColumnDensitiesResults} shows the simulated column densities for \HI, \CII, and \SiIII\, as a function of normalized impact parameter, $b /R_{200c}$. Each figure column is the 2-D probability histogram for all simulated galaxies within a given stellar mass bin (denoted at the top of each column), and each row shows the results for a given ion. The solid black line represents the median column density for the \MM\, sample. Observational data from \citet{Zheng_24} and a sub-sample of \citet{Mishra_24} (described in Section \ref{sec:Simulations}) are shown as points on each figure. Black points represent detections in a given ion, while white points with a downward-pointing arrow are non-detections and are treated as upper limits. Black points with an upward pointing arrow are saturations and are treated as lower limits. Black bars are intervals provided in \citet{Mishra_24} that show the upper and lower limits for saturated systems. The marker style of each point signifies the publication of the observation (legend located at the bottom of the figure), and the key follows the same style as Figure 4 from \citet{Zheng_24} for comparison sake.\footnote{ Key differences between Figures \ref{fig:LowIonColumnDensitiesResults} and \ref{fig:HighIonColumnDensitiesResults} and Figure 4 of \citet{Zheng_24}: \citet{Zheng_24} utilizes the virial definition of 200 times the matter density of the Universe. As we use 200 times the critical density of the Universe, we have shifted the observed points to $b/R_{\bf{200c}}$ and sorted the observations by stellar mass.} Table \ref{tab:N_stats} quantifies the agreement with observations by presenting the number of detections encompassed by 1 and 2 standard deviations ($\sigma$) from the total \MM\, sample median (black lines in Figures \ref{fig:LowIonColumnDensitiesResults} and \ref{fig:HighIonColumnDensitiesResults}).

\textit{HI Column Densities:} We find broad agreement between the \MM\, sample and observations of \HI\, column densities (\NHI). For $\text{log}(M_*/M_{\odot}) > 7.5$ galaxies, the synthetic \NHI\, distributions encompass $70\%$ (10/14) of observations within $2\sigma$. For $\text{log}(M_*/M_{\odot}) < 7.5$ galaxies, the \MM\, sample reproduce all observations (2/2) within $2\sigma$, however, there are only two detections available. Further, there are no \NHI\, observations of the inner CGM of $6.0< \text{log}(M_*/M_{\odot}) \leq 7.5$ galaxies that meet our selection criteria.

\NHI\, values fall off steeply in the inner CGM $b/R_{200c}<0.5 $, a feature also seen in Figure \ref{fig:N_maps}. Currently, only observations of $\text{log}(M_*/M_{\odot}) > 7.5$ galaxies probe this inner region of the CGM, and there is a mixture of detections and lower limits. Simulated galaxies of the same mass exhibit inner \NHI\, profiles generally above these observations, which agrees with the published lower limits while having the detections on the edge of the distribution. We find that with decreasing galaxy mass, the inner \NHI\, profile becomes less steep, a feature discussed further in Section \ref{sec:Discussion}. For $b/R_{200c} >0.5 $, the \NHI\, profiles become much flatter and plateau at \NHI$\sim 10^{14.0} \text{cm}^{-2}$ \citep[See also][]{Mina2021,Cook_24} and Figure \ref{fig:N_maps}. Interestingly, the \NHI\, value that the profiles plateau to shows little dependence on host galaxy mass, a physical interpretation for this is also provided in Section \ref{sec:Discussion}. 

\textit{Low Metal Ion Column Densities:} We find that all metal ion column densities fall off with increasing impact parameter, and median values decrease with decreasing host stellar mass. Low ions (\CII, \SiII, \SiIII) fall off steeply (also seen in Figure \ref{fig:N_maps}), dropping below detectable limits by $0.5 R_{200c}$. Within this inner CGM, the \MM\, sample reproduces 2/4 existing \CII\, detections and 7/8 \SiIII\, detections within $2\sigma$. Beyond $0.5 R_{200c}$, these low ions have fallen well below detectable limits and match the published upper limits. Thus the \MM\, sample shows similar trends to observations, and the majority of detections are within $2\sigma$.

\textit{Intermediate Metal Ion Column Densities:} Figure \ref{fig:HighIonColumnDensitiesResults} is the same as Figure \ref{fig:LowIonColumnDensitiesResults} but for ions of higher ionization potential---\SiIV, \CIV, and \OVI. For these ions, column densities decrease with increasing impact parameters. This decline in column density is shallower than the low ions while also falling below detection limits by $0.5 R_{200c}$, agreeing with published upper limits. Within $0.5 R_{200c}$, \SiIV\, column densities show broad agreement with observations, with 3/4 of detections within $2\sigma$. However, \CIV\, column densities show weaker agreement with observations. With existing detections, 5/8 detections lie within $2\sigma$ and 0/8 detections lie within $1\sigma$. This likely indicates the \MM\, sample underpredicts \CIV\, and we present a brief analysis on what may be driving this lack of \CIV\, in Section \ref{subsec:Discussion_OVI}.

\textit{High Metal Ion Column Densities:} Column densities for \OVI\, (\NOVI) are provided in Figure \ref{fig:HighIonColumnDensitiesResults} and are compared with observations from \citet{Johnson_17} and \citet{Mishra_24}. No observations from \citet{Zheng_24} are shown because the \OVI\, $\lambda\lambda1032,1038$\AA\AA\, transition is not observable at $z\sim0$. Thus, we note that the \NOVI\, observations shown do have a median redshift of $z\sim0.15$ and are, therefore, at systematically higher redshifts than the \MM\ sample. Despite this difference in redshift, we find the \MM\, sample produces \NOVI\, values below all observed detections by $\sim1-1.5$dex and near the observational threshold of $N\sim10^{13} \text{cm}^{-2}$. Further, the \MM\, sample reproduces 0/6 observations within $2\sigma$. Thus, the \MM\, sample underestimates \NOVI\, compared to observations. A brief investigation into \OVI\, at higher redshifts and an interpretation for this underprediction are provided in Section \ref{subsec:Discussion_OVI}.

\textit{Column Density Scatter:} At a given impact parameter, the distributions shown in Figures \ref{fig:LowIonColumnDensitiesResults} and  \ref{fig:HighIonColumnDensitiesResults} can range $>1$ dex in column densities. Given the number of galaxies in the \MM\, sample, we can quantify the scatter in column densities for a single CGM  and the galaxy-galaxy scatter to assess which is driving this range. For \NHI, \NCIV, and \NOVI, the difference between the 16th and 84th percentiles in a bin size of $0.02\,b/R_{200c}$ is on the order of $0.5$ dex, while the spread in \NCII, \NSiII, \NSiIII, and \NSiIV\, tends to be larger with typical values between $0.6-1$ dex. While this scatter is non-negligible, we see a greater spread when considering galaxy-galaxy scatter. In particular, median column density profiles for individual galaxies tend to range $0.5-2$ dex. These findings can be seen in the Appendix (Figures \ref{fig:Appendix_N_I} and \ref{fig:Appendix_N_II}) where we provide the median column density profiles and their 16th and 84th percentiles for each galaxy in the \MM\, sample. Ultimately, there is a larger effect on the scatter in column densities from including a large number of galaxies in the \MM\, sample, however, the spread in column densities of a single CGM also increases this scatter.

\begin{table}
    \centering
    \begin{tabular}{c|c|c|c}
      & \textbf{Low-$\mathbf{M_*}$}  & \textbf{Mid-$\mathbf{M_*}$}       & \textbf{High-$\mathbf{M_*}$}   \\
    & $10^{6.0-7.5} ~M_{{\odot}}$ & $10^{7.5-8.5} ~M_{{\odot}}$    & $10^{8.5-9.5} ~M_{{\odot}}$ \\
    & 1$\sigma$,2$\sigma$,(All)     & 1$\sigma$,2$\sigma$,(All)      & 1$\sigma$,2$\sigma$,(All)  \\

             \hline
        \HI     & 1, 2, (2) & 3, 4, (7) & 4, 6, (7)   \\
         \CII   & 0, 0, (0) & 1, 1, (3)   &  1, 1, (1) \\
         \SiIII    & 0, 0, (0) & 4, 6, (7)  & 1, 1, (1)\\
         \SiIV   & 0, 0, (0) & 2, 2, (3)  & 1, 1, (1)\\
        \CIV    & 0, 0, (1) & 0, 3, (5)  & 0, 2, (2) \\
        \OVI    & 0, 0, (1)  & 0, 0, (5)  & 0, 0, (0)\\

    \end{tabular}
    \caption{Summary of the number of detections within $1\sigma$ and $2\sigma$ of the simulated distribution medians relative to the total number of observed detections for a given ion around a given galaxy mass. Each entry consists of 3 numbers and is ordered as: detections within $1\sigma$, detections within $2\sigma$, (total number of observed detections).}
    \label{tab:N_stats}
\end{table}

\section{CGM Budget}\label{sec:CGMBudgets}
Motivated by the agreement in most ions between existing observations and the \MM\, sample, the following sections aim to contextualize the column densities and their profiles based on the properties of the CGM. The lack of low ion detections seen observationally and small low ion column densities provided in Section \ref{sec:CGMObservables} likely indicate either a) dwarf galaxies do not effectively retain metals/mass in the outer CGM, or b) there is a prominent warmer phase where ion fractions for low ions are small. To differentiate between these possibilities, the following sections present the halo mass retention (Section \ref{subsec:MassBudget}) and metal retention (Section \ref{subsec:MetalBudget}). In Section \ref{subsec:PhaseBudgets}, we assess how the mass and metal retention rates vary by CGM gas phase. 

\subsection{Mass Budget}\label{subsec:MassBudget}
The halo mass retention can be described by the contributions of all baryons to the baryonic budget.  The baryonic budget provides insight into the fraction of baryons present in a given halo with respect to the ``expected'' or cosmological abundance of baryons. The cosmological baryonic budget ($f_{b,Cosmo}$) is defined as

\begin{equation}\label{eqn:f_bCosmo}
    f_{b,Cosmo}=\frac{m}{M_{vir} \cdot \Omega_{b}/\Omega_{M}},
\end{equation}

\noindent where the denominator represents the ``expected'' abundance of baryons for a given halo mass based on the cosmological baryon ($\Omega_{b}$) and matter density ($\Omega_{M}$). The numerator ($m$) represents any mass of interest (e.g., CGM mass, ISM mass, etc.). Deviations from a baryonic budget of unity suggest mass loss from the halo via feedback \citep[e.g.,][]{2009ASPC..419..347D}.

\begin{figure}[t]
    \centering
    \includegraphics[width=8.5cm]{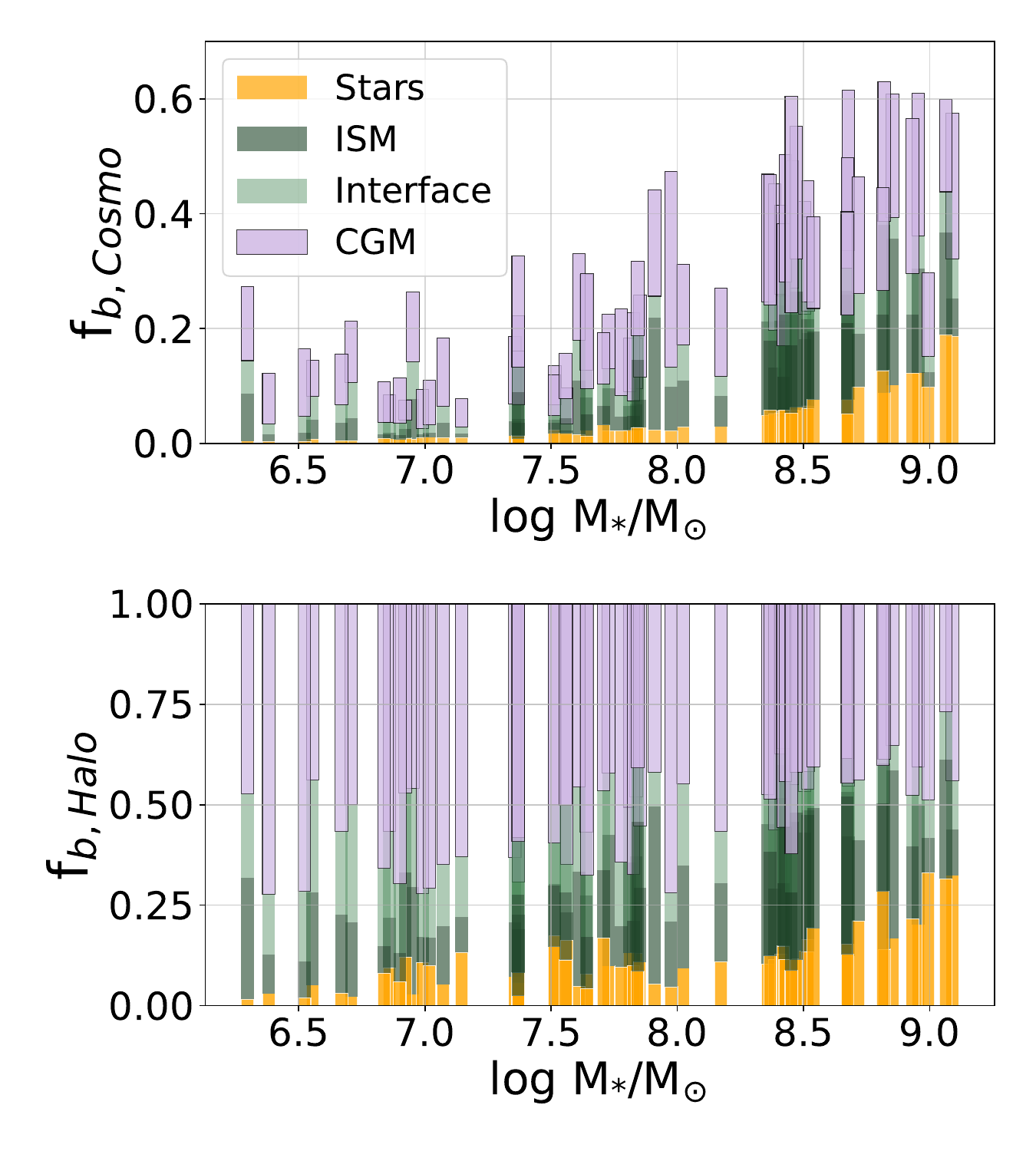}
    \caption{Contributions to the cosmic baryonic budget (\textbf{top panel}, defined as $f_{b,cosmo}=M/(\Omega_b/\Omega_m M_{200c})$) and halo baryonic budget (\textbf{bottom panel}, defined as $f_{b,halo}=M/M_{b}(<R_{200c})$) from the CGM (uppermost bar, purple) beyond $15\% \,R_{200c}$, the disk-CGM interface (light green), the \rHI\, selected ISM (dark green), and stars (bottommost bar, yellow). With decreasing stellar mass, dwarf galaxies show a declining cosmic baryonic budget, yet, for all masses, the CGM represents a significant ($48\%\pm11\%$) reservoir of halo baryons.}
    \label{fig:BaryonicBudget}
\end{figure}

The upper panel of Figure \ref{fig:BaryonicBudget} shows $f_{b,Cosmo}$ for each galaxy in the \MM\, sample with respect to stellar mass. For a given galaxy, we divide the total baryonic budget into its individual contributions from the stellar component (orange), ISM (dark green), disk-CGM interface (light green), and the CGM (light purple bar). We define the ISM mass as 1.4 times the total \HI\, mass (to determine the total $\text{H}+\text{He}$ mass) within a radius at which the surface density of \HI\, (as measured when the galaxy is oriented face-on) falls below $1~M_{\odot}~\text{pc}^{-2}$, hereafter we define this radius as \rHI. This ISM mass definition and surface density limit were chosen to match common observational methods and detection limits of \HI\, surveys \citep[e.g.,][]{2024Deg}. Once again, we define the CGM as all gas particles within $0.15<r/R_{200c}<1$. To fully account for all baryons in the halo, we include the disk-CGM interface, which is the spherical shell between the ISM and CGM, \rHI$< r <0.15 R_{200c}$.

\begin{figure}[t]
    \centering
    \includegraphics[width=8.5cm]{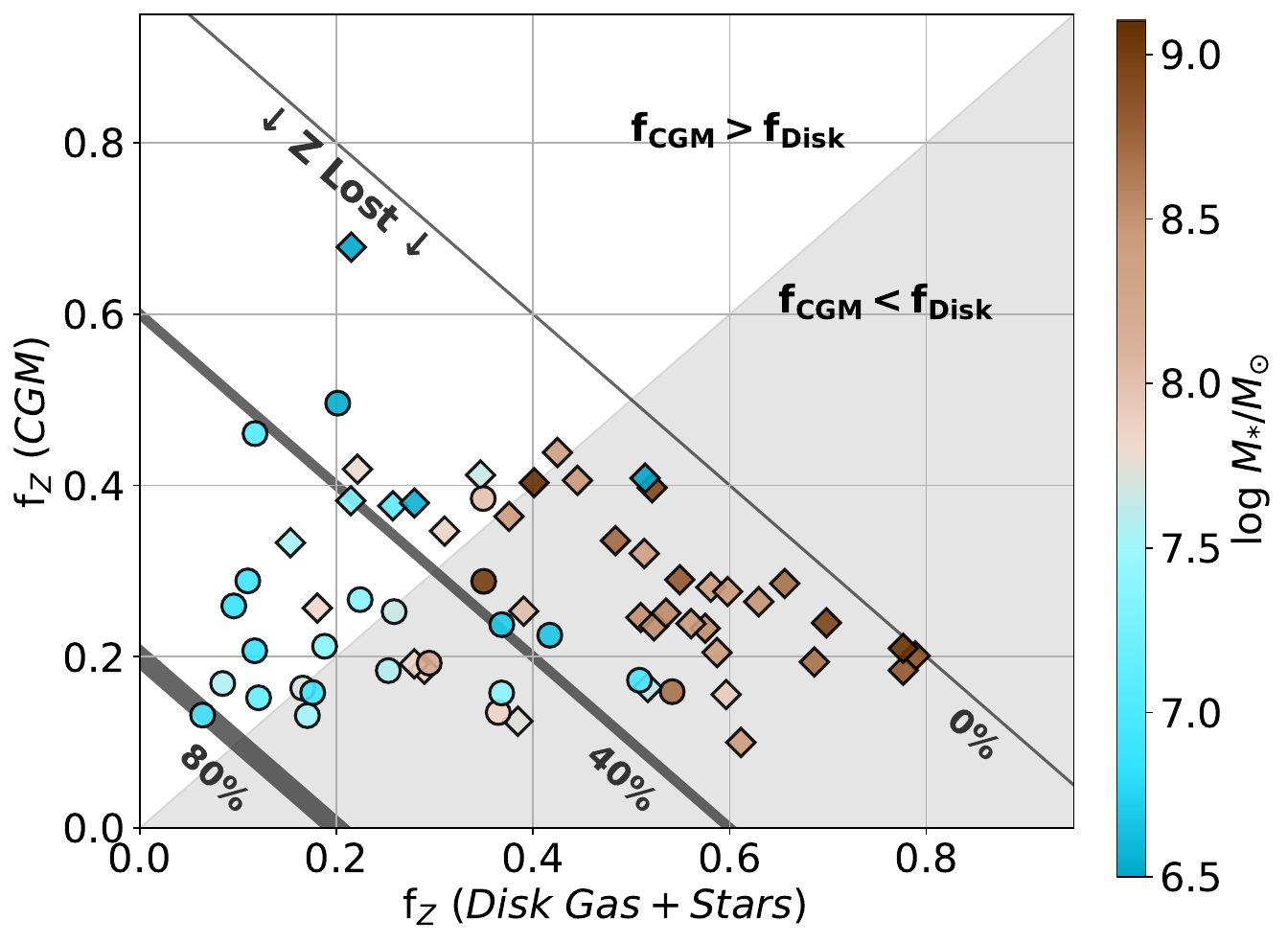}
    \caption{Total halo metal retention for each halo. The horizontal axis shows the fraction of metals retained in the disk (gas and stars), defined by \rHI. The vertical axis shows the fraction of metals retained in all gas beyond the disk (extended CGM and disk-CGM interface) out to $R_{200c}$. Galaxies are differentiated by simulation suite by marker type (i.e., Marvel: circles, Massive: diamonds) and color-coded by stellar mass.  If a halo lies in the lower right, grey-shaded region, then the disk holds more metals than the CGM; the reverse is true in the upper-left white region. The solid grey lines denote the fraction of metals lost relative to the metal mass predicted by the SFH. For example, if a halo lies on the 40\% line, then 40\% of the metals are not accounted for and are likely outside of the halo. With decreasing stellar mass, dwarf galaxies are less effective at retaining metals in their disk and more effective at losing metals from their halo, however, the CGM metal retention tends to remain within $0.1-0.45$.
    \label{fig:f_zret_DiskCGM}}
\end{figure}
We find that with declining stellar mass, the total fraction of baryons retained in the halo declines, resulting in dwarf galaxies becoming increasingly dark-matter-dominated. This decline also indicates an increased sensitivity to feedback with decreasing mass, resulting in more of the dwarf galaxy's baryons being lost from the halo via feedback processes or prevented from accreting onto the halo due to the photoionizing background. The median total baryonic budget and its standard deviation for $\text{log}(M_*/M_{\odot}) > 8.0$ (28/64 halos in the \MM\, sample) halos is $0.467 \pm0.050$, which falls to $0.183 \pm0.048$ for $\text{log}(M_*/M_{\odot}) < 8.0$ halos. Intuitively, as the total baryonic budget decreases, the contributions from the stars, ISM, and CGM also decline. High-mass dwarf galaxies ($\text{log}(M_*/M_{\odot}) > 8.0$) in the \MM\, sample exhibit median CGM contributions of $0.21\pm0.03$ while for low-mass dwarf galaxies ($\text{log}(M_*/M_{\odot}) < 8.0$) this median drops to $0.09\pm0.03$. In a similar trend, high-mass dwarf galaxies exhibit median ISM and stellar baryonic budgets of $0.14\pm0.03$ and $0.06\pm0.02$, respectively, which fall to $0.03\pm0.02$ and $0.01\pm0.01$ for low-mass dwarf galaxies.

The lower panel of Figure \ref{fig:BaryonicBudget} shows the fraction of all baryons within $R_{200c}$ ($f_{b,Halo}$) that reside in the same components (e.g., stars, CGM, etc) as the upper panel. $f_{b,Halo}$ is defined by the following equation: 

\begin{equation}\label{eqn:f_bHalo}
    f_{b,Halo}=\frac{m}{M_{Gas}(<R_{200c}) + M_{*}(<R_{200c})},
\end{equation}

\noindent where the denominator is simply the sum of all baryons within the halo in the form of gas ($M_{Gas}$) and stars $M_{*}$. Across stellar masses, we find that the CGM of low-mass galaxies tends to contain a comparable, and frequently dominant, fraction of baryons within the halo, as compared to the ISM. For $\text{log}(M_*/M_{\odot}) > 8.0$, 23/28 galaxies have halo baryonic budgets dominated by the CGM over the ISM. In this mass range, the median $f_{b,Halo}$ values for the CGM and ISM are respectively $0.44\pm 0.07$ and $0.30\pm 0.10$. Similarly, 35/36 $\text{log}(M_*/M_{\odot}) < 8.0$ galaxies have CGM-dominated baryonic budgets and the median $f_{b,Halo}$ for the CGM and ISM are $0.57\pm 0.10$ and $0.16\pm 0.09$. Given these results, the \MM\, sample predicts there is a significant portion of baryons in the CGM-DG.

\subsection{Metal Budget}\label{subsec:MetalBudget}

In addition to the total baryonic mass, the rate of metals retained in the CGM is pertinent for understanding column densities. We follow the methods presented in \citet{Sanchez2024} to calculate the metal retention for our galaxies, defined in Equation \ref{eqn:fZ} as

\begin{equation}\label{eqn:fZ}
    f_Z = M_Z \biggl/ M_{Z, form},
\end{equation}

\noindent where $M_Z$ is the total mass in metals present within a given volume (e.g., disk, CGM, halo) and $M_{Z,form}$ is the total mass of metals formed by the galaxy across cosmic time. To calculate $M_{Z, form}$, we use \pynbody\, \citep{pynbody} to replicate the metal production calculations from SNIa and SNII (done by the simulation during runtime) for all star particles within the galaxy at $z = 0$. That is, we calculate the total mass of metals formed by the galaxy's stars throughout runtime. The fraction of metals retained ($f_Z$) is then found by dividing these two masses. This analysis provides insight into how effectively dwarf galaxies eject their metals from their disks and halos.

Figure \ref{fig:f_zret_DiskCGM} presents the results for metal retention within the volume of the galaxy ($f_{Z}(\text{Disk Gas+Stars})$) versus the CGM ($f_{Z} (CGM)$). The disk of a galaxy is defined using \rHI, the same definition as the baryonic budget analysis. The horizontal axis of Figure \ref{fig:f_zret_DiskCGM} shows the metal retention for all disk gas plus the stellar content. The vertical axis shows the value of $f_{Z}$ for the remainder of the halo out to $R_{200c}$ (\rHI$< r < R_{200c}$); this method includes the disk-CGM interface as part of the CGM (see below for the $f_{Z}$ of the interface region). Each galaxy is shown as a circle or diamond point to differentiate between the Marvel and Massive suites, respectively, and is color-coded by stellar mass. If a given galaxy exists in the grey-shaded region in the bottom right of the plot, then the disk of the galaxy holds more metals than the CGM, while if a galaxy exists in the upper left, white region, the CGM holds more metals than the disk. 

As this method tallies all metals present with $R_{200c}$, the position of each galaxy also indicates an estimate of the fraction of the galaxy's metals that have been lost from the halo ($f_{Z,\,\text{lost}}$). Each solid line denotes this value, where the thinnest line labeled ``$0\%$'' denotes that $0\%$ of metals have been lost.

We find that galaxies with higher stellar masses also tend to have more metals within their disk region and have lost $f_{Z,\,\text{lost}}\sim10\%$ of their metals from the halo. For $\text{log}(M_*/M_{\odot}) > 8.0$ galaxies, the disk gas contains a median value of $f_{Z,\text{disk}}=0.47\pm0.13$ which exceeds the stellar contribution of $f_{Z,\text{star}}=0.08\pm0.05$. Lower-mass galaxies ($\text{log}(M_*/M_{\odot}) < 8.0$) exhibit smaller retention rates for disk gas, $f_{Z,\text{disk}}=0.21\pm0.13$, and stars, $f_{Z,\text{star}}=0.02\pm0.01$.  As stellar mass decreases, the disks of galaxies tend to have lost a larger fraction of their metals, and $f_{Z,\,\text{lost}}$ tends to increase (shift left horizontally). This results in more metals residing in the CGM than the disk of lower-mass galaxies, and these halos losing $f_{Z,\,\text{lost}}=40 - 80\%$ of metals produced. Interestingly, despite the decrease in $f_{Z, \text{disk}}$ and increase in $f_{Z,\,\text{lost}}$ with declining stellar mass, the median fraction of metals retained in the CGM tends to remain constant across these stellar masses. The CGM metal retention for the \MM\, sample remains near 0.25 ($\text{log}(M_*/M_{\odot}) > 8.0$ galaxies: $f_{Z,\,\text{CGM}}=0.25\pm0.08$, $\text{log}(M_*/M_{\odot}) < 8.0$ galaxies: $f_{Z,\,\text{CGM}}=0.23\pm0.13$), however, with a non-negligible degree of scatter. In the \MM\, sample, 60/64 ($94\%$) of galaxies exhibit CGM retention rates between $0.1-0.45$.

\textit{Disk-CGM Interface Retention:}
The fraction of metals retained in the disk-CGM interface is not shown explicitly in Figure \ref{fig:f_zret_DiskCGM}, as this region is included in the CGM definition for this analysis. We find that the little variation of $f_{Z,\,\text{CGM}}$ with respect to stellar mass explained above persists regardless of whether this interface region is included or excluded from the CGM. However, for completeness, we report these values here. For $\text{log}(M_*/M_{\odot}) > 8.0$ galaxies, the disk-CGM interface typically retains $f_{Z, \text{interface}}= 0.069 \pm 0.035$, similarly, for $\text{log}(M_*/M_{\odot}) < 8.0$ galaxies, this interface typically retains $f_{Z, \text{interface}}= 0.081 \pm 0.043$. If we exclude this region from the CGM, this lowers the CGM retention to $0.19\pm0.06$ for $\text{log}(M_*/M_{\odot}) > 8.0$ galaxies and $0.15\pm0.09$ for $\text{log}(M_*/M_{\odot}) < 8.0$ galaxies.

\subsection{Budgets By CGM Temperature Phases}\label{subsec:PhaseBudgets}

The previous sections have quantified the mass and metal budget for the entire halo in the \MM\, sample. This section seeks to understand how the CGM mass and metal budgets are partitioned by temperature phases. For this section and the rest of this work, we define specific temperature regimes as follows:

\begin{itemize}
    \item Cold Gas: $T<10^{3.8}K$
    \item Cool Gas: $10^{3.8}<T<10^{4.5}K$
    \item Warm Gas: $10^{4.5}<T<10^{5.5}K$
    \item Hot Gas: $T>10^{5.5}K$
\end{itemize}

We adopt these definitions based on the physical properties of the CGM in the \MM\, sample (see Section \ref{sec:CGMPhase}, Figure \ref{fig:CGMPhases_Profiles}) and previous theoretical CGM work \citep[e.g.,][]{2019MNRAS.484.3625R} and reviews \citep[e.g.,][]{TumCGMreview}. 

Additionally, the previous section aimed to understand the contributions to the mass and metal budget from all regions within the halo (i.e., the disk, disk-CGM interface, and CGM). This section solely focuses on the CGM ($0.15R_{200c}>r>1.0R_{200c}$).

\begin{figure}[t]
    \centering
    \includegraphics[width=8.5cm]{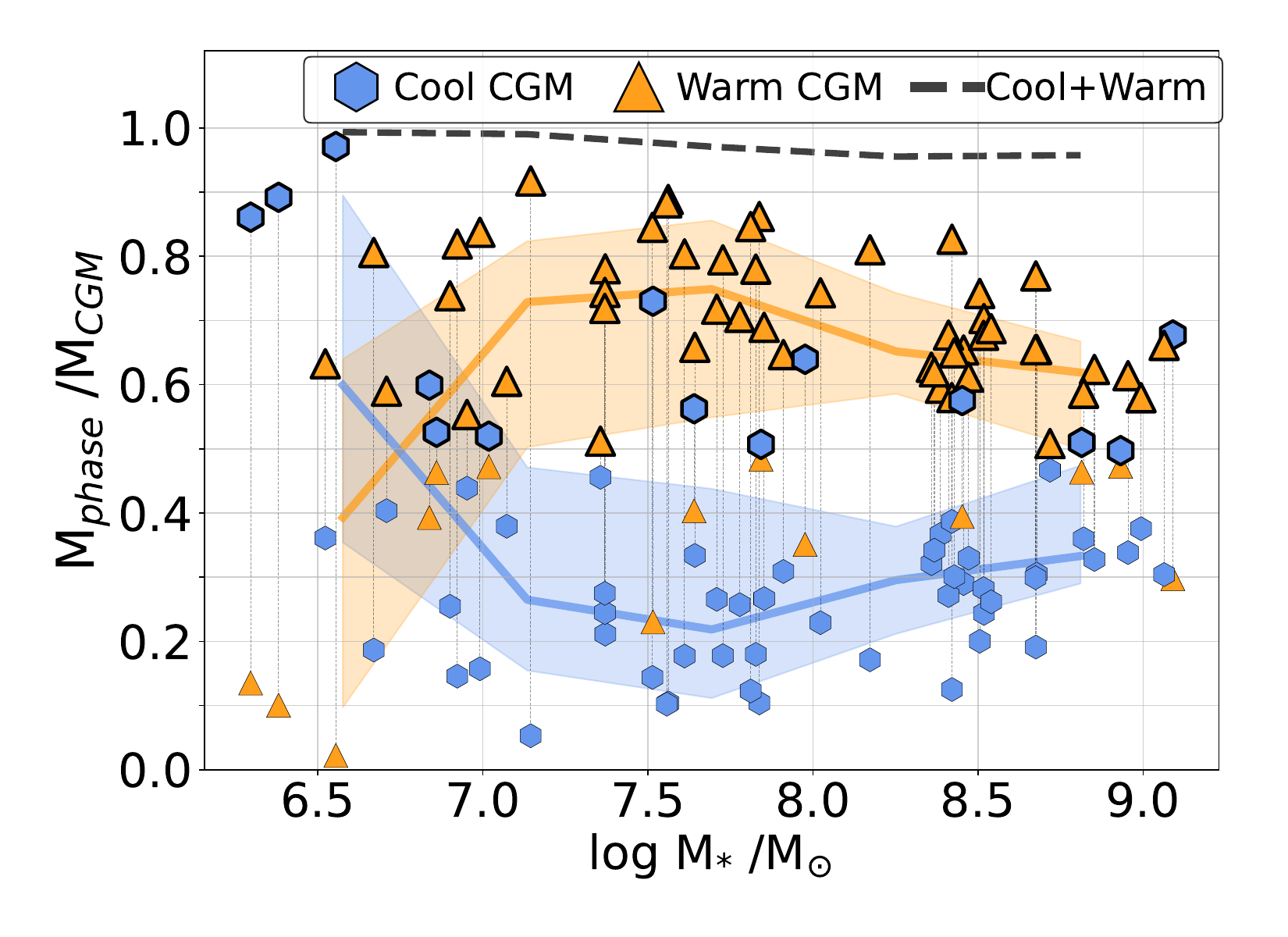}
    \caption{ Fraction of CGM mass in the cool phase, $T=10^{3.8}-10^{4.5}$ K, (blue hexagons) and the warm phase, $T=10^{4.5}-10^{5.5}$ K, (orange triangles), as a function of stellar mass. For a given halo, the two phases are connected by a grey dashed line, and the dominant phase is bolded. The solid line and shaded region show the median CGM percent value and 16th-84th percentile for the cool or warm phases in a bin size of 0.6 dex, plotted at the center of the bin. The dashed black line represents the sum of the median phase fractions, indicating the typical fraction of CGM mass not residing in the cool or warm gas phases (i.e., the cold or hot phases). The majority of CGM mass ($>95\%$) resides in the cool and warm phases, and, for log$M_*/M_{\odot}>7$ galaxies, the cool and warm CGM comprise a median value of 30\% and 70\% of CGM mass, respectively.}
    \label{fig:CoolWarm}
\end{figure}
\begin{figure}[t]
    \centering
    \includegraphics[width=1\linewidth]{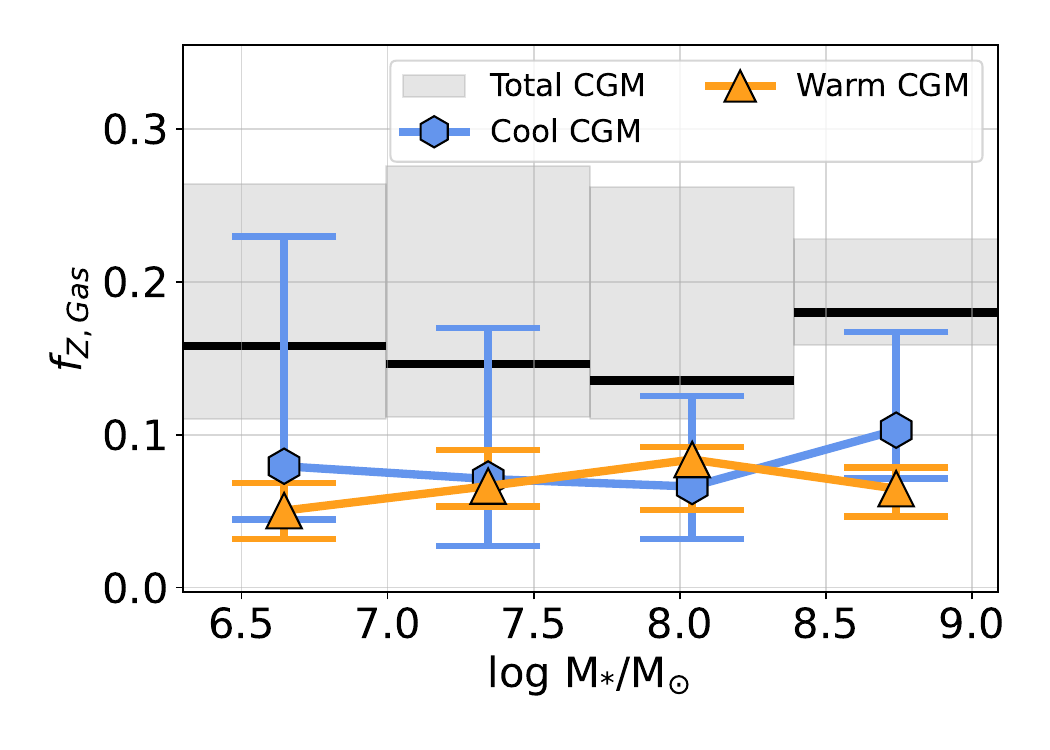}
    \caption{Metal retention in the cool (blue) and warm (orange) gas phases in the CGM ($r>0.15R_{200c}$). The total CGM retention is shown as black bars and grey shaded regions. Shown are the median values for galaxies binned by stellar mass; the error bars/shaded regions denote the 16th-84th percentile in that bin. We find similar fractions ($5-10\%$) are retained in the cool and warm phases and, across stellar masses, a relatively consistent fraction is retained in the two phases and total CGM.}
    \label{fig:f_zret_PHASES}
\end{figure}
\textit{CGM Mass Budget:} We find that in the \MM\, sample, the vast majority of CGM mass resides in either the cool or warm phase. Figure \ref{fig:CoolWarm} shows how the CGM mass is divided between these two dominant phases. For each galaxy, the percentage of CGM mass that is in the cool (blue hexagon) or warm phase (orange triangle) is shown as a function of stellar mass. We also show the sum of these median percentages of the two phases, which indicates the typical fraction of CGM mass that does not reside in either of these phases. 

For $\text{log}(M_*/M_{\odot}) > 8.0$ halos, $\sim95\%$ of CGM mass exists in the cool or warm phase while the remaining mass is at hotter temperatures ($T>10^{5.5}$ K). At the low-mass end ($\text{log}(M_*/M_{\odot}) < 8.0$), less than $2\%$ of CGM mass resides in this hotter, $T>10^{5.5}$ K phase, and $>98\%$ CGM mass resides in the cool or warm phase. The lack of hot gas in low-mass halos is likely due to their shallower gravitational potential wells, which prevent the halo from effectively retaining this hot gas.

In this mass range, 50/64 galaxies in the \MM\, sample have CGM dominated by the warm phase. For high and intermediate virial masses ($M_{vir} \sim 10^{10-11} M_{\odot}$), a relatively constant fraction of $75\%$ of the CGM resides in this warm phase. However, at the highest masses, the two phases begin to approach an equal contribution. This is likely due to the disk extending closer to the $0.15R_{200c}$ boundary with increasing galaxy mass, which results in more cool dense gas in the inner CGM (see Figure \ref{fig:CGMPhases_Profiles} and Appendix \ref{appendix:CGMDef}.) Additionally, at all masses, there are halos with cool-dominated CGM. At lower masses, the fraction of warm CGM mass decreases considerably, and the cool phase approaches 100\%. The sharp increase in cool phase percent is driven by the decrease in the virial temperature at this mass and the definitions of the CGM phases. Specifically, $M_*\leq 10^{6.5} M_{\odot}$ galaxies tend to reside in halos of virial masses below $M_{vir} \sim 10^{9.8} M_{\odot}$ which corresponds to virial temperatures $T_{vir} \lessapprox 10^{4.5}$ K. With these virial temperatures, the CGM is generally unable to retain gas above this temperature (seen also in Figure \ref{fig:CGMPhases_Profiles}) and the bulk of CGM mass resides in the cool phase. Ultimately, across virial masses, the bulk of CGM mass resides near the virial temperature. Using particle tracking, we find the majority of this $T\sim T_{vir}$ phase was beyond $R_{200c}$ at previous times, indicating this phase is primarily comprised of IGM accretion.

\textit{CGM Metal Budget:} The same method of determining metal retention introduced in Section \ref{subsec:MetalBudget} can also be done by further dividing the CGM into its two dominant temperature phases. Figure \ref{fig:f_zret_PHASES} shows the metal retention in the cool (blue squares) and warm (orange squares) CGM phases, and the total CGM retention (black lines) across stellar masses. The median values for galaxies binned by stellar mass are shown. For each median, the 16th-84th percentile for the bin is shown as errorbars (cold and warm phase) or shaded regions (total CGM). This figure selects all particles within $0.15R_{200c}<r<1.0R_{200c}$, differing from Figure \ref{fig:f_zret_DiskCGM} which selected all particles $\text{\rHI}<r<1.0R_{200c}$ (including the disk-CGM interface).

We find that although there is variation in the median total CGM retention at different masses, there is no clear trend with stellar mass. Rather, the \MM\, sample shows a relatively flat CGM metal retention rate between $15-20\%$. Similarly, the cold and warm median metal retention rates also remain between $5-10\%$ across stellar masses. Notably, for each bin, the median rates for the two phases are comparable, despite the two phases comprising different mass fractions of the CGM (Figure \ref{fig:CoolWarm}). The difference in mass fractions and similarity in metal retention indicates the cool phase tends to be higher metallicity material than the warm phase (also seen in Section \ref{sec:CGMPhase}, Figure \ref{fig:CGMPhases_Profiles}), likely due to metals enhancing cooling rates into the cool phase and the warm phase being diluted by pristine IGM accretion.



\section{Physical Properties of the CGM}\label{sec:CGMPhase}

The previous sections have demonstrated that dwarf galaxies in the \MM\, sample (1) contain a significant fraction of baryons in their CGM ($\sim50\%$) relative to their ISM and (2) retain $15-20\%$ of their metals in their CGM and (3) that there is a prominent warm phase that constitutes the majority of the CGM mass budget. Having established that there is sufficient mass and metals for detection in the CGM, the physical conditions of this gas (temperature and density) will determine the ion fractions, therefore setting the abundance for a given species. Also, how the phase of the CGM depends on the distance from the galaxy will contextualize the column density profiles seen in Section \ref{subsec:CGMObservables}. In this section, we show the phase diagram of the total CGM (gas density-temperature diagram) and radial profiles for these properties, along with metallicity. Finally, in Section \ref{subsec:UVObs}, we present the total mass accessible by UV spectroscopy and in individual ions. 

\begin{figure*}
    \centering
    \includegraphics[width=1\textwidth]{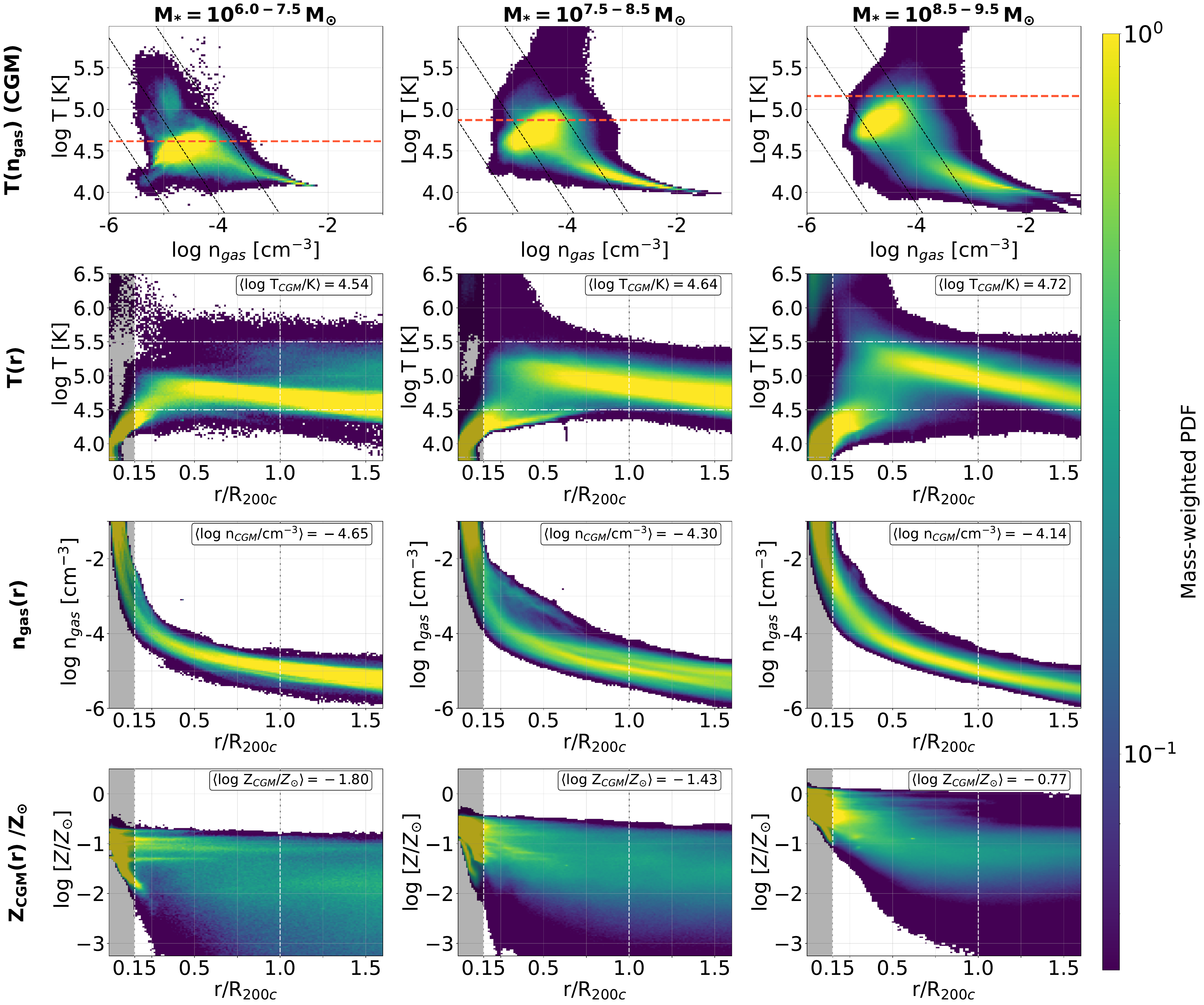}
    \caption{Physical properties of the gas within the halos of the \MM\, sample. Particle data halo is binned by stellar mass and plotted in the corresponding column (labeled on the top of each column) according to the following ranges $M_{*}=10^{6.0 - 7.5}M_{\odot}$ (\textbf{left column}), $M_{*}=10^{7.5 - 8.5}M_{\odot}$ (\textbf{center column}), $M_{*}=10^{8.5 - 9.5}M_{\odot}$ (\textbf{right column}).
    \textbf{Top row:} gas density versus temperature diagram ($T(n_{gas})$) for only CGM particles ($0.15-1.0R_{200c}$). The orange dashed line represents the median virial temperature, while the black lines represent isobaric lines of $10^{-17}, 10^{-16}, 10^{-15}$ Ba in thermal gas pressure. Also included are radial profiles as a function of normalized radial distance, $r/R_{200c}$, of gas temperature \textbf{($T(r)$, upper-middle row)}, gas density \textbf{($n_{gas}(r)$, lower-middle row)}, and gas metallicity relative to solar \textbf{($Z(r)/Z_{\odot}$, bottom row)}. The grey-shaded region denotes what is removed from the halo as the ``disk,'' i.e., not included in the phase diagram. The color bar for all panels shows the mass-weighted probability density. The definitions for the various CGM temperature ranges are also shown in the $T(r)$ panels. Median conditions of the CGM ($r=0.15-1.0R_{200c}$) are shown in each panel, we find that with decreasing mass, typical CGM conditions become more diffuse, cooler, and less metal-enriched.}
    \label{fig:CGMPhases_Profiles}
\end{figure*}

The top row of Figure \ref{fig:CGMPhases_Profiles} shows the gas density-temperature phase diagrams. The three columns show particle data around galaxies in a given stellar mass bin according to the following ranges $M_{*}=10^{6.0 - 7.5}M_{\odot}$ (left column), $M_{*}=10^{7.5 - 8.5}M_{\odot}$ (center column), $M_{*}=10^{8.5 - 9.5}M_{\odot}$ (right column); the same stellar mass bins as used in Figure \ref{fig:LowIonColumnDensitiesResults} and \ref{fig:HighIonColumnDensitiesResults}.

The phase diagrams in the top row Figure \ref{fig:CGMPhases_Profiles} show the mass-weighted probability density for all gas particles around galaxies of the corresponding mass and within $0.15 -1.0R_{200c}$. The orange dashed line shows the median virial temperature for the galaxy mass bin, and black dashed curves show lines of constant pressure at $10^{-17}, 10^{-16}, 10^{-15}$ Ba.

In the highest mass galaxies, we find a substantial fraction of mass in a cool and dense phase ($n_{gas}\sim10^{-3}$ cm$^{-3}$, $T\sim10^4$ K) and a warmer, more diffuse phase. This bimodality is consistent with theoretical expectations for higher-mass galaxies \citep[e.g., TNG50: see Figure 6 of][]{2020MNRAS.498.2391N} and existing work on lower-mass galaxies  \citep[e.g., FIRE: see Figure 11 of ][]{DwarfCGM_in_FIRE},  in which the cooler gas is photoionized and the warmer component is a virialized component. Furthermore, the virialized component is largely comprised of low-metallicity gas, likely due to dilution from pristine IGM accretion, as shown in the radial profiles described below. 

Radial profiles with respect to normalized galactocentric radius  ($r/R_{200c}$) of temperature, density, and metallicity are shown in the bottom three rows of Figure \ref{fig:CGMPhases_Profiles}. The colorbar shows the mass-weighted probability density (the same as the one used in the phase diagrams in the leftmost column). We highlight the CGM region ($0.15 < r/R_{200c}<1.0$), which is demarcated by dashed lines, but also include the disk region ($r/R_{200c}<0.15$) and the region immediately beyond the halo ($1.0 < r/R_{200c}<1.6$)\footnote{We choose to extend out to $1.6R_{200c}$ which is approximately $R_{\bf{200m}}$ because some works choose to implement the $R_{\bf{200m}}$ virial definition as opposed to $R_{200c}$.}. For clarity, we also annotate each panel with the median properties of all particles within $0.15 -1.0R_{200c}$.

As shown in the temperature and density radial profiles (upper-middle and lower-middle rows of Figure \ref{fig:CGMPhases_Profiles}), we find the cool component of the CGM is largely centralized to the inner regions of the halo, while the warmer phase exists in the outskirts. Similar to Auriga \citep{Cook_24}, FIRE-2 \citep{Stern2021}, and IllustrisTNG \citep{IllustrisTNG_CRS_Ramesh24}, we find a sharp transition from the cool to the warm phase at $r/R_{200c}\sim 0.4$. This transition in temperature roughly corresponds to the location where the cooling time becomes shorter than the free fall time ($t_{\rm cool}<t_{\rm ff}$). This allows gas to cool to temperatures of $\sim10^4$ K before falling into the galaxy on a free-fall time \citep[e.g.,][]{Stern2021}.  The outer regions generally follow the virial temperature, which is in line with previous theoretical work predicting gas falling into a halo will be shock-heated to near the virial temperature \citep[e.g.,][]{Brooks2009, 2012MNRAS.423.2991V}. Also, the cooling times in this outer region are longer than the free-fall times, which keeps the gas at near virial temperatures before falling into the galaxy. We find that the behavior of cooling and free-fall times in the \MM\, sample generally follows the behaviors seen in \citet{Stern2021}; however, future work will investigate this more explicitly. Lastly, for all three galaxy mass bins, we find the hottest gas to have been recently heated by SNe feedback.  

With decreasing mass, we find the inner CGM remains near $T\sim 10^4$ K, and the median gas density decreases. This occurs because, with decreasing virial mass, the denser gas moves interior to our $0.15 R_{200c}$ CGM definition (grey shaded region in Figure \ref{fig:CGMPhases_Profiles}). We discuss the implications of our radial definitions in Appendix \ref{appendix:CGMDef}. In the intermediate regions, decreasing halo mass leads to the transition between the cool and warm temperature phases moving inward towards the galaxy. In the outer regions, decreasing halo mass results in cooler temperatures in the outskirts owing to the lower virial temperatures. We also find that the gas density in the outskirts of the CGM remains near $n_{gas} \sim 10^{-5}-10^{-5.5}$ cm$^{-3}$ for all halos. 

Finally, the bottom row of Figure \ref{fig:CGMPhases_Profiles} shows metallicity profiles for the CGM in the \MM\, sample, relative to solar metallicity ($Z_{\odot}=0.0134$; \citet{Asplund2009}). We find the inner regions are the most metal-enriched for all galaxies, and the median CGM metallicity decreases with decreasing halo mass. Median gas metallicities relative to solar (log $Z/Z_\odot$) at $r/R_{200c}= [0.25, 0.5, 1]$ are:

\begin{itemize}
    \item $M_{*}=10^{6.0 - 7.5}M_{\odot} : [-1.23,-1.65,-2.55]$
    \item $M_{*}=10^{7.5 - 8.5}M_{\odot} : [-1.11,-1.44,-1.90]$
    \item $M_{*}=10^{8.5 - 9.5}M_{\odot} : [-0.50,-0.88,-1.26]$
\end{itemize}

We note that the metallicity of gas immediately inside and just beyond $R_{200c}$ is similar and that there is no clear change in metallicity at either of these boundaries. The extent of metal-enriched gas implies that feedback is efficient at driving metals out of the disk, into the outer CGM and beyond the halo, likely beyond $1.6R_{200c}$ as well.

\subsection{Ionization Fractions}
\begin{figure}
    \centering
    \includegraphics[width=7.75cm]{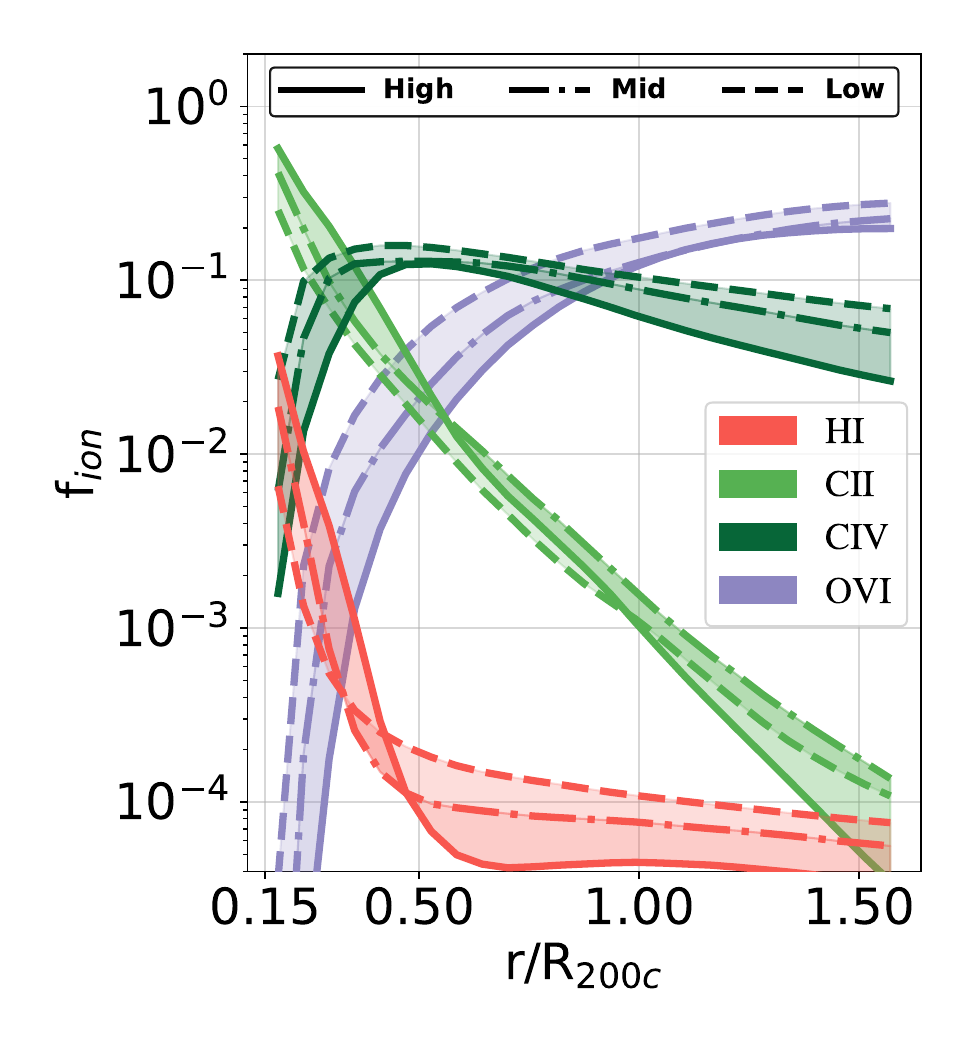}
    \caption{Ion fractions ($f_{ion}$) for \HI\, (salmon) and a representative low ion (\CII, light green), intermediate ion (\CIV, dark green), and high ion (\OVI, purple) as a function of radial separation from the galaxy. For each ion, we show the median values for the same three virial mass bins as Figure \ref{fig:CGMPhases_Profiles}, with solid, dash-dot, and dashed lines representing high-, intermediate-, and low-mass galaxies. We see little variation in the shape of $f_{ion}$ radial profiles with halo mass. However, with increasing distance from a galaxy, low ions (\HI\, and \CII) become increasingly less abundant and intermediate/high ions (\CIV\, and \OVI) become the dominant ionization stages---this is due to radial dependence of CGM physical properties (Figure \ref{fig:CGMPhases_Profiles}).  }
    \label{fig:CGM_ionfractions}
\end{figure}

The physical conditions of the CGM gas determine the ion fractions for each species. Figure \ref{fig:CGM_ionfractions} shows the median ion fractions ($f_{ion}$) calculated using \cloudy\, (assuming the \citet{HM12} UVB) for \HI\, as well as a low metal ion, \CII, intermediate metal ion, \CIV, and high metal ion, \OVI. Each ion shows three curves corresponding to the same three mass bins used in Figure \ref{fig:CGMPhases_Profiles}. We find that the most suitable conditions (cool, $T<10^{4.5}$ K, and dense, $n_{gas}>10^{-3}$cm$^{-3}$) for species with lower ionization potentials (e.g. \HI\, and \CII) are in the inner CGM. Both $f_{\text{HI}}$ and $f_{\text{CII}}$ fall off steeply with increasing values of $r/R_{200c}$, however for $r/R_{200c}>0.5$ $f_{\text{HI}}$ remains near a consistent value.

We find the intermediate ions (\CIV) are most prominent in the intermediate regions ($r/R_{200c}\sim0.5$) and show a flatter ion fraction profile. $f_{\text{CIV}}$ is greatest in regions of intermediate densities ($10^{-3}<n_{gas}<10^{-5}$cm$^{-3}$) and warmer temperatures ($10^{4.5}<T<10^{5.0}$ K). For the higher ion, $f_{\text{OVI}}$ is greatest in regions of similar temperatures as $f_{\text{CIV}}$ ($10^{4.5}<T<10^{5.0}$ K) but more diffuse ($n_{gas}<10^{-5}$cm$^{-3}$). Additionally, $f_{\text{OVI}}$ peaks beyond $R_{200c}$, which indicates there is likely a significant portion of \OVI\, mass just outside the halo. Observations of \OVI\, column densities are thus likely to detect a significant portion of \OVI\, that exists beyond $R_{200c}$.  These results are consistent with observed radial trends \citep[e.g.,][]{Johnson2015, Mishra_24}, and comparable to trends found in FIRE simulations \citep{DwarfCGM_in_FIRE}.

When comparing galaxies of different virial masses, we find little variation in the qualitative behavior of $f_{ion}$ (i.e., $f_{\text{CII}}$ always peaks in the innermost regions while $f_{\text{OVI}}$ has a maximum beyond $R_{200c}$). However, the conditions in which the various ions are most prominent by mass do vary slightly with virial mass based on the availability of material at certain densities and temperatures. For instance, high-mass galaxies have more mass in \CIV\, at $T>10^{5.0}$ K than low-mass galaxies, given that low-mass galaxies have very little CGM mass at $T>10^{5.0}$ K. Nonetheless, the similarity of $f_{ion}$ across virial masses indicates that the change in column densities with respect to galaxy mass (comparing panels within a given row) seen in Figure \ref{fig:LowIonColumnDensitiesResults} and \ref{fig:HighIonColumnDensitiesResults} is more strongly driven by the change in CGM gas and metal mass rather than the changes in the phase of the CGM phase seen in Figure \ref{fig:CGMPhases_Profiles}. However, the change in column densities from ion to ion within a given mass bin (comparing panels within a given column) is driven by the multiphase structure of the CGM seen in Figure \ref{fig:CGMPhases_Profiles}.

\subsection{UV Observable Budget}\label{subsec:UVObs}
   
\begin{figure*}
    \centering
    \includegraphics[width=1\textwidth]{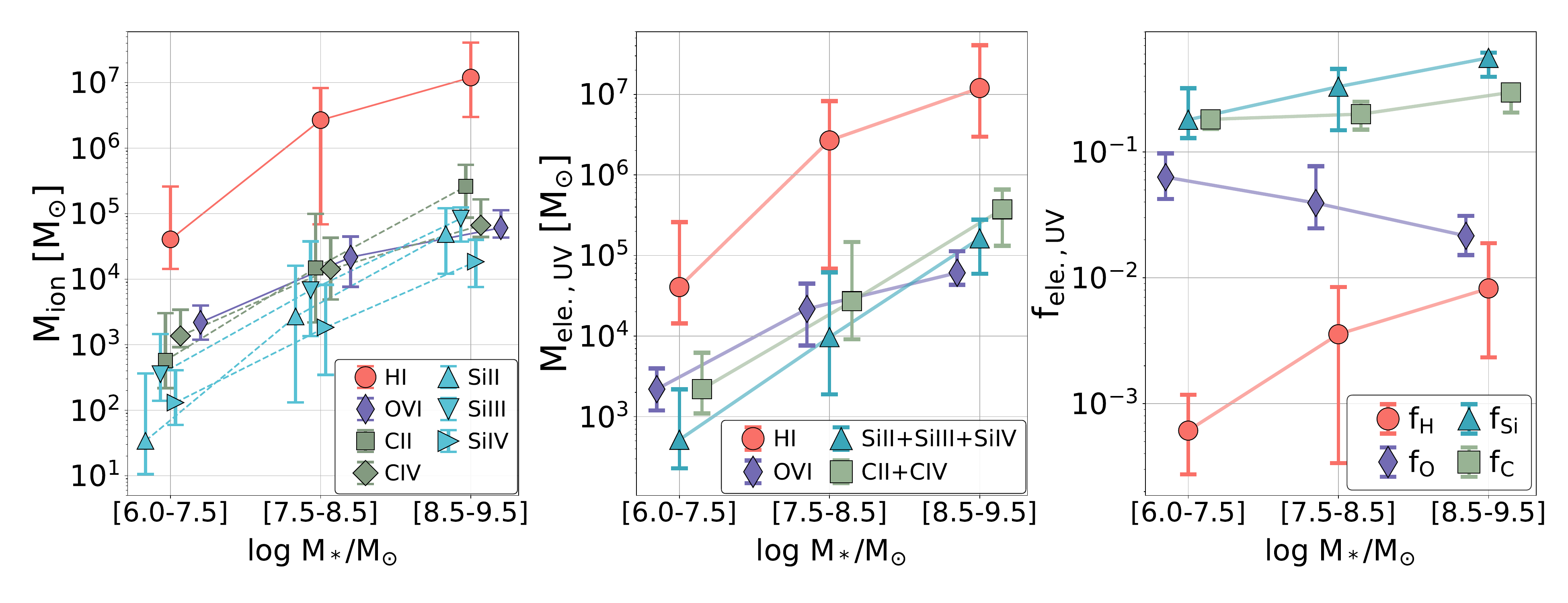}
    \caption{CGM UV observable (at $z=0-0.3$) budgets for the \MM\, sample. \textbf{Left panel:} Median masses of individual ions with UV-detectable transitions in the CGM. \textbf{Center panel:} Median total UV-observable mass in the CGM ($M_{\text{ele., UV}}$), defined as the summed mass of ions with UV-detectable transitions (explicitly written in the legend and plotted in the left panel). \textbf{Right panel:} Corresponding UV-observable fraction/ionization correction ($f_{\text{ele., UV}}$), defined as the UV-observable mass divided by total element mass ($f_{\text{ele., UV}}=M_{\text{ele., UV}}/M_{\text{ele.}}$). Each panel shows results for three stellar mass bins (log $M_*/M_\odot$ = 6.0–7.5, 7.5–8.5, 8.5–9.5), and the errorbars show the 16th-84th percentiles; markers are horizontally offset for clarity. Elements are represented as follows: hydrogen (salmon circles), silicon (blue squares), carbon (green triangles), and oxygen (purple diamonds). In the left panel, markers are rotated to distinguish between ions.}
    \label{fig:UVmass}
\end{figure*}

Ultimately, the total gas mass and ion fractions of the CGM will determine the total mass of a given ion. In this section, we characterize the total mass of individual ions in the CGM and the total mass that is accessible in the UV. This analysis provides benchmarks for estimating how much of the CGM mass UV absorption spectroscopy is probing and allows for better comparisons to other simulations.

The left panel of Figure \ref{fig:UVmass} shows the mass of individual ions (\HI, \CII, \CIV, \SiII, \SiIII, \SiIV, \OVI) in the CGM in the \MM\, sample. Each point represents the median ion mass in the CGM around a galaxy within a given stellar mass bin, and the errorbars represent the 16th-84th percentiles. The three mass bins are log $M_*/M_{\odot}={8.5-9.5}$, log $M_*/M_{\odot}={7.5-8.5}$, and log $M_*/M_{\odot}={6.0-7.5}$ and are denoted on the horizontal axis. The mass in each ion trends with the stellar mass of the galaxy, in that the highest mass galaxies tend to also have the most mass in each ion studied. We note, however, that, especially for the intermediate- and low-mass bins, there is substantial scatter in the total mass of ions, some as great as 1-2 dex.

The center panel of Figure \ref{fig:UVmass} shows the median total mass in UV observable hydrogen, silicon, carbon, and oxygen in the CGM for the same three stellar mass bins denoted on the horizontal axis. We define the UV observable mass as the sum of ion masses that have transitions accessible by UV absorption spectroscopy at $z=0-0.3$. More specifically, that is the total mass of \HI\, for hydrogen; \SiII, \SiIII, and \SiIV\, for silicon; \CII\, and \CIV\, for carbon; and \OVI\, for oxygen. The right panel of Figure \ref{fig:UVmass} shows the fraction of the total element mass that is in UV observable ions (i.e. $f_C=(M_{CII}+M_{CIV}) / M_C$) for the same elements and stellar mass bins. The fraction of UV observable mass can also act as an ionization correction for observational studies that estimate the total element mass using a set of UV observable ions \citep[e.g.,][]{Zheng_24}. For convenience, we list the median values and associated 16th-84th percentile ranges for the fraction of UV observable mass in Table \ref{tab:f_UV}.

We find that higher-mass galaxies tend to hold the most mass accessible in the UV. With decreasing stellar mass, the total element mass and total UV observable mass in each element decrease. Similarly, the percentages generally tend to also decline. The fraction of oxygen that is accessible in the UV does increase towards the low-mass end, however, the change is on the $1\%$ level. The oxygen UV observable budget is small ($<10\%$) for galaxies across stellar masses due to a large fraction of \OVI\, mass residing at or beyond the $R_{200c}$ boundary and the majority of the oxygen mass exisiting in other ionization stages (see Section \ref{subsec:Discussion_OVI} for more details).


\section{Discussion}\label{sec:Discussion}

\subsection{Underpredicted Column Densities}\label{subsec:Discussion_OVI}

In this subsection, we present a brief investigation into the ions that are in the weakest agreement with observations (\OVI\, and \CIV) and assess what may be driving the underprediction of these ions seen in Figure \ref{fig:HighIonColumnDensitiesResults}.

\textit{OVI Column Densities:} 
\NOVI\, shows the weakest agreement with observations across all ions covered in this work, with 0/6 detections lying within $2\sigma$ of the \MM\, sample medians. As noted in Section \ref{sec:CGMObservables}, the \OVI\, observations included in this work are at systematically higher redshifts. A brief investigation into the simulated \NOVI\, across redshifts ranging from $z=0-1$ revealed that these column densities can indeed change with redshift; however, for a given galaxy, the columns may increase or decrease within $0.5$ dex with decreasing redshift. Notably, an increase of $0.5$ dex is too small a change to match the observations shown in Figure \ref{fig:HighIonColumnDensitiesResults} and thus the \MM\, sample likely underestimates \OVI\, even at redshifts consistent with current observations. 

Since the change in \NOVI\, with redshift cannot produce columns in agreement with observations, the underprediction of \NOVI\, is likely due to one or both of a) a lack of metals in the outer CGM and b) a lack of material in the correct phase. Previous work for high-mass galaxies has shown that BH feedback can strongly shape the \OVI\, content in the CGM and enhance column densities in \OVI\, by pushing more metal-enriched material out to the CGM outskirts \cite[e.g.][]{Nelson2018, Sanchez2019,Sanchez2021}. The change in simulated \NOVI\, values revealed by this brief investigation implies a dependence on both redshift and galaxy activity. Thus, a similar explanation may apply to our simulations, such that increased SNe feedback in our dwarf galaxies could lead to \OVI\, column densities in better agreement with observations.

The underprediction of \OVI\, and agreement of low ions and \SiIV\, may also point to our simulations underproducing a certain gas phase. We investigate this by finding the mass in individual ionization states within $140\,\text{kpc}$ of each galaxy (approximately the observed region probed by the synthetic sightlines). Relative to \OVI\, mass, we find that there is significant mass in all ionization states \OII\, through \OVII. For $\text{log}(M_*/M_{\odot}) > 8.0$ galaxies, there is $1-3\times$ as much mass in \OV\, and \OVII\, as there is in \OVI\, (e.g., $M_{\text{OV}}/M_{\text{OVI}} = 1-3$) and $1-7\times$ as much mass in \OII, \OIII, and \OIV. For $\text{log}(M_*/M_{\odot}) < 8.0$ galaxies, this ratio declines, but there is still $1-2\times$ as much mass in \OV, \OIV, and \OVII. These results suggest that the underprediction of \NOVI\, is likely strongly driven by a bulk of CGM mass residing in different ionization states than \OVI.

\textit{CIV Column Densities:} 
Similarly, our \NCIV\, results tend to be below current observations (0/8 \&\, 5/8 detections within $1\sigma$ and $2\sigma$, respectively). We perform a similar calculation as oxygen for carbon and determine which ionization states hold the most mass within $140\,\text{kpc}$ of each galaxy. For carbon, we also find significant mass in ionization states other than \CIV. Specifically, across galaxy masses, we find there is $4-6\times$ as much mass in \CV\, and $1-3\times$ as much mass in \CVI, relative to \CIV ~(e.g., $M_{\text{CV}}/M_{\text{CIV}} = 4-6$). Also, there is significant mass in \CIII, with $3-6\times$ and $1-2\times$ more mass than \CIV\, for high-mass and low-mass galaxies, respectively.

Relative to lower ions, \OVI\, and \CIV\, are typically expected to be prominent in warmer and more diffuse material (see Figure \ref{fig:N_maps}). Thus, the \MM\, sample likely underpredicts warm and diffuse material (where ionization fractions peak in \OVI\, and \CIV) in the outer regions of the halo while still producing enough cooler material in the central regions of the halo to match the observations of \HI, low ions, and \SiIV. However, previous work (see above) has also pointed to a correlation between \NOVI\, and feedback processes, suggesting that stronger feedback may also increase \NOVI\, values in the \MM\, sample. 

Future works will investigate the properties of \OVI\, in the \MM\, sample at higher redshifts and the effects on the CGM-DG from including additional physics, aiming to further disentangle which of the above scenarios is producing the low \OVI\, column densities seen in this work.

{\renewcommand{\arraystretch}{1.2}
\begin{table}
    \centering
    \begin{tabular}{c|c|c|c}
      & \textbf{Low-$\mathbf{M_*}$}  & \textbf{Mid-$\mathbf{M_*}$}       & \textbf{High-$\mathbf{M_*}$}   \\
    & $10^{6.0-7.5} ~M_{{\odot}}$ & $10^{7.5-8.5} ~M_{{\odot}}$    & $10^{8.5-9.5} ~M_{{\odot}}$ \\

             \hline
    Hydrogen & $0.06^{+0.06} _{-0.03}\, \%$  &  $0.36^{+0.49} _{-0.32}\, \%$  &  $0.82^{+1.06} _{-0.59}\, \%$   \\
    Carbon   & $18.2^{+1.4} _{-2.9}\, \%$  &  $20.0^{+5.1} _{-4.9}\, \%$  &  $29.8^{+3.9} _{-9.2}\, \%$  \\
    Oxygen   & $6.3^{+3.4} _{-2.1}\, \%$  &  $3.9^{+3.8} _{-1.5}\, \%$  &  $2.2^{+0.9} _{-0.6}\, \%$\\
    Silicon  & $18.0^{+14.1} _{-5.1}\, \%$  &  $33.0^{+12.7} _{-18.1}\, \%$  &  $55.9^{+5.5} _{-16.4}\, \%$\\

    \end{tabular}
    \caption{Summary of $f_{\rm UV}$ for hydrogen, carbon, oxygen, and silicon. Each value presented is the median and 16th-84th percentile ranges. The values provided here are the same as those plotted in the left panel of Figure \ref{fig:UVmass} and duplicated for convenience.
    }
    \label{tab:f_UV}
\end{table}}
\subsection{Tying CGM Column Densities to CGM Structure }\label{subsec:Discussion_CGMStructure}

In addition to comparing our mock column densities to observed values, we interpret them in the context of the CGM structure. In Section \ref{sec:CGMObservables}, Figures \ref{fig:N_maps}, \ref{fig:LowIonColumnDensitiesResults}, and \ref{fig:HighIonColumnDensitiesResults}, we observe a steep inner \NHI\, profile followed by a plateau in the outer CGM---most pronounced for galaxies with $\text{log}(M_*/M_{\odot}) > 7.5$. We argue that this behavior results from the radial temperature and density profile of the CGM, which is evident when determining the individual contributions from the cool and warm phases to each \NHI\ value. By selecting all particles along a single LOS within a given temperature range, we calculate the \NHI\, from that single phase. We find the cool CGM contribution dominates over the warm phase within $b/R_{200c} < 0.5$, reaching \NHI$= 10^{18}~\text{cm}^{-2}$.  However, with increasing $b/R_{200c}$, the cool phase column density falls off steeply. This decline mirrors the behavior of $f_{\text{HI}}$ in Figure \ref{fig:CGM_ionfractions}, which is driven by the declining cool gas mass with increasing $r/R_{200c}$ (Figure \ref{fig:CGMPhases_Profiles}). Beyond $\sim 0.5 R_{200c}$, the absence of dense, $T\sim10^4$ K gas causes the cold phase contribution to fall off and the warm phase contribution to dominate. 

Across all $b/R_{200c}$, the warm phase contributes $\sim 10^{14}$ $\text{cm}^{-2}$ to each \NHI. We can interpret the relatively constant warm phase contribution to \NHI\ using Figure \ref{fig:CGM_ionfractions}, where $f_{\text{HI}}$ plateaus for $r/R_{200c} > 0.5$ and the warm phase dominates. 
The combination of this flat $f_{\text{HI}}$ and the substantial mass in the warm phase (Figure \ref{fig:CoolWarm}) leads to the warm phase contributing $\sim 10^{14}~\text{cm}^{-2}$ to the total \NHI\, across all impact parameters. Once the cool phase contribution falls off at greater $b/R_{200c}$, the warm phase's consistent \NHI$\sim 10^{14}~\text{cm}^{-2}$ dominates and produces the \NHI\, plateau.

Similarly, low ions like \CII, \SiII, and \SiIII\, also peak in cool, dense gas, and thus the same physical reason for the steep inner profile of \NHI\, also gives rise to the steep low ion profiles. These ions have ion fractions that fall off rapidly with $r/R_{200c}$ (Figure \ref{fig:CGM_ionfractions}); however, due to these ions having significantly less mass than \HI\, the column density does not plateau above detectable limits. 

By mass, the intermediate ion \CIV\, is most abundant in the intermediate regions of the CGM ($b/R_{200c} \sim 0.5$), with the majority of \CIV\, mass residing between $0.25-0.5R_{200c}$. Given that the \CIV\, is most abundant in the intermediate regions, the \NCIV\, profile is less steep than lower ions and shows a flatter decline with $b/R_{200c}$, seen in Figure \ref{fig:N_maps} and \ref{fig:HighIonColumnDensitiesResults}. Finally, a similar explanation can also be used to explain the flat \NOVI\, profile, since the \OVI\, ion fractions peak beyond the halo.

\subsection{Comparison to Previous Work}

Previous work studying the simulated CGM-DG has found qualitatively similar column density profiles as we find here. Specifically, the \NHI\, profiles with a clear plateau beyond $b\sim0.5 R_{200c}$ has been found in previous work \citep[e.g.][]{Gutcke2017, Mina2021,Cook_24}, as well as the steep \NSiII\, profiles dropping below detectable limits before $b\sim0.5 R_{200c}$ \citep[e.g.][]{Mina2021,Cook_24}. We find similar \OVI\, column densities below current observations as NIHAO \citep{Gutcke2017}, which investigated the \HI\, and \OVI\, CGM content in a wide range of galaxies. However, at $b\sim0.5 R_{200c}$, we find lower \OVI\, column densities by $\sim0.5$ dex compared to \citet{Mina2021} (which studied the ``Seven Dwarfs'' simulations ran with \gasolineII~\citep{Wadsley2017}), despite our works showing similar numerical results when comparing \HI\, and \CIV\, column densities. \citet{Mina2021} examined differences in the CGM  for a set of simulated galaxies\footnote{In fact, the ``Seven Dwarf'' simulations presented in \citet{Mina2021} use the same initial conditions as our ``cptmarvel,'' part of the Marvel-ous Dwarfs suite, but ran with \gasolineII\, instead of \changa.} run with both blastwave and superbubble feedback models.  They find that the blastwave \OVI\, column densities are $\sim$1 dex lower than the superbubble \OVI\, column densities. Although we do not see a clear trend in \OVI\, between suites in the \MM\, sample (see Appendix \ref{appendix:SuitebySuite}), future work will investigate further how the CGM changes when altering SNe feedback.

Similarly, at $b\sim0.5 R_{200c}$ around $\text{log}(M_*/M_{\odot})>8.5$ dwarf galaxies the \MM\, sample produces higher column densities than Auriga \citep{Cook_24} in both \CIV\, (\MM: $\text{\NCIV}\sim10^{13}\text{cm}^{-2}\,$; Auriga: $\text{\NCIV}<10^{11}\text{cm}^{-2}$) and \OVI\, (\MM: $\text{\NOVI}\sim10^{13}\text{cm}^{-2}\,$; Auriga: $\text{\NOVI}<10^{11}\text{cm}^{-2}$). These lower values of \NCIV\, and \NOVI\, are likely due to the Auriga galaxies having much lower metallicities in the outer CGM, where these ions would preferentially exist, than the \MM\, sample. Specifically, \citet{Cook_24} shows that beyond $r/R_{200c} \sim 0.25$ the CGM  of $M_{200c}=10^{10.0}M_{\odot}$ and $10^{10.5}M_{\odot}$ galaxies in Auriga has metallicities $\text{log }Z/Z_{\odot} < -3$. The authors note that the low metallicity in the outer regions of the halo is caused by the outflows from the dwarf galaxies not reaching larger distances, as evidenced by an average outflow radial velocity of $\sim$0 km s$^{-1}$ at $r/R_{200c}\sim0.25$. This implies that the \MM\, sample dwarf galaxies are more efficient at dispersing metals into the full extent of the CGM and beyond the halo, and this fact is critical in producing greater \NCIV\, and \NOVI. 

We also compare our mass and metal retentions to previous works. The total halo mass retention rates for the \MM\, sample are roughly in agreement with those presented in \citet{Hafen2019} (FIRE-2) and \citet{Christensen_2016}, peaking at around $\sim60\%$ for the highest mass dwarf galaxies and dropping to $\sim10\%$ at the lowest masses. However, metal retention rates do differ between these works. \citet{Hafen2019} finds in FIRE-2 that at all halo masses, $z=0.25$ dwarf galaxies have lost $\sim40\%$ of their metals from the halo. Notably, this work claims a relatively flat fraction of metals lost from the halo, $f_{Z,lost}$. However, the \MM\, sample and \citet{Christensen_2018}, show an \textit{increasing} $f_{Z,lost}$ with decreasing mass (as indicated by the ``Z Lost'' diagonal lines in Figure \ref{fig:f_zret_DiskCGM}).\footnote{We note that \citet{Christensen_2018} used \gasolineI\, \citep{Wadsley2004},  which is a predecessor to \changa. \gasolineI\, and \changa\, implement similar star formation parameters and a blastwave feedback model. However, \changa\, is designed to scale more efficiently and uses the hydrodynamics methods from \gasolineII\, \citep{Wadsley2017}.} For the same range of halo masses as \citet{Hafen2019}, \citet{Christensen_2018} finds $f_{Z,lost}=30-70\%$ while in the \MM\, sample $f_{Z,lost}=0-70\%$. Although the \MM\, sample loss rates differ at the high-mass end compared to \citet{Christensen_2018}, we note that we utilize a larger number of galaxies and observe a fair degree of scatter in metal loss rates. When instead considering the median loss rates for the \MM\, sample, we find $f_{Z,lost}=19\pm10 \%$ for $M_{200c}\sim10^{10.5-11.0}M_{\odot}$ galaxies and $f_{Z,lost}=51\pm15 \%$ for $M_{200c}\sim10^{10.0-10.5}M_{\odot}$ galaxies; more in agreement with \citet{Christensen_2018}. As noted in \citet{Christensen_2018} and \citet{Hafen2019}, the differences in $f_{Z,lost}$ across simulations are likely due to differences in feedback prescriptions, as the feedback mechanisms will have a dominant role in dispersing metals into the CGM and IGM. However, further work is needed to understand the origin of the differences in detail.

In agreement with the \MM\, sample, the CGM-DG in EAGLE \citep{Zheng_24} exhibits an approximately equal division in metal mass by temperature ranges, despite also constituting different masses (we find $\sim3$x as much mass in the warm phase as cool, while EAGLE contains $> 7$x in the warm phase). In both the \MM\, sample and EAGLE, the cold phase retains $\sim10\%$ of the metals produced by the galaxy. This roughly equal distribution in metals between gas phases is not found in FIRE \citep{Muratov2017}, which finds that most metals reside in the cooler temperature range, $10^{4}<T<10^{4.7}$ K. This may suggest that more cool gas is ejected in the FIRE runs than in the \MM\, sample or EAGLE.  Note that \citet{Muratov2017} studied FIRE-1 galaxies, which did not include metal mixing, and might lead to higher metallicity in ejected cool gas and enhanced cooling in the CGM.  \citet{Hafen2019} notes that the overall metal content of the CGM is consistent between FIRE-1 and FIRE-2, however.

In terms of the physical properties of the CGM, the \MM\, sample, FIRE-2 \citep{DwarfCGM_in_FIRE} and EAGLE \citep{Zheng_24} find a prominent warm (T $>10^{4.5}$ K) and diffuse (n $<10^{-3}$cm$^{-3}$) component for $r>0.5R_{200c}$ CGM gas around $7.5\leq \text{log}(M_*/M_{\odot}) \leq 9.5$ galaxies. However, in the same regions, FIRE also produces a cool (T$\leq10^{4.0}$ K) and dense (n$>10^{-1}$cm$^{-3}$) component, which is not present in the \MM\, sample. Again, this effect may indicate that feedback in FIRE-2 galaxies couples stronger with the ISM gas and is more effective at pushing cool, dense gas into the outer regions of the CGM. Across galaxy masses, the majority of CGM mass in the \MM\, sample exists at temperatures within T$\sim10^{4.5}-10^{5.5}$ K.  In contrast to these other studies, \citet{2024arXiv241216440T}, which studied dwarf galaxies in an increased resolution run of IllustrisTNG ($\sim100\times$ higher resolution than the fiducial suite),  finds the majority of CGM gas exists within T$\sim10^3-10^4$ K. The authors suggest this preferentially colder CGM is due to the higher resolution of their simulations, increasingly resolving cold gas structures, though it is surprising that a substantial amount of gas is below the virial temperature of the halos.

\subsection{Limitations}\label{subsec:Discussion_Limitations}
In this section, we discuss the potential limitations of our results based on the effects of the force and mass resolution limits and feedback physics not included in our simulations.

Both the Marvelous Dwarfs and Marvelous Massive Dwarfs suites have comparable force resolutions, both $<$ 100 pc. As our analysis of the CGM-DG tends to focus on global properties integrated over halo scales ($\sim100$ kpc), the results presented in this work are operating well above the physical length scale resolution limits. 

In regards to the mass resolution, previous observational works have posited that the CGM may comprise low ions existing in sub-kiloparsec clumps of cool/cold gas embedded in a warmer medium \citep[e.g.,][]{Hsu2011, McCourt2018, Zahedy2019, Augustin2021, Zheng_24}. Previous theoretical works from other simulations found that increased mass resolution increases the number of these cool clouds and affects their geometry \citep[e.g.,][]{FOGGIE_19,Gible24} and increases the overall cool gas content of the CGM \citep[e.g.,][]{FOGGIE_19,vandeVoort19,Hummels_2019,2024MNRAS.528.5412R}. Among the effects of increased mass resolution on the cold CGM, the GIBLE project \citep[IllustrisTNG;][]{Gible24} finds the cold gas fraction and mass in the CGM are converged properties when improving the cell mass resolution in the CGM from $m_{gas,CGM}\sim 8.5\times10^{5}M_{\odot}$ to $m_{gas,CGM}\sim 1.8\times10^{3}M_{\odot}$---comparable gas mass resolutions to the Marvelous Dwarfs and Marvelous Massive Dwarfs suites, and comparable to \citet{Hummels_2019}. 

With increasing mass resolution, the column densities in low ions may increase as much as a factor of 2 \citep[e.g., \HI\, in ][]{vandeVoort19, Hummels_2019}; however, work using FOGGIE suggests these column densities only marginally change (within $10\%–20\%$) when increasing CGM resolution \citep[e.g.,][]{FOGGIE_19}. For this work, the little variation in column densities and cool/warm gas fractions between suites in the \MM\, sample, despite differences in gas mass resolution, suggests that increased mass resolution could marginally increase our column densities but will have a greater effect on the substructure of the CGM, similar to results in \citet{FOGGIE_19} and \citet{Gible24}. 

Additionally, theoretical work focusing on Milky Way-mass galaxies has demonstrated that the inclusion of cosmic ray (CR) feedback can alter the CGM \citep[e.g.][]{2022ApJ...935...69B}. Although our simulations do not include CR physics, work done in the FIRE group predicts that the pressure support from CRs likely only becomes important for $M_{vir}\geq 10^{12} M_{\odot}$ galaxies \citep{Ji20}. Furthermore, \citet{DwarfCGM_in_FIRE} finds the inclusion of CRs in FIRE dwarfs only slightly enhanced the equivalent width in \OVI.  In IllustrisTNG, it has been shown that CRs largely do not alter the density, temperature, and total pressure in the CGM-DG. However, varying strengths of CR feedback can alter star formation \citep{IllustrisTNG_CRS_Ramesh24}. In both IllustrisTNG and in FIRE, the effect is strongest for MW galaxies (suppressing star formation by a factor of $\sim$10), but for $M_* \sim 10^9 M_{\odot}$ dwarf galaxies star formation can be suppressed by a factor of $\sim$2, and approaches unity with decreasing mass \citep{IllustrisTNG_CRS_Ramesh24}. Although the \MM\, sample lacks CR feedback, the aforementioned studies suggest that this feedback mechanism plays a less significant role in the low-mass CGM/galaxy evolution compared to its impact on higher-mass systems.

Lastly, in generating ionization fractions, we model the UV  photoionizing background (UVB) using the \citet{HM12} metagalactic radiation field. Previous work has demonstrated that the \citet{HM12} field underpredicts \HI\, photoionization \citep[e.g.,][]{2020MNRAS.493.1614F}, as such, adopting a different photoionizing background would likely increase the low ion content in our simulations. Further, we assume that the radiation from the central galaxy is negligible. Although this assumption is motivated by the low star formation rates seen in Figure \ref{fig:ScalingRelations}, existing literature has begun to indicate that explicitly modeling ionizing radiation from stars can affect CGM properties around Milky-Way mass galaxies. For instance, \citep{Baumschlager2024} implemented an ``on-the-fly'' radiative transfer (RT) model in \gasolineII\, which discretizes the radiation field at discrete photon energies. The authors find the inclusion of RT in modeling the radiation from the UVB and local stars results in a significant increase in \HI\, mass in the ISM and the CGM \citep[see figure 1 in ][]{Baumschlager2024}. Furthermore, Baumschlager et al.~(2025, in prep) finds that for the same dwarf galaxies as used in \citet{Mina2021}, RT resulted in a non-negligible impact on ions like CIV. Ultimately, the inclusion of RT in the \MM\, sample could alter the abundance of species like \HI, \CIV, and \OVI. 

\subsection{Observability of the CGM of dwarf galaxies}\label{subsec:Discussion_Observability}

The \MM\, sample can be leveraged to provide theoretical expectations for future CGM-DG observations. In this section, we cover the halo and stellar mass dependence of the CGM and its implications for observations. We organize the following section by galaxy mass, beginning with our interpretations for low-mass dwarf galaxies and moving up the mass scale.

\paragraph{\textbf{Low-mass Dwarf Galaxies}}
   We find the CGM of $M_*=10^{6.0-7.5} M_{\odot}$ galaxies exhibits a less multiphase structure compared to its higher-mass counterparts (Figure \ref{fig:CGMPhases_Profiles}). The majority of mass beyond $0.15R_{200c}$ resides in cool ($10^{4}<T/\text{K}<10^{4.6}$), diffuse ($n_{gas}<10^{-4}\text{cm}^{-3}$), and low metallicity $Z$ material in the outskirts of the CGM ($r>0.5R_{200c}$). Since this material in the outer CGM is very diffuse, the ion fractions for low ions are small (Figure \ref{fig:CGM_ionfractions}) despite being at cooler temperatures. For higher ions like \CIV\, and \OVI, the gas conditions are suitable for high ion fractions, but the lack of mass and metals (Figures \ref{fig:BaryonicBudget} and \ref{fig:f_zret_DiskCGM}) in these systems limits the amount of UV observable mass. Specifically, the CGM in this mass regime only has $\sim10^3M_{\odot}$ of silicon, carbon, or oxygen mass that is accessible with UV spectroscopy (Figure \ref{fig:UVmass}). Ultimately, the small amount of mass in UV-accessible ions leads to columns below the detection threshold in the vast majority of ions (Figures \ref{fig:LowIonColumnDensitiesResults} and \ref{fig:HighIonColumnDensitiesResults}). Thus, metals in the CGM of $M_*=10^{6.0-7.5} M_{\odot}$ galaxies are largely below current detection limits. However, within $b/R_{200c} < 0.2$, metal column densities may be detectable. Assuming the \MM\, sample does produce realistic CGM, future observations targeting $M_*=10^{6.0-7.5} M_{\odot}$ galaxies will likely need to observe the CGM within $\sim 4-5$ kpc of the disk to detect metal ions such as \CII, \CIV, \SiII, and \SiIII.

\paragraph{\textbf{High-mass Dwarf Galaxies}}
 The CGM of $M_*=10^{8.5-9.5} M_{\odot}$ dwarf galaxies shows a highly multiphase structure with $60-75\%$ of its mass in the warm/virialized phase, $30-40\%$ in the cool/photoionized phase, and $<5\%$ in a hot phase comprised of recently SNe heated and ejected gas (Figures \ref{fig:CoolWarm} and \ref{fig:CGMPhases_Profiles}). These galaxies have typically retained a larger fraction of the metals produced ($80-100\%$) within their halo, as compared to lower-mass galaxies (Figure \ref{fig:f_zret_DiskCGM}). Although the majority of the metals produced are within the disk region ($40-80\%$), and primarily in the gas phase, the CGM of these galaxies retains a non-negligible fraction of metals ($20-40\%$). We find the budget of UV-observable CGM mass is greatest for these galaxies, with $\sim32\%$ of carbon mass and $\sim60\%$ of silicon mass existing in UV-accessible ions (Figure \ref{fig:UVmass} and Table \ref{tab:f_UV}). However, only $\sim2\%$ of oxygen mass is in \OVI, while there is $>2\times$ as much mass in other ionization stages like \OII\, through \OV\, and \OVII\, (see Section \ref{subsec:Discussion_OVI}). Given the masses of UV observable ions, most ions studied are above detectable limits within $0.5R_{200c}$, but at greater distances, column densities (especially in low ions) are below detectable limits, in agreement with observations. For low ions, the decrease in column densities is due to declining ion fractions because of warmer and more diffuse conditions in the outer CGM regions. For this mass range, the \MM\, sample produces CGM column densities in broad agreement with observations ($\geq90\%$ of detections are within $2\sigma$). Assuming the \MM\, sample does produce realistic CGM, we predict future observations will continue to find detections of low ions and \CIV\, within 40-50 kpc of the galaxy. However, these lower ions do not probe the majority of the CGM mass residing in the virialized phase, which retains a comparable fraction of metals to the cool phase ($\sim10\%$, Figure \ref{fig:f_zret_PHASES}).

\section{Summary}\label{sec:summary}
In this work, we use two suites of cosmological zoom simulations to study the CGM of dwarf galaxies. We select a sample of 64 isolated, $z=0$ dwarf galaxies within the stellar mass range $6.0\leq \text{log}(M_*/M_{\odot}) \leq 9.5$, referred to as the \MM\, sample. We compare to current observations by deriving synthetic column densities in various ions and characterize the CGM by investigating its mass and metal retention, and physical properties. Ultimately, we find:

\begin{enumerate}
    \item The \MM\, sample produces synthetic column densities in broad agreement with current observations included in \citet{Zheng_24} and a sub-sample of those presented in \citet{Mishra_24}. Generally, $M_*>10^8 M_{\odot}$ dwarf galaxies have CGM that can be observed in both \HI\, and metal ions. While the CGM of lower mass dwarf galaxies can be observed in \HI, the metal ions studied show column densities below detectable limits, likely due to large rates of metal loss from the CGM (Figure \ref{fig:f_zret_DiskCGM}).
    
    \begin{enumerate}
        \item In the most massive dwarf galaxies, \NHI\, shows a steeply declining inner profile that plateaus at $\sim 10^{14}$ cm$^{-2}$. With decreasing mass, the inner profile becomes less steep while the outer plateau value remains around $\sim 10^{14}$ cm$^{-2}$. We argue this profile is due to the density and temperature profile of the CGM setting the ionization fractions of \HI\, (see Section \ref{subsec:Discussion_CGMStructure} and Figures \ref{fig:CGMPhases_Profiles} and \ref{fig:CGM_ionfractions}).

        \item Low ions (\SiII, \SiIII, \CII) have a steeply declining profile that falls well below detectable limits by $0.5R_{200c}$. This trend is in agreement with the published observations where these ions are only observable at low impact parameters ($b/R_{200c}<0.5$). We argue this steep decline in column density is due to the CGM becoming warmer at greater values of $r/R_{200c}$, leading to the decline in ionization fractions of these low ions (Figures \ref{fig:CGMPhases_Profiles} and \ref{fig:CGM_ionfractions}). 

        \item The intermediate ion \CIV\, does not have a column density profile that falls off as rapidly as lower ions. The shallower \NCIV\, profile is due to \CIV\,  preferentially existing in the warmer, intermediate regions of the halo ($r/R_{200c}\sim0.5$) where its ionization fractions peak (Figure \ref{fig:CGM_ionfractions}). The \MM\, sample reproduce 0/8 and 5/8 published detections within $1\sigma$ and $2\sigma$ respectively. The lack of detections lying within $1\sigma$ of the simulated median indicates the \MM\, sample likely underproduces \CIV\, and overproduces higher ionization states of carbon (Section \ref{subsec:Discussion_OVI}).

        \item The highest ion we include, \OVI, has a flat column density profile because it exists predominantly beyond $R_{200c}$ where ionization conditions are most suitable (Figure \ref{fig:CGM_ionfractions}). We find \NOVI\, values are less than all published detections at higher redshifts by $\sim$1 dex, indicating the \MM\, sample underpredicts the \OVI-bearing phase. Similar to carbon, we find substantial mass in other ionization states of oxygen.
        
    \end{enumerate}
    
    \item The CGM ($r/R_{200c}=0.15-1.0$) in the \MM\, sample dominates the total baryonic budget  for 58/64 galaxies (Figure \ref{fig:BaryonicBudget}). Across halo masses, the CGM tends to retain a consistent $15-20\%$ of metals produced (Figure \ref{fig:f_zret_PHASES}). Although the CGM metal retention shows little variation with respect to the galaxy mass, the disk retention decreases with mass, and correspondingly, the fraction of metals lost increases (Figure \ref{fig:f_zret_DiskCGM}). With decreasing halo mass, the median CGM conditions become preferentially cooler, more diffuse, and lower metallicity (Figure \ref{fig:CGMPhases_Profiles}).     

    \item For all galaxies in the \MM\, sample, there are two prominent temperature phases that comprise $>94\%$ of the CGM mass.  
    \begin{enumerate}
        \item The warm phase ($10^{4.5} < T < 10^{5.5}$K) constitutes the majority of CGM mass ($\sim75\%$) for 50/64 galaxies in the \MM\, sample (Figure \ref{fig:CoolWarm}). This phase traces the virial temperature of the halo and exists primarily in the outer regions at $r/R_{200c}\geq 0.5$. In terms of metal retention, the warm phase harbors between $5-10\%$ of all metals produced by the dwarf galaxy (Figure \ref{fig:f_zret_PHASES}). 
        
        \item The cool phase ($10^{3.8} < T < 10^{4.5}$K) tends to comprise $\sim25\%$ of the CGM mass, except at the low virial masses, $M_{200c}<10^{9.9}$M$_{\odot}$. At the low mass end, the virial temperature of the halo approaches $T\sim 10^{4.5}$K, resulting in $\sim100\%$ of CGM mass being within the cool phase. For $M_{200c}>10^{9.9}$M$_{\odot}$ galaxies, the cool phase exists in the inner regions of the halo $r\leq 0.5 R_{vir}$. It represents the primarily photoionized phase where \HI\, and low metal ions' ionization fractions are greatest in the CGM. The cool phase tends to harbor between $5-10\%$ of all metals produced by the dwarf galaxy (Figure \ref{fig:f_zret_PHASES}), notably a comparable fraction to the warm phase.     
    \end{enumerate}
    
\end{enumerate}

Our work is focused on the $z=0$ simulated universe and useful in comparing a large sample of simulated dwarf galaxies to Local observations. Future work can expand upon this study by including a range of redshifts and comparing to more observational work, such as \citet{Mishra_24} and \citet{Dutta24}. Additionally, studies investigating the properties of \OVI, how this ion relates to dwarf galaxy activity, and, more generally, how the CGM of dwarf galaxies depends on the modeling choices built in modern simulations, will likely prove useful in advancing our understanding of simulated feedback physics and the baryon cycle in the low-mass regime.

\bigskip
We thank Kristen B. W. McQuinn and Iryna S. Butsky for helpful conversations during this work. Resources supporting this work were provided by the NASA High-End Computing (HEC) Program through the NASA Advanced Supercomputing (NAS) Division at Ames Research Center. Some of the simulations were performed using resources made available by the Flatiron Institute. The Flatiron Institute is a division of the Simons Foundation. This work used Stampede2 at the Texas Advanced Computing Center (TACC) through allocation MCA94P018 from the Advanced Cyberinfrastructure Coordination Ecosystem: Services \& Support (ACCESS) program, which is supported by U.S. National Science Foundation grants \#2138259, \#2138286, \#2138307, \#2137603, and \#2138296. D.R.P and A.M.B are supported by NASA Grant 80NSSC24K0894. A.M.B acknowledges support by grant FI-CCA-Research-00011826 from the Simons Foundation. J.W. is supported by a grant from NSERC (National Science and Engineering Research Council) Canada. C.C. was supported by the NSF under CAREER grant AST-1848107, and this work was performed in part at Aspen Center for Physics, which is supported by NSF grant PHY-2210452. N.N.S. was supported by the National Science Foundation MPS-Ascend award ID 2212959.

\bibliography{sample7}{}

@ARTICLE{Zheng_24,
       author = {{Zheng}, Yong and {Faerman}, Yakov and {Oppenheimer}, Benjamin D. and {Putman}, Mary E. and {McQuinn}, Kristen B.~W. and {Kirby}, Evan N. and {Burchett}, Joseph N. and {Telford}, O. Grace and {Werk}, Jessica K. and {Kim}, Doyeon A.},
        title = "{A Comprehensive Investigation of Metals in the Circumgalactic Medium of Nearby Dwarf Galaxies}",
      journal = {\apj},
         year = 2024,
        month = jan,
       volume = {960},
       number = {1},
          eid = {55},
        pages = {55},
          doi = {10.3847/1538-4357/acfe6b},
archivePrefix = {arXiv},
       eprint = {2301.12233},
 primaryClass = {astro-ph.GA},
       adsurl = {https://ui.adsabs.harvard.edu/abs/2024ApJ...960...55Z}
}

@ARTICLE{yt,
   author = {{Turk}, M.~J. and {Smith}, B.~D. and {Oishi}, J.~S. and {Skory}, S. and
     {Skillman}, S.~W. and {Abel}, T. and {Norman}, M.~L.},
    title = "{yt: A Multi-code Analysis Toolkit for Astrophysical Simulation Data}",
  journal = {The Astrophysical Journal Supplement Series},
archivePrefix = "arXiv",
   eprint = {1011.3514},
 primaryClass = "astro-ph.IM",
     year = 2011,
    month = jan,
   volume = 192,
      eid = {9},
    pages = {9},
      doi = {10.1088/0067-0049/192/1/9},
   adsurl = {https://ui.adsabs.harvard.edu/abs/2011ApJS..192....9T}
}

@ARTICLE{2021ApJ...923...35M,
       author = {{Munshi}, Ferah and {Brooks}, Alyson M. and {Applebaum}, Elaad and {Christensen}, Charlotte R. and {Quinn}, T. and {Sligh}, Serena},
        title = "{Quantifying Scatter in Galaxy Formation at the Lowest Masses}",
      journal = {\apj},
         year = 2021,
        month = dec,
       volume = {923},
       number = {1},
          eid = {35},
        pages = {35},
          doi = {10.3847/1538-4357/ac0db6},
archivePrefix = {arXiv},
       eprint = {2101.05822},
 primaryClass = {astro-ph.GA},
       adsurl = {https://ui.adsabs.harvard.edu/abs/2021ApJ...923...35M}
}

@ARTICLE{2007ApJS..170..377S,
       author = {{Spergel}, D.~N. and {Bean}, R. and {Dor{\'e}}, O. and {Nolta}, M.~R. and {Bennett}, C.~L. and {Dunkley}, J. and {Hinshaw}, G. and {Jarosik}, N. and {Komatsu}, E. and {Page}, L. and {Peiris}, H.~V. and {Verde}, L. and {Halpern}, M. and {Hill}, R.~S. and {Kogut}, A. and {Limon}, M. and {Meyer}, S.~S. and {Odegard}, N. and {Tucker}, G.~S. and {Weiland}, J.~L. and {Wollack}, E. and {Wright}, E.~L.},
        title = "{Three-Year Wilkinson Microwave Anisotropy Probe (WMAP) Observations: Implications for Cosmology}",
      journal = {\apjs},
         year = 2007,
        month = jun,
       volume = {170},
       number = {2},
        pages = {377-408},
          doi = {10.1086/513700},
archivePrefix = {arXiv},
       eprint = {astro-ph/0603449},
 primaryClass = {astro-ph},
       adsurl = {https://ui.adsabs.harvard.edu/abs/2007ApJS..170..377S}
}

@ARTICLE{CHANGA,
       author = {{Menon}, Harshitha and {Wesolowski}, Lukasz and {Zheng}, Gengbin and {Jetley}, Pritish and {Kale}, Laxmikant and {Quinn}, Thomas and {Governato}, Fabio},
        title = "{Adaptive techniques for clustered N-body cosmological simulations}",
      journal = {Computational Astrophysics and Cosmology},
         year = 2015,
        month = mar,
       volume = {2},
          eid = {1},
        pages = {1},
          doi = {10.1186/s40668-015-0007-9},
archivePrefix = {arXiv},
       eprint = {1409.1929},
 primaryClass = {astro-ph.IM},
       adsurl = {https://ui.adsabs.harvard.edu/abs/2015ComAC...2....1M}
}

@ARTICLE{HM12,
       author = {{Haardt}, Francesco and {Madau}, Piero},
        title = "{Radiative Transfer in a Clumpy Universe. IV. New Synthesis Models of the Cosmic UV/X-Ray Background}",
      journal = {\apj},
         year = 2012,
        month = feb,
       volume = {746},
       number = {2},
          eid = {125},
        pages = {125},
          doi = {10.1088/0004-637X/746/2/125},
archivePrefix = {arXiv},
       eprint = {1105.2039},
 primaryClass = {astro-ph.CO},
       adsurl = {https://ui.adsabs.harvard.edu/abs/2012ApJ...746..125H}
}

@ARTICLE{Stinson06-BW,
       author = {{Stinson}, Greg and {Seth}, Anil and {Katz}, Neal and {Wadsley}, James and {Governato}, Fabio and {Quinn}, Tom},
        title = "{Star formation and feedback in smoothed particle hydrodynamic simulations - I. Isolated galaxies}",
      journal = {\mnras},
         year = 2006,
        month = dec,
       volume = {373},
       number = {3},
        pages = {1074-1090},
          doi = {10.1111/j.1365-2966.2006.11097.x},
archivePrefix = {arXiv},
       eprint = {astro-ph/0602350},
 primaryClass = {astro-ph},
       adsurl = {https://ui.adsabs.harvard.edu/abs/2006MNRAS.373.1074S}
}

@ARTICLE{Keller14-SB,
       author = {{Keller}, B.~W. and {Wadsley}, J. and {Benincasa}, S.~M. and {Couchman}, H.~M.~P.},
        title = "{A superbubble feedback model for galaxy simulations}",
      journal = {\mnras},
         year = 2014,
        month = aug,
       volume = {442},
       number = {4},
        pages = {3013-3025},
          doi = {10.1093/mnras/stu1058},
archivePrefix = {arXiv},
       eprint = {1405.2625},
 primaryClass = {astro-ph.GA},
       adsurl = {https://ui.adsabs.harvard.edu/abs/2014MNRAS.442.3013K}
}

@ARTICLE{TumCGMreview,
       author = {{Tumlinson}, Jason and {Peeples}, Molly S. and {Werk}, Jessica K.},
        title = "{The Circumgalactic Medium}",
      journal = {\araa},
         year = 2017,
        month = aug,
       volume = {55},
       number = {1},
        pages = {389-432},
          doi = {10.1146/annurev-astro-091916-055240},
archivePrefix = {arXiv},
       eprint = {1709.09180},
 primaryClass = {astro-ph.GA},
       adsurl = {https://ui.adsabs.harvard.edu/abs/2017ARA&A..55..389T}
}

@ARTICLE{Bordoloi_14,
       author = {{Bordoloi}, Rongmon and {Tumlinson}, Jason and {Werk}, Jessica K. and {Oppenheimer}, Benjamin D. and {Peeples}, Molly S. and {Prochaska}, J. Xavier and {Tripp}, Todd M. and {Katz}, Neal and {Dav{\'e}}, Romeel and {Fox}, Andrew J. and {Thom}, Christopher and {Ford}, Amanda Brady and {Weinberg}, David H. and {Burchett}, Joseph N. and {Kollmeier}, Juna A.},
        title = "{The COS-Dwarfs Survey: The Carbon Reservoir around Sub-L* Galaxies}",
      journal = {\apj},
         year = 2014,
        month = dec,
       volume = {796},
       number = {2},
          eid = {136},
        pages = {136},
          doi = {10.1088/0004-637X/796/2/136},
archivePrefix = {arXiv},
       eprint = {1406.0509},
 primaryClass = {astro-ph.GA},
       adsurl = {https://ui.adsabs.harvard.edu/abs/2014ApJ...796..136B}
}

@ARTICLE{Christensen_2018,
       author = {{Christensen}, Charlotte R. and {Dav{\'e}}, Romeel and {Brooks}, Alyson and {Quinn}, Thomas and {Shen}, Sijing},
        title = "{Tracing Outflowing Metals in Simulations of Dwarf and Spiral Galaxies}",
      journal = {\apj},
         year = 2018,
        month = nov,
       volume = {867},
       number = {2},
          eid = {142},
        pages = {142},
          doi = {10.3847/1538-4357/aae374},
archivePrefix = {arXiv},
       eprint = {1808.07872},
 primaryClass = {astro-ph.GA},
       adsurl = {https://ui.adsabs.harvard.edu/abs/2018ApJ...867..142C}
}

@ARTICLE{Bordoloi_18,
       author = {{Bordoloi}, Rongmon and {Prochaska}, J. Xavier and {Tumlinson}, Jason and {Werk}, Jessica K. and {Tripp}, Todd M. and {Burchett}, Joseph N.},
        title = "{On the CGM Fundamental Plane: The Halo Mass Dependency of Circumgalactic H I}",
      journal = {\apj},
         year = 2018,
        month = sep,
       volume = {864},
       number = {2},
          eid = {132},
        pages = {132},
          doi = {10.3847/1538-4357/aad8ac},
archivePrefix = {arXiv},
       eprint = {1712.02348},
 primaryClass = {astro-ph.GA},
       adsurl = {https://ui.adsabs.harvard.edu/abs/2018ApJ...864..132B}
}

@ARTICLE{LiangChen_14,
       author = {{Liang}, Cameron J. and {Chen}, Hsiao-Wen},
        title = "{Mining circumgalactic baryons in the low-redshift universe}",
      journal = {\mnras},
         year = 2014,
        month = dec,
       volume = {445},
       number = {2},
        pages = {2061-2081},
          doi = {10.1093/mnras/stu1901},
archivePrefix = {arXiv},
       eprint = {1402.3602},
 primaryClass = {astro-ph.CO},
       adsurl = {https://ui.adsabs.harvard.edu/abs/2014MNRAS.445.2061L}
}

@ARTICLE{Tchernyshyov_22,
       author = {{Tchernyshyov}, Kirill and {Werk}, Jessica K. and {Wilde}, Matthew C. and {Prochaska}, J. Xavier and {Tripp}, Todd M. and {Burchett}, Joseph N. and {Bordoloi}, Rongmon and {Howk}, J. Christopher and {Lehner}, Nicolas and {O'Meara}, John M. and {Tejos}, Nicolas and {Tumlinson}, Jason},
        title = "{The CGM$^{2}$ Survey: Circumgalactic O VI from Dwarf to Massive Star-forming Galaxies}",
      journal = {\apj},
         year = 2022,
        month = mar,
       volume = {927},
       number = {2},
          eid = {147},
        pages = {147},
          doi = {10.3847/1538-4357/ac450c},
archivePrefix = {arXiv},
       eprint = {2110.13167},
 primaryClass = {astro-ph.GA},
       adsurl = {https://ui.adsabs.harvard.edu/abs/2022ApJ...927..147T}
}

@ARTICLE{Johnson_17,
       author = {{Johnson}, Sean D. and {Chen}, Hsiao-Wen and {Mulchaey}, John S. and {Schaye}, Joop and {Straka}, Lorrie A.},
        title = "{The Extent of Chemically Enriched Gas around Star-forming Dwarf Galaxies}",
      journal = {\apjl},
         year = 2017,
        month = nov,
       volume = {850},
       number = {1},
          eid = {L10},
        pages = {L10},
          doi = {10.3847/2041-8213/aa9370},
archivePrefix = {arXiv},
       eprint = {1710.06441},
 primaryClass = {astro-ph.GA},
       adsurl = {https://ui.adsabs.harvard.edu/abs/2017ApJ...850L..10J}
}

@ARTICLE{2022ApJ...935...69B,
       author = {{Butsky}, Iryna S. and {Werk}, Jessica K. and {Tchernyshyov}, Kirill and {Fielding}, Drummond B. and {Breneman}, Joseph and {Piacitelli}, Daniel R. and {Quinn}, Thomas R. and {Sanchez}, N. Nicole and {Cruz}, Akaxia and {Hummels}, Cameron B. and {Burchett}, Joseph N. and {Tremmel}, Michael},
        title = "{The Impact of Cosmic Rays on the Kinematics of the Circumgalactic Medium}",
      journal = {\apj},
         year = 2022,
        month = aug,
       volume = {935},
       number = {2},
          eid = {69},
        pages = {69},
          doi = {10.3847/1538-4357/ac7ebd},
archivePrefix = {arXiv},
       eprint = {2106.14889},
 primaryClass = {astro-ph.GA},
       adsurl = {https://ui.adsabs.harvard.edu/abs/2022ApJ...935...69B}
}

@ARTICLE{Burchett_16,
       author = {{Burchett}, Joseph N. and {Tripp}, Todd M. and {Bordoloi}, Rongmon and {Werk}, Jessica K. and {Prochaska}, J. Xavier and {Tumlinson}, Jason and {Willmer}, C.~N.~A. and {O'Meara}, John and {Katz}, Neal},
        title = "{A Deep Search for Faint Galaxies Associated with Very Low Redshift C IV Absorbers. III. The Mass- and Environment-dependent Circumgalactic Medium}",
      journal = {\apj},
         year = 2016,
        month = dec,
       volume = {832},
       number = {2},
          eid = {124},
        pages = {124},
          doi = {10.3847/0004-637X/832/2/124},
archivePrefix = {arXiv},
       eprint = {1512.00853},
 primaryClass = {astro-ph.GA},
       adsurl = {https://ui.adsabs.harvard.edu/abs/2016ApJ...832..124B}
}

@ARTICLE{2015ApJ...815L..17M,
       author = {{McQuinn}, Kristen B.~W. and {Skillman}, Evan D. and {Dolphin}, Andrew and {Cannon}, John M. and {Salzer}, John J. and {Rhode}, Katherine L. and {Adams}, Elizabeth A.~K. and {Berg}, Danielle and {Giovanelli}, Riccardo and {Haynes}, Martha P.},
        title = "{Leo P: How Many Metals Can a Very Low Mass, Isolated Galaxy Retain?}",
      journal = {\apjl},
         year = 2015,
        month = dec,
       volume = {815},
       number = {2},
          eid = {L17},
        pages = {L17},
          doi = {10.1088/2041-8205/815/2/L17},
archivePrefix = {arXiv},
       eprint = {1512.00459},
 primaryClass = {astro-ph.GA},
       adsurl = {https://ui.adsabs.harvard.edu/abs/2015ApJ...815L..17M}
}

@ARTICLE{QuBregman_22,
       author = {{Qu}, Zhijie and {Bregman}, Joel N.},
        title = "{Absorption Line Search through Three Local Group Dwarf Galaxy Halos}",
      journal = {\apj},
         year = 2022,
        month = mar,
       volume = {927},
       number = {2},
          eid = {228},
        pages = {228},
          doi = {10.3847/1538-4357/ac51df},
archivePrefix = {arXiv},
       eprint = {2203.08246},
 primaryClass = {astro-ph.GA},
       adsurl = {https://ui.adsabs.harvard.edu/abs/2022ApJ...927..228Q}
}

@ARTICLE{Christensen_2016,
       author = {{Christensen}, Charlotte R. and {Dav{\'e}}, Romeel and {Governato}, Fabio and {Pontzen}, Andrew and {Brooks}, Alyson and {Munshi}, Ferah and {Quinn}, Thomas and {Wadsley}, James},
        title = "{In-N-Out: The Gas Cycle from Dwarfs to Spiral Galaxies}",
      journal = {\apj},
         year = 2016,
        month = jun,
       volume = {824},
       number = {1},
          eid = {57},
        pages = {57},
          doi = {10.3847/0004-637X/824/1/57},
archivePrefix = {arXiv},
       eprint = {1508.00007},
 primaryClass = {astro-ph.GA},
       adsurl = {https://ui.adsabs.harvard.edu/abs/2016ApJ...824...57C}
}

@ARTICLE{Mina2021,
       author = {{Mina}, Mattia and {Shen}, Sijing and {Keller}, Benjamin Walter and {Mayer}, Lucio and {Madau}, Piero and {Wadsley}, James},
        title = "{The baryon cycle of Seven Dwarfs with superbubble feedback}",
      journal = {\aap},
         year = 2021,
        month = nov,
       volume = {655},
          eid = {A22},
        pages = {A22},
          doi = {10.1051/0004-6361/202039420},
archivePrefix = {arXiv},
       eprint = {2009.06646},
 primaryClass = {astro-ph.GA},
       adsurl = {https://ui.adsabs.harvard.edu/abs/2021A&A...655A..22M}
}

@ARTICLE{Trident,
       author = {{Hummels}, Cameron B. and {Smith}, Britton D. and {Silvia}, Devin W.},
        title = "{Trident: A Universal Tool for Generating Synthetic Absorption Spectra from Astrophysical Simulations}",
      journal = {\apj},
         year = 2017,
        month = sep,
       volume = {847},
       number = {1},
          eid = {59},
        pages = {59},
          doi = {10.3847/1538-4357/aa7e2d},
archivePrefix = {arXiv},
       eprint = {1612.03935},
 primaryClass = {astro-ph.IM},
       adsurl = {https://ui.adsabs.harvard.edu/abs/2017ApJ...847...59H}
}

@ARTICLE{Asplund2009,
       author = {{Asplund}, Martin and {Grevesse}, Nicolas and {Sauval}, A. Jacques and {Scott}, Pat},
        title = "{The Chemical Composition of the Sun}",
      journal = {\araa},
         year = 2009,
        month = sep,
       volume = {47},
       number = {1},
        pages = {481-522},
          doi = {10.1146/annurev.astro.46.060407.145222},
archivePrefix = {arXiv},
       eprint = {0909.0948},
 primaryClass = {astro-ph.SR},
       adsurl = {https://ui.adsabs.harvard.edu/abs/2009ARA&A..47..481A}
}

@ARTICLE{Ferland17,
   author = {{Ferland}, G.~J. and {Chatzikos}, M. and {Guzm{\'a}n}, F. and {Lykins}, M.~L. and {van Hoof}, P.~A.~M. and {Williams}, R.~J.~R. and {Abel}, N.~P. and {Badnell}, N.~R. and {Keenan}, F.~P. and {Porter}, R.~L. and {Stancil}, P.~C.},
    title = "{The 2017 Release Cloudy}",
  journal = {\rmxaa},
archivePrefix = "arXiv",
   eprint = {1705.10877},
     year = 2017,
    month = oct,
   volume = 53,
    pages = {385-438},
   adsurl = {http://adsabs.harvard.edu/abs/2017RMxAA..53..385F}}

@ARTICLE{Qu_24,
       author = {{Qu}, Zhijie and {Chen}, Hsiao-Wen and {Johnson}, Sean D. and {Rudie}, Gwen C. and {Zahedy}, Fakhri S. and {DePalma}, David and {Schaye}, Joop and {Boettcher}, Erin T. and {Cantalupo}, Sebastiano and {Chen}, Mandy C. and {Faucher-Gigu{\`e}re}, Claude-Andr{\'e} and {Li}, Jennifer I-Hsiu and {Mulchaey}, John S. and {Petitjean}, Patrick and {Rafelski}, Marc},
        title = "{The Cosmic Ultraviolet Baryon Survey (CUBS) VII: on the warm-hot circumgalactic medium probed by O VI and Ne VIII at 0.4 $\lesssim$ z $\lesssim$ 0.7}",
      journal = {arXiv e-prints},
         year = 2024,
        month = feb,
          eid = {arXiv:2402.08016},
        pages = {arXiv:2402.08016},
          doi = {10.48550/arXiv.2402.08016},
archivePrefix = {arXiv},
       eprint = {2402.08016},
 primaryClass = {astro-ph.GA},
       adsurl = {https://ui.adsabs.harvard.edu/abs/2024arXiv240208016Q}
}

@ARTICLE{2012MNRAS.425.3058C,
       author = {{Christensen}, Charlotte and {Quinn}, Thomas and {Governato}, Fabio and {Stilp}, Adrienne and {Shen}, Sijing and {Wadsley}, James},
        title = "{Implementing molecular hydrogen in hydrodynamic simulations of galaxy formation}",
      journal = {\mnras},
         year = 2012,
        month = oct,
       volume = {425},
       number = {4},
        pages = {3058-3076},
          doi = {10.1111/j.1365-2966.2012.21628.x},
archivePrefix = {arXiv},
       eprint = {1205.5567},
 primaryClass = {astro-ph.CO},
       adsurl = {https://ui.adsabs.harvard.edu/abs/2012MNRAS.425.3058C}
}

@ARTICLE{2010MNRAS.407.1581S,
       author = {{Shen}, S. and {Wadsley}, J. and {Stinson}, G.},
        title = "{The enrichment of the intergalactic medium with adiabatic feedback - I. Metal cooling and metal diffusion}",
      journal = {\mnras},
         year = 2010,
        month = sep,
       volume = {407},
       number = {3},
        pages = {1581-1596},
          doi = {10.1111/j.1365-2966.2010.17047.x},
archivePrefix = {arXiv},
       eprint = {0910.5956},
 primaryClass = {astro-ph.CO},
       adsurl = {https://ui.adsabs.harvard.edu/abs/2010MNRAS.407.1581S}
}

@ARTICLE{Kroupa2001_IMF,
       author = {{Kroupa}, Pavel},
        title = "{On the variation of the initial mass function}",
      journal = {\mnras},
         year = 2001,
        month = apr,
       volume = {322},
       number = {2},
        pages = {231-246},
          doi = {10.1046/j.1365-8711.2001.04022.x},
archivePrefix = {arXiv},
       eprint = {astro-ph/0009005},
 primaryClass = {astro-ph},
       adsurl = {https://ui.adsabs.harvard.edu/abs/2001MNRAS.322..231K}
}

@ARTICLE{2016A&A...594A..13P,
       author = {{Planck Collaboration} and {Ade}, P.~A.~R. and {Aghanim}, N. and {Arnaud}, M. and {Ashdown}, M. and {Aumont}, J. and {Baccigalupi}, C. and {Banday}, A.~J. and {Barreiro}, R.~B. and {Bartlett}, J.~G. and {Bartolo}, N. and {Battaner}, E. and {Battye}, R. and {Benabed}, K. and {Beno{\^\i}t}, A. and {Benoit-L{\'e}vy}, A. and {Bernard}, J. -P. and {Bersanelli}, M. and {Bielewicz}, P. and {Bock}, J.~J. and {Bonaldi}, A. and {Bonavera}, L. and {Bond}, J.~R. and {Borrill}, J. and {Bouchet}, F.~R. and {Boulanger}, F. and {Bucher}, M. and {Burigana}, C. and {Butler}, R.~C. and {Calabrese}, E. and {Cardoso}, J. -F. and {Catalano}, A. and {Challinor}, A. and {Chamballu}, A. and {Chary}, R. -R. and {Chiang}, H.~C. and {Chluba}, J. and {Christensen}, P.~R. and {Church}, S. and {Clements}, D.~L. and {Colombi}, S. and {Colombo}, L.~P.~L. and {Combet}, C. and {Coulais}, A. and {Crill}, B.~P. and {Curto}, A. and {Cuttaia}, F. and {Danese}, L. and {Davies}, R.~D. and {Davis}, R.~J. and {de Bernardis}, P. and {de Rosa}, A. and {de Zotti}, G. and {Delabrouille}, J. and {D{\'e}sert}, F. -X. and {Di Valentino}, E. and {Dickinson}, C. and {Diego}, J.~M. and {Dolag}, K. and {Dole}, H. and {Donzelli}, S. and {Dor{\'e}}, O. and {Douspis}, M. and {Ducout}, A. and {Dunkley}, J. and {Dupac}, X. and {Efstathiou}, G. and {Elsner}, F. and {En{\ss}lin}, T.~A. and {Eriksen}, H.~K. and {Farhang}, M. and {Fergusson}, J. and {Finelli}, F. and {Forni}, O. and {Frailis}, M. and {Fraisse}, A.~A. and {Franceschi}, E. and {Frejsel}, A. and {Galeotta}, S. and {Galli}, S. and {Ganga}, K. and {Gauthier}, C. and {Gerbino}, M. and {Ghosh}, T. and {Giard}, M. and {Giraud-H{\'e}raud}, Y. and {Giusarma}, E. and {Gjerl{\o}w}, E. and {Gonz{\'a}lez-Nuevo}, J. and {G{\'o}rski}, K.~M. and {Gratton}, S. and {Gregorio}, A. and {Gruppuso}, A. and {Gudmundsson}, J.~E. and {Hamann}, J. and {Hansen}, F.~K. and {Hanson}, D. and {Harrison}, D.~L. and {Helou}, G. and {Henrot-Versill{\'e}}, S. and {Hern{\'a}ndez-Monteagudo}, C. and {Herranz}, D. and {Hildebrandt}, S.~R. and {Hivon}, E. and {Hobson}, M. and {Holmes}, W.~A. and {Hornstrup}, A. and {Hovest}, W. and {Huang}, Z. and {Huffenberger}, K.~M. and {Hurier}, G. and {Jaffe}, A.~H. and {Jaffe}, T.~R. and {Jones}, W.~C. and {Juvela}, M. and {Keih{\"a}nen}, E. and {Keskitalo}, R. and {Kisner}, T.~S. and {Kneissl}, R. and {Knoche}, J. and {Knox}, L. and {Kunz}, M. and {Kurki-Suonio}, H. and {Lagache}, G. and {L{\"a}hteenm{\"a}ki}, A. and {Lamarre}, J. -M. and {Lasenby}, A. and {Lattanzi}, M. and {Lawrence}, C.~R. and {Leahy}, J.~P. and {Leonardi}, R. and {Lesgourgues}, J. and {Levrier}, F. and {Lewis}, A. and {Liguori}, M. and {Lilje}, P.~B. and {Linden-V{\o}rnle}, M. and {L{\'o}pez-Caniego}, M. and {Lubin}, P.~M. and {Mac{\'\i}as-P{\'e}rez}, J.~F. and {Maggio}, G. and {Maino}, D. and {Mandolesi}, N. and {Mangilli}, A. and {Marchini}, A. and {Maris}, M. and {Martin}, P.~G. and {Martinelli}, M. and {Mart{\'\i}nez-Gonz{\'a}lez}, E. and {Masi}, S. and {Matarrese}, S. and {McGehee}, P. and {Meinhold}, P.~R. and {Melchiorri}, A. and {Melin}, J. -B. and {Mendes}, L. and {Mennella}, A. and {Migliaccio}, M. and {Millea}, M. and {Mitra}, S. and {Miville-Desch{\^e}nes}, M. -A. and {Moneti}, A. and {Montier}, L. and {Morgante}, G. and {Mortlock}, D. and {Moss}, A. and {Munshi}, D. and {Murphy}, J.~A. and {Naselsky}, P. and {Nati}, F. and {Natoli}, P. and {Netterfield}, C.~B. and {N{\o}rgaard-Nielsen}, H.~U. and {Noviello}, F. and {Novikov}, D. and {Novikov}, I. and {Oxborrow}, C.~A. and {Paci}, F. and {Pagano}, L. and {Pajot}, F. and {Paladini}, R. and {Paoletti}, D. and {Partridge}, B. and {Pasian}, F. and {Patanchon}, G. and {Pearson}, T.~J. and {Perdereau}, O. and {Perotto}, L. and {Perrotta}, F. and {Pettorino}, V. and {Piacentini}, F. and {Piat}, M. and {Pierpaoli}, E. and {Pietrobon}, D. and {Plaszczynski}, S. and {Pointecouteau}, E. and {Polenta}, G. and {Popa}, L. and {Pratt}, G.~W. and {Pr{\'e}zeau}, G. and {Prunet}, S. and {Puget}, J. -L. and {Rachen}, J.~P. and {Reach}, W.~T. and {Rebolo}, R. and {Reinecke}, M. and {Remazeilles}, M. and {Renault}, C. and {Renzi}, A. and {Ristorcelli}, I. and {Rocha}, G. and {Rosset}, C. and {Rossetti}, M. and {Roudier}, G. and {Rouill{\'e} d'Orfeuil}, B. and {Rowan-Robinson}, M. and {Rubi{\~n}o-Mart{\'\i}n}, J.~A. and {Rusholme}, B. and {Said}, N. and {Salvatelli}, V. and {Salvati}, L. and {Sandri}, M. and {Santos}, D. and {Savelainen}, M. and {Savini}, G. and {Scott}, D. and {Seiffert}, M.~D. and {Serra}, P. and {Shellard}, E.~P.~S. and {Spencer}, L.~D. and {Spinelli}, M. and {Stolyarov}, V. and {Stompor}, R. and {Sudiwala}, R. and {Sunyaev}, R. and {Sutton}, D. and {Suur-Uski}, A. -S. and {Sygnet}, J. -F. and {Tauber}, J.~A. and {Terenzi}, L. and {Toffolatti}, L. and {Tomasi}, M. and {Tristram}, M. and {Trombetti}, T. and {Tucci}, M. and {Tuovinen}, J. and {T{\"u}rler}, M. and {Umana}, G. and {Valenziano}, L. and {Valiviita}, J. and {Van Tent}, F. and {Vielva}, P. and {Villa}, F. and {Wade}, L.~A. and {Wandelt}, B.~D. and {Wehus}, I.~K. and {White}, M. and {White}, S.~D.~M. and {Wilkinson}, A. and {Yvon}, D. and {Zacchei}, A. and {Zonca}, A.},
        title = "{Planck 2015 results. XIII. Cosmological parameters}",
      journal = {\aap},
         year = 2016,
        month = sep,
       volume = {594},
          eid = {A13},
        pages = {A13},
          doi = {10.1051/0004-6361/201525830},
archivePrefix = {arXiv},
       eprint = {1502.01589},
 primaryClass = {astro-ph.CO},
       adsurl = {https://ui.adsabs.harvard.edu/abs/2016A&A...594A..13P}
}

@ARTICLE{Stern2021,
       author = {{Stern}, Jonathan and {Faucher-Gigu{\`e}re}, Claude-Andr{\'e} and {Fielding}, Drummond and {Quataert}, Eliot and {Hafen}, Zachary and {Gurvich}, Alexander B. and {Ma}, Xiangcheng and {Byrne}, Lindsey and {El-Badry}, Kareem and {Angl{\'e}s-Alc{\'a}zar}, Daniel and {Chan}, T.~K. and {Feldmann}, Robert and {Kere{\v{s}}}, Du{\v{s}}an and {Wetzel}, Andrew and {Murray}, Norman and {Hopkins}, Philip F.},
        title = "{Virialization of the Inner CGM in the FIRE Simulations and Implications for Galaxy Disks, Star Formation, and Feedback}",
      journal = {\apj},
         year = 2021,
        month = apr,
       volume = {911},
       number = {2},
          eid = {88},
        pages = {88},
          doi = {10.3847/1538-4357/abd776},
archivePrefix = {arXiv},
       eprint = {2006.13976},
 primaryClass = {astro-ph.GA},
       adsurl = {https://ui.adsabs.harvard.edu/abs/2021ApJ...911...88S}
}

@ARTICLE{2019MNRAS.484.3625R,
       author = {{Roca-F{\`a}brega}, S. and {Dekel}, A. and {Faerman}, Y. and {Gnat}, O. and {Strawn}, C. and {Ceverino}, D. and {Primack}, J. and {Macci{\`o}}, A.~V. and {Dutton}, A.~A. and {Prochaska}, J.~X. and {Stern}, J.},
        title = "{CGM properties in VELA and NIHAO simulations; the OVI ionization mechanism: dependence on redshift, halo mass, and radius}",
      journal = {\mnras},
         year = 2019,
        month = apr,
       volume = {484},
       number = {3},
        pages = {3625-3645},
          doi = {10.1093/mnras/stz063},
archivePrefix = {arXiv},
       eprint = {1808.09973},
 primaryClass = {astro-ph.GA},
       adsurl = {https://ui.adsabs.harvard.edu/abs/2019MNRAS.484.3625R}
}

@ARTICLE{DwarfCGM_in_FIRE,
       author = {{Li}, Fei and {Rahman}, Mubdi and {Murray}, Norman and {Hafen}, Zachary and {Faucher-Gigu{\`e}re}, Claude-Andr{\'e} and {Stern}, Jonathan and {Hummels}, Cameron B. and {Hopkins}, Philip F. and {El-Badry}, Kareem and {Kere{\v{s}}}, Du{\v{s}}an},
        title = "{Probing the CGM of low-redshift dwarf galaxies using FIRE simulations}",
      journal = {\mnras},
         year = 2021,
        month = jan,
       volume = {500},
       number = {1},
        pages = {1038-1053},
          doi = {10.1093/mnras/staa3322},
archivePrefix = {arXiv},
       eprint = {2010.13606},
 primaryClass = {astro-ph.GA},
       adsurl = {https://ui.adsabs.harvard.edu/abs/2021MNRAS.500.1038L}
}

@ARTICLE{Mishra_24,
       author = {{Mishra}, Nishant and {Johnson}, Sean D. and {Rudie}, Gwen C. and {Chen}, Hsiao-Wen and {Schaye}, Joop and {Qu}, Zhijie and {Zahedy}, Fakhri S. and {Boettcher}, Erin T. and {Cantalupo}, Sebastiano and {Chen}, Mandy C. and {Faucher-Gigu{\`e}re}, Claude-Andr{\'e} and {Greene}, Jenny E. and {Li}, Jennifer I-Hsiu and {Zhuoqi} and {Liu} and {Lopez}, Sebastian and {Petitjean}, Patrick},
        title = "{The Cosmic Ultraviolet Baryon Survey (CUBS) IX: The enriched circumgalactic and intergalactic medium around star-forming field dwarf galaxies traced by O VI absorption}",
      journal = {arXiv e-prints},
         year = 2024,
        month = aug,
          eid = {arXiv:2408.11151},
        pages = {arXiv:2408.11151},
          doi = {10.48550/arXiv.2408.11151},
archivePrefix = {arXiv},
       eprint = {2408.11151},
 primaryClass = {astro-ph.GA},
       adsurl = {https://ui.adsabs.harvard.edu/abs/2024arXiv240811151M}
}

@ARTICLE{Cook_24,
       author = {{Cook}, Andrew W.~S. and {van de Voort}, Freeke and {Pakmor}, R{\"u}diger and {Grand}, Robert J.~J.},
        title = "{The halo mass dependence of physical and observable properties in the circumgalactic medium at z = 0}",
      journal = {\mnras},
         year = 2025,
        month = oct,
       volume = {543},
       number = {2},
        pages = {1224-1238},
          doi = {10.1093/mnras/staf1537},
archivePrefix = {arXiv},
       eprint = {2409.05578},
 primaryClass = {astro-ph.GA},
       adsurl = {https://ui.adsabs.harvard.edu/abs/2025MNRAS.543.1224C}
}

@ARTICLE{Azartash-Namin24,
       author = {{Azartash-Namin}, Bianca and {Engelhardt}, Anna and {Munshi}, Ferah and {Keller}, B.~W. and {Brooks}, Alyson M. and {Van Nest}, Jordan and {Christensen}, Charlotte R. and {Quinn}, Tom and {Wadsley}, James},
        title = "{Bursting with Feedback: The Relationship between Feedback Model and Bursty Star Formation Histories in Dwarf Galaxies}",
      journal = {\apj},
         year = 2024,
        month = jul,
       volume = {970},
       number = {1},
          eid = {40},
        pages = {40},
          doi = {10.3847/1538-4357/ad49a5},
archivePrefix = {arXiv},
       eprint = {2401.06041},
 primaryClass = {astro-ph.GA},
       adsurl = {https://ui.adsabs.harvard.edu/abs/2024ApJ...970...40A}
}

@ARTICLE{Thielemann86,
       author = {{Thielemann}, F. -K. and {Nomoto}, K. and {Yokoi}, K.},
        title = "{Explosive nucleosynthesis in carbon deflagration models of Type I supernovae}",
      journal = {\aap},
         year = 1986,
        month = apr,
       volume = {158},
       number = {1-2},
        pages = {17-33},
       adsurl = {https://ui.adsabs.harvard.edu/abs/1986A&A...158...17T}
}

@ARTICLE{Raiteri96,
       author = {{Raiteri}, C.~M. and {Villata}, M. and {Navarro}, J.~F.},
        title = "{Simulations of Galactic chemical evolution. I. O and Fe abundances in a simple collapse model.}",
      journal = {\aap},
         year = 1996,
        month = nov,
       volume = {315},
        pages = {105-115},
       adsurl = {https://ui.adsabs.harvard.edu/abs/1996A&A...315..105R}
}

@ARTICLE{Weidemann_87,
       author = {{Weidemann}, V.},
        title = "{The initial-final mass relation : galactic disk and Magellanic Clouds.}",
      journal = {\aap},
         year = 1987,
        month = dec,
       volume = {188},
        pages = {74-84},
       adsurl = {https://ui.adsabs.harvard.edu/abs/1987A&A...188...74W}
}

@ARTICLE{2020MNRAS.498.2391N,
       author = {{Nelson}, Dylan and {Sharma}, Prateek and {Pillepich}, Annalisa and {Springel}, Volker and {Pakmor}, R{\"u}diger and {Weinberger}, Rainer and {Vogelsberger}, Mark and {Marinacci}, Federico and {Hernquist}, Lars},
        title = "{Resolving small-scale cold circumgalactic gas in TNG50}",
      journal = {\mnras},
         year = 2020,
        month = oct,
       volume = {498},
       number = {2},
        pages = {2391-2414},
          doi = {10.1093/mnras/staa2419},
archivePrefix = {arXiv},
       eprint = {2005.09654},
 primaryClass = {astro-ph.GA},
       adsurl = {https://ui.adsabs.harvard.edu/abs/2020MNRAS.498.2391N}
}

@ARTICLE{2012MNRAS.423.2991V,
       author = {{van de Voort}, Freeke and {Schaye}, Joop},
        title = "{Properties of gas in and around galaxy haloes}",
      journal = {\mnras},
         year = 2012,
        month = jul,
       volume = {423},
       number = {4},
        pages = {2991-3010},
          doi = {10.1111/j.1365-2966.2012.20949.x},
archivePrefix = {arXiv},
       eprint = {1111.5039},
 primaryClass = {astro-ph.CO},
       adsurl = {https://ui.adsabs.harvard.edu/abs/2012MNRAS.423.2991V}
}

@ARTICLE{McConnachie12,
       author = {{McConnachie}, Alan W.},
        title = "{The Observed Properties of Dwarf Galaxies in and around the Local Group}",
      journal = {\aj},
         year = 2012,
        month = jul,
       volume = {144},
       number = {1},
          eid = {4},
        pages = {4},
          doi = {10.1088/0004-6256/144/1/4},
archivePrefix = {arXiv},
       eprint = {1204.1562},
 primaryClass = {astro-ph.CO},
       adsurl = {https://ui.adsabs.harvard.edu/abs/2012AJ....144....4M}
}

@ARTICLE{Bellovary21,
       author = {{Bellovary}, Jillian M. and {Hayoune}, Sarra and {Chafla}, Katheryn and {Vincent}, Donovan and {Brooks}, Alyson and {Christensen}, Charlotte R. and {Munshi}, Ferah D. and {Tremmel}, Michael and {Quinn}, Thomas R. and {Van Nest}, Jordan and {Sligh}, Serena K. and {Luzuriaga}, Michelle},
        title = "{The origins of off-centre massive black holes in dwarf galaxies}",
      journal = {\mnras},
         year = 2021,
        month = aug,
       volume = {505},
       number = {4},
        pages = {5129-5141},
          doi = {10.1093/mnras/stab1665},
archivePrefix = {arXiv},
       eprint = {2102.09566},
 primaryClass = {astro-ph.GA},
       adsurl = {https://ui.adsabs.harvard.edu/abs/2021MNRAS.505.5129B}
}

@ARTICLE{Christensen24,
       author = {{Christensen}, Charlotte R. and {Brooks}, Alyson M. and {Munshi}, Ferah and {Riggs}, Claire and {Van Nest}, Jordan and {Akins}, Hollis and {Quinn}, Thomas R. and {Chamberland}, Lucas},
        title = "{Environment Matters: Predicted Differences in the Stellar Mass{\textendash}Halo Mass Relation and History of Star Formation for Dwarf Galaxies}",
      journal = {\apj},
         year = 2024,
        month = feb,
       volume = {961},
       number = {2},
          eid = {236},
        pages = {236},
          doi = {10.3847/1538-4357/ad0c5a},
archivePrefix = {arXiv},
       eprint = {2311.04975},
 primaryClass = {astro-ph.GA},
       adsurl = {https://ui.adsabs.harvard.edu/abs/2024ApJ...961..236C}
}

@ARTICLE{Riggs24,
       author = {{Riggs}, Claire L. and {Brooks}, Alyson M. and {Munshi}, Ferah and {Christensen}, Charlotte R. and {Cohen}, Roger E. and {Quinn}, Thomas R. and {Wadsley}, James},
        title = "{Testable predictions of outside-in age gradients in dwarf galaxies of all types}",
      journal = {arXiv e-prints},
         year = 2024,
        month = aug,
          eid = {arXiv:2408.10379},
        pages = {arXiv:2408.10379},
          doi = {10.48550/arXiv.2408.10379},
archivePrefix = {arXiv},
       eprint = {2408.10379},
 primaryClass = {astro-ph.GA},
       adsurl = {https://ui.adsabs.harvard.edu/abs/2024arXiv240810379R}
}

@ARTICLE{Ji20,
       author = {{Ji}, Suoqing and {Chan}, T.~K. and {Hummels}, Cameron B. and {Hopkins}, Philip F. and {Stern}, Jonathan and {Kere{\v{s}}}, Du{\v{s}}an and {Quataert}, Eliot and {Faucher-Gigu{\`e}re}, Claude-Andr{\'e} and {Murray}, Norman},
        title = "{Properties of the circumgalactic medium in cosmic ray-dominated galaxy haloes}",
      journal = {\mnras},
         year = 2020,
        month = aug,
       volume = {496},
       number = {4},
        pages = {4221-4238},
          doi = {10.1093/mnras/staa1849},
archivePrefix = {arXiv},
       eprint = {1909.00003},
 primaryClass = {astro-ph.GA},
       adsurl = {https://ui.adsabs.harvard.edu/abs/2020MNRAS.496.4221J}
}

@ARTICLE{IllustrisTNG_CRS_Ramesh24,
       author = {{Ramesh}, Rahul and {Nelson}, Dylan and {Girichidis}, Philipp},
        title = "{IllustrisTNG + Cosmic Rays with a Simple Transport Model: From Dwarfs to L$^\star$ Galaxies}",
      journal = {arXiv e-prints},
         year = 2024,
        month = sep,
          eid = {arXiv:2409.18238},
        pages = {arXiv:2409.18238},
          doi = {10.48550/arXiv.2409.18238},
archivePrefix = {arXiv},
       eprint = {2409.18238},
 primaryClass = {astro-ph.GA},
       adsurl = {https://ui.adsabs.harvard.edu/abs/2024arXiv240918238R}
}

@ARTICLE{Gible24,
       author = {{Ramesh}, Rahul and {Nelson}, Dylan},
        title = "{Zooming in on the circumgalactic medium with GIBLE: Resolving small-scale gas structure in cosmological simulations}",
      journal = {\mnras},
         year = 2024,
        month = feb,
       volume = {528},
       number = {2},
        pages = {3320-3339},
          doi = {10.1093/mnras/stae237},
archivePrefix = {arXiv},
       eprint = {2307.11143},
 primaryClass = {astro-ph.GA},
       adsurl = {https://ui.adsabs.harvard.edu/abs/2024MNRAS.528.3320R}
}

@ARTICLE{Hsu2011,
       author = {{Hsu}, W. -H. and {Putman}, M.~E. and {Heitsch}, F. and {Stanimirovi{\'c}}, S. and {Peek}, J.~E.~G. and {Clark}, S.~E.},
        title = "{Physical Properties of Complex C Halo Clouds}",
      journal = {\aj},
         year = 2011,
        month = feb,
       volume = {141},
       number = {2},
          eid = {57},
        pages = {57},
          doi = {10.1088/0004-6256/141/2/57},
archivePrefix = {arXiv},
       eprint = {1011.0011},
 primaryClass = {astro-ph.GA},
       adsurl = {https://ui.adsabs.harvard.edu/abs/2011AJ....141...57H}
}

@ARTICLE{Zahedy2019,
       author = {{Zahedy}, Fakhri S. and {Chen}, Hsiao-Wen and {Johnson}, Sean D. and {Pierce}, Rebecca M. and {Rauch}, Michael and {Huang}, Yun-Hsin and {Weiner}, Benjamin J. and {Gauthier}, Jean-Ren{\'e}},
        title = "{Characterizing circumgalactic gas around massive ellipticals at z {\ensuremath{\sim}} 0.4 - II. Physical properties and elemental abundances}",
      journal = {\mnras},
         year = 2019,
        month = apr,
       volume = {484},
       number = {2},
        pages = {2257-2280},
          doi = {10.1093/mnras/sty3482},
archivePrefix = {arXiv},
       eprint = {1809.05115},
 primaryClass = {astro-ph.GA},
       adsurl = {https://ui.adsabs.harvard.edu/abs/2019MNRAS.484.2257Z}
}

@ARTICLE{FOGGIE_19,
       author = {{Peeples}, Molly S. and {Corlies}, Lauren and {Tumlinson}, Jason and {O'Shea}, Brian W. and {Lehner}, Nicolas and {O'Meara}, John M. and {Howk}, J. Christopher and {Earl}, Nicholas and {Smith}, Britton D. and {Wise}, John H. and {Hummels}, Cameron B.},
        title = "{Figuring Out Gas \& Galaxies in Enzo (FOGGIE). I. Resolving Simulated Circumgalactic Absorption at 2 {\ensuremath{\leq}} z {\ensuremath{\leq}} 2.5}",
      journal = {\apj},
         year = 2019,
        month = mar,
       volume = {873},
       number = {2},
          eid = {129},
        pages = {129},
          doi = {10.3847/1538-4357/ab0654},
archivePrefix = {arXiv},
       eprint = {1810.06566},
 primaryClass = {astro-ph.GA},
       adsurl = {https://ui.adsabs.harvard.edu/abs/2019ApJ...873..129P}
}

@ARTICLE{Augustin2021,
       author = {{Augustin}, Ramona and {P{\'e}roux}, C{\'e}line and {Hamanowicz}, Aleksandra and {Kulkarni}, Varsha and {Rahmani}, Hadi and {Zanella}, Anita},
        title = "{Clumpiness of observed and simulated cold circumgalactic gas}",
      journal = {\mnras},
         year = 2021,
        month = aug,
       volume = {505},
       number = {4},
        pages = {6195-6205},
          doi = {10.1093/mnras/stab1673},
archivePrefix = {arXiv},
       eprint = {2105.11480},
 primaryClass = {astro-ph.GA},
       adsurl = {https://ui.adsabs.harvard.edu/abs/2021MNRAS.505.6195A}
}

@ARTICLE{vandeVoort19,
       author = {{van de Voort}, Freeke and {Springel}, Volker and {Mandelker}, Nir and {van den Bosch}, Frank C. and {Pakmor}, R{\"u}diger},
        title = "{Cosmological simulations of the circumgalactic medium with 1 kpc resolution: enhanced H I column densities}",
      journal = {\mnras},
         year = 2019,
        month = jan,
       volume = {482},
       number = {1},
        pages = {L85-L89},
          doi = {10.1093/mnrasl/sly190},
archivePrefix = {arXiv},
       eprint = {1808.04369},
 primaryClass = {astro-ph.GA},
       adsurl = {https://ui.adsabs.harvard.edu/abs/2019MNRAS.482L..85V}
}

@ARTICLE{Sharma2020,
       author = {{Sharma}, Ray S. and {Brooks}, Alyson M. and {Somerville}, Rachel S. and {Tremmel}, Michael and {Bellovary}, Jillian and {Wright}, Anna C. and {Quinn}, Thomas R.},
        title = "{Black Hole Growth and Feedback in Isolated ROMULUS25 Dwarf Galaxies}",
      journal = {\apj},
         year = 2020,
        month = jul,
       volume = {897},
       number = {1},
          eid = {103},
        pages = {103},
          doi = {10.3847/1538-4357/ab960e},
archivePrefix = {arXiv},
       eprint = {1912.06646},
 primaryClass = {astro-ph.GA},
       adsurl = {https://ui.adsabs.harvard.edu/abs/2020ApJ...897..103S}
}

@ARTICLE{Tremmel2017,
       author = {{Tremmel}, M. and {Karcher}, M. and {Governato}, F. and {Volonteri}, M. and {Quinn}, T.~R. and {Pontzen}, A. and {Anderson}, L. and {Bellovary}, J.},
        title = "{The Romulus cosmological simulations: a physical approach to the formation, dynamics and accretion models of SMBHs}",
      journal = {\mnras},
         year = 2017,
        month = sep,
       volume = {470},
       number = {1},
        pages = {1121-1139},
          doi = {10.1093/mnras/stx1160},
archivePrefix = {arXiv},
       eprint = {1607.02151},
 primaryClass = {astro-ph.GA},
       adsurl = {https://ui.adsabs.harvard.edu/abs/2017MNRAS.470.1121T}
}

@ARTICLE{Karachentsev2013,
       author = {{Karachentsev}, Igor D. and {Makarov}, Dmitry I. and {Kaisina}, Elena I.},
        title = "{Updated Nearby Galaxy Catalog}",
      journal = {\aj},
         year = 2013,
        month = apr,
       volume = {145},
       number = {4},
          eid = {101},
        pages = {101},
          doi = {10.1088/0004-6256/145/4/101},
archivePrefix = {arXiv},
       eprint = {1303.5328},
 primaryClass = {astro-ph.CO},
       adsurl = {https://ui.adsabs.harvard.edu/abs/2013AJ....145..101K}
}

@ARTICLE{Lee2011,
       author = {{Lee}, Janice C. and {Gil de Paz}, Armando and {Kennicutt}, Jr., Robert C. and {Bothwell}, Matthew and {Dalcanton}, Julianne and {Jos{\'e} G. Funes S.}, J. and {Johnson}, Benjamin D. and {Sakai}, Shoko and {Skillman}, Evan and {Tremonti}, Christy and {van Zee}, Liese},
        title = "{A GALEX Ultraviolet Imaging Survey of Galaxies in the Local Volume}",
      journal = {\apjs},
         year = 2011,
        month = jan,
       volume = {192},
       number = {1},
          eid = {6},
        pages = {6},
          doi = {10.1088/0067-0049/192/1/6},
archivePrefix = {arXiv},
       eprint = {1009.4705},
 primaryClass = {astro-ph.CO},
       adsurl = {https://ui.adsabs.harvard.edu/abs/2011ApJS..192....6L}
}

@ARTICLE{Pace2024LVDB,
       author = {{Pace}, Andrew B},
        title = "{The Local Volume Database: a library of the observed properties of nearby dwarf galaxies and star clusters}",
      journal = {The Open Journal of Astrophysics},
         year = 2025,
        month = sep,
       volume = {8},
          eid = {142},
        pages = {142},
          doi = {10.33232/001c.144859},
archivePrefix = {arXiv},
       eprint = {2411.07424},
 primaryClass = {astro-ph.GA},
       adsurl = {https://ui.adsabs.harvard.edu/abs/2025OJAp....8E.142P}
}

@ARTICLE{2020ApJ...891..181M,
       author = {{McQuinn}, Kristen. B.~W. and {Berg}, Danielle A. and {Skillman}, Evan D. and {Adams}, Elizabeth A.~K. and {Cannon}, John M. and {Dolphin}, Andrew E. and {Salzer}, John J. and {Giovanelli}, Riccardo and {Haynes}, Martha P. and {Hirschauer}, Alec S. and {Janoweicki}, Steven and {Klapkowski}, Myles and {Rhode}, Katherine L.},
        title = "{The Leoncino Dwarf Galaxy: Exploring the Low-metallicity End of the Luminosity-Metallicity and Mass-Metallicity Relations}",
      journal = {\apj},
         year = 2020,
        month = mar,
       volume = {891},
       number = {2},
          eid = {181},
        pages = {181},
          doi = {10.3847/1538-4357/ab7447},
archivePrefix = {arXiv},
       eprint = {2002.11723},
 primaryClass = {astro-ph.GA},
       adsurl = {https://ui.adsabs.harvard.edu/abs/2020ApJ...891..181M}
}

@ARTICLE{2023ApJ...951..138L,
       author = {{Lin}, Yu-Heng and {Scarlata}, Claudia and {Mehta}, Vihang and {Skillman}, Evan and {Hayes}, Matthew and {McQuinn}, Kristen B.~W. and {Fortson}, Lucy and {Chworowsky}, Katherine and {Clarke}, Leonardo},
        title = "{Low-metallicity Galaxies from the Dark Energy Survey}",
      journal = {\apj},
         year = 2023,
        month = jul,
       volume = {951},
       number = {2},
          eid = {138},
        pages = {138},
          doi = {10.3847/1538-4357/acd181},
archivePrefix = {arXiv},
       eprint = {2211.02094},
 primaryClass = {astro-ph.GA},
       adsurl = {https://ui.adsabs.harvard.edu/abs/2023ApJ...951..138L}
}

@ARTICLE{2007ApJ...658..941D,
       author = {{Dalcanton}, Julianne J.},
        title = "{The Metallicity of Galaxy Disks: Infall versus Outflow}",
      journal = {\apj},
         year = 2007,
        month = apr,
       volume = {658},
       number = {2},
        pages = {941-959},
          doi = {10.1086/508913},
archivePrefix = {arXiv},
       eprint = {astro-ph/0608590},
 primaryClass = {astro-ph},
       adsurl = {https://ui.adsabs.harvard.edu/abs/2007ApJ...658..941D}
}

@ARTICLE{2007ApJS..173..293W,
       author = {{Wyder}, Ted K. and {Martin}, D. Christopher and {Schiminovich}, David and {Seibert}, Mark and {Budav{\'a}ri}, Tam{\'a}s and {Treyer}, Marie A. and {Barlow}, Tom A. and {Forster}, Karl and {Friedman}, Peter G. and {Morrissey}, Patrick and {Neff}, Susan G. and {Small}, Todd and {Bianchi}, Luciana and {Donas}, Jos{\'e} and {Heckman}, Timothy M. and {Lee}, Young-Wook and {Madore}, Barry F. and {Milliard}, Bruno and {Rich}, R. Michael and {Szalay}, Alex S. and {Welsh}, Barry Y. and {Yi}, Sukyoung K.},
        title = "{The UV-Optical Galaxy Color-Magnitude Diagram. I. Basic Properties}",
      journal = {\apjs},
         year = 2007,
        month = dec,
       volume = {173},
       number = {2},
        pages = {293-314},
          doi = {10.1086/521402},
archivePrefix = {arXiv},
       eprint = {0706.3938},
 primaryClass = {astro-ph},
       adsurl = {https://ui.adsabs.harvard.edu/abs/2007ApJS..173..293W}
}

@ARTICLE{2024arXiv241216440T,
       author = {{Tung}, Pei-Cheng and {Chen}, Ke-Jung},
        title = "{Coevolution of Dwarf Galaxies and Their Circumgalactic Medium Across Cosmic Time}",
      journal = {arXiv e-prints},
         year = 2024,
        month = dec,
          eid = {arXiv:2412.16440},
        pages = {arXiv:2412.16440},
          doi = {10.48550/arXiv.2412.16440},
archivePrefix = {arXiv},
       eprint = {2412.16440},
 primaryClass = {astro-ph.GA},
       adsurl = {https://ui.adsabs.harvard.edu/abs/2024arXiv241216440T}
}

@ARTICLE{Hafen2019,
       author = {{Hafen}, Zachary and {Faucher-Gigu{\`e}re}, Claude-Andr{\'e} and {Angl{\'e}s-Alc{\'a}zar}, Daniel and {Stern}, Jonathan and {Kere{\v{s}}}, Du{\v{s}}an and {Hummels}, Cameron and {Esmerian}, Clarke and {Garrison-Kimmel}, Shea and {El-Badry}, Kareem and {Wetzel}, Andrew and {Chan}, T.~K. and {Hopkins}, Philip F. and {Murray}, Norman},
        title = "{The origins of the circumgalactic medium in the FIRE simulations}",
      journal = {\mnras},
         year = 2019,
        month = sep,
       volume = {488},
       number = {1},
        pages = {1248-1272},
          doi = {10.1093/mnras/stz1773},
archivePrefix = {arXiv},
       eprint = {1811.11753},
 primaryClass = {astro-ph.GA},
       adsurl = {https://ui.adsabs.harvard.edu/abs/2019MNRAS.488.1248H}
}

@ARTICLE{Hummels_2019,
       author = {{Hummels}, Cameron B. and {Smith}, Britton D. and {Hopkins}, Philip F. and {O'Shea}, Brian W. and {Silvia}, Devin W. and {Werk}, Jessica K. and {Lehner}, Nicolas and {Wise}, John H. and {Collins}, David C. and {Butsky}, Iryna S.},
        title = "{The Impact of Enhanced Halo Resolution on the Simulated Circumgalactic Medium}",
      journal = {\apj},
         year = 2019,
        month = sep,
       volume = {882},
       number = {2},
          eid = {156},
        pages = {156},
          doi = {10.3847/1538-4357/ab378f},
archivePrefix = {arXiv},
       eprint = {1811.12410},
 primaryClass = {astro-ph.GA},
       adsurl = {https://ui.adsabs.harvard.edu/abs/2019ApJ...882..156H}
}

@ARTICLE{GenetIC2021,
       author = {{Stopyra}, Stephen and {Pontzen}, Andrew and {Peiris}, Hiranya and {Roth}, Nina and {Rey}, Martin P.},
        title = "{GenetIC{\textemdash}A New Initial Conditions Generator to Support Genetically Modified Zoom Simulations}",
      journal = {\apjs},
         year = 2021,
        month = feb,
       volume = {252},
       number = {2},
          eid = {28},
        pages = {28},
          doi = {10.3847/1538-4365/abcd94},
archivePrefix = {arXiv},
       eprint = {2006.01841},
 primaryClass = {astro-ph.IM},
       adsurl = {https://ui.adsabs.harvard.edu/abs/2021ApJS..252...28S}
}

@article{Onorbe_2013,
   title={How to zoom: bias, contamination and Lagrange volumes in multimass cosmological simulations},
   volume={437},
   ISSN={1365-2966},
   url={http://dx.doi.org/10.1093/mnras/stt2020},
   DOI={10.1093/mnras/stt2020},
   number={2},
   journal={Monthly Notices of the Royal Astronomical Society},
   publisher={Oxford University Press (OUP)},
   author={Onorbe, J. and Garrison-Kimmel, S. and Maller, A. H. and Bullock, J. S. and Rocha, M. and Hahn, O.},
   year={2013},
   month=nov, pages={1894–1908} }

@ARTICLE{Brooks2009,
       author = {{Brooks}, A.~M. and {Governato}, F. and {Quinn}, T. and {Brook}, C.~B. and {Wadsley}, J.},
        title = "{The Role of Cold Flows in the Assembly of Galaxy Disks}",
      journal = {\apj},
         year = 2009,
        month = mar,
       volume = {694},
       number = {1},
        pages = {396-410},
          doi = {10.1088/0004-637X/694/1/396},
archivePrefix = {arXiv},
       eprint = {0812.0007},
 primaryClass = {astro-ph},
       adsurl = {https://ui.adsabs.harvard.edu/abs/2009ApJ...694..396B}
}

@ARTICLE{2013ApJ...766...56M,
       author = {{Munshi}, Ferah and {Governato}, F. and {Brooks}, A.~M. and {Christensen}, C. and {Shen}, S. and {Loebman}, S. and {Moster}, B. and {Quinn}, T. and {Wadsley}, J.},
        title = "{Reproducing the Stellar Mass/Halo Mass Relation in Simulated {\ensuremath{\Lambda}}CDM Galaxies: Theory versus Observational Estimates}",
      journal = {\apj},
         year = 2013,
        month = mar,
       volume = {766},
       number = {1},
          eid = {56},
        pages = {56},
          doi = {10.1088/0004-637X/766/1/56},
archivePrefix = {arXiv},
       eprint = {1209.1389},
 primaryClass = {astro-ph.CO},
       adsurl = {https://ui.adsabs.harvard.edu/abs/2013ApJ...766...56M}
}

@ARTICLE{Muratov2017,
       author = {{Muratov}, Alexander L. and {Kere{\v{s}}}, Du{\v{s}}an and {Faucher-Gigu{\`e}re}, Claude-Andr{\'e} and {Hopkins}, Philip F. and {Ma}, Xiangcheng and {Angl{\'e}s-Alc{\'a}zar}, Daniel and {Chan}, T.~K. and {Torrey}, Paul and {Hafen}, Zachary H. and {Quataert}, Eliot and {Murray}, Norman},
        title = "{Metal flows of the circumgalactic medium, and the metal budget in galactic haloes}",
      journal = {\mnras},
         year = 2017,
        month = jul,
       volume = {468},
       number = {4},
        pages = {4170-4188},
          doi = {10.1093/mnras/stx667},
archivePrefix = {arXiv},
       eprint = {1606.09252},
 primaryClass = {astro-ph.GA},
       adsurl = {https://ui.adsabs.harvard.edu/abs/2017MNRAS.468.4170M}
}

@INPROCEEDINGS{2009ASPC..419..347D,
       author = {{Dav{\'e}}, R.},
        title = "{Missing Halo Baryons and Galactic Outflows}",
    booktitle = {Galaxy Evolution: Emerging Insights and Future Challenges},
         year = 2009,
       editor = {{Jogee}, S. and {Marinova}, I. and {Hao}, L. and {Blanc}, G.~A.},
       series = {Astronomical Society of the Pacific Conference Series},
       volume = {419},
        month = dec,
        pages = {347},
          doi = {10.48550/arXiv.0901.3149},
archivePrefix = {arXiv},
       eprint = {0901.3149},
 primaryClass = {astro-ph.CO},
       adsurl = {https://ui.adsabs.harvard.edu/abs/2009ASPC..419..347D}
}

@ARTICLE{2024MNRAS.528.5412R,
       author = {{Rey}, Martin P. and {Katz}, Harley B. and {Cameron}, Alex J. and {Devriendt}, Julien and {Slyz}, Adrianne},
        title = "{Boosting galactic outflows with enhanced resolution}",
      journal = {\mnras},
         year = 2024,
        month = mar,
       volume = {528},
       number = {3},
        pages = {5412-5431},
          doi = {10.1093/mnras/stae388},
archivePrefix = {arXiv},
       eprint = {2302.08521},
 primaryClass = {astro-ph.GA},
       adsurl = {https://ui.adsabs.harvard.edu/abs/2024MNRAS.528.5412R}
}

@ARTICLE{Keller_2016,
       author = {{Keller}, B.~W. and {Wadsley}, J. and {Couchman}, H.~M.~P.},
        title = "{Cosmological galaxy evolution with superbubble feedback - II. The limits of supernovae}",
      journal = {\mnras},
         year = 2016,
        month = dec,
       volume = {463},
       number = {2},
        pages = {1431-1445},
          doi = {10.1093/mnras/stw2029},
archivePrefix = {arXiv},
       eprint = {1604.08244},
 primaryClass = {astro-ph.GA},
       adsurl = {https://ui.adsabs.harvard.edu/abs/2016MNRAS.463.1431K}
}

@ARTICLE{Keller_2020,
       author = {{Keller}, Benjamin W. and {Kruijssen}, J.~M. Diederik and {Wadsley}, James W.},
        title = "{Entropy-driven winds: Outflows and fountains lifted gently by buoyancy}",
      journal = {\mnras},
         year = 2020,
        month = apr,
       volume = {493},
       number = {2},
        pages = {2149-2170},
          doi = {10.1093/mnras/staa380},
archivePrefix = {arXiv},
       eprint = {1909.00815},
 primaryClass = {astro-ph.GA},
       adsurl = {https://ui.adsabs.harvard.edu/abs/2020MNRAS.493.2149K}
}

@ARTICLE{2024Deg,
       author = {{Deg}, N. and {Arora}, N. and {Spekkens}, K. and {Halloran}, R. and {Catinella}, B. and {Jones}, M.~G. and {Courtois}, H. and {Glazebrook}, K. and {Bosma}, A. and {Cortese}, L. and {D{\'e}nes}, H. and {Elagali}, A. and {For}, B. -Q. and {Kamphuis}, P. and {Koribalski}, B.~S. and {Lee-Waddell}, K. and {Mancera Pi{\~n}a}, P.~E. and {Mould}, J. and {Rhee}, J. and {Shao}, L. and {Staveley-Smith}, L. and {Wang}, J. and {Westmeier}, T. and {Wong}, O.~I.},
        title = "{WALLABY Pilot Survey: Gas-rich Galaxy Scaling Relations from Marginally Resolved Kinematic Models}",
      journal = {\apj},
         year = 2024,
        month = dec,
       volume = {976},
       number = {2},
          eid = {159},
        pages = {159},
          doi = {10.3847/1538-4357/ad84ba},
archivePrefix = {arXiv},
       eprint = {2411.06993},
 primaryClass = {astro-ph.GA},
       adsurl = {https://ui.adsabs.harvard.edu/abs/2024ApJ...976..159D}
}

@ARTICLE{Johnson2015,
       author = {{Johnson}, Sean D. and {Chen}, Hsiao-Wen and {Mulchaey}, John S.},
        title = "{On the possible environmental effect in distributing heavy elements beyond individual gaseous haloes}",
      journal = {\mnras},
         year = 2015,
        month = may,
       volume = {449},
       number = {3},
        pages = {3263-3273},
          doi = {10.1093/mnras/stv553},
archivePrefix = {arXiv},
       eprint = {1503.04199},
 primaryClass = {astro-ph.GA},
       adsurl = {https://ui.adsabs.harvard.edu/abs/2015MNRAS.449.3263J}
}

@ARTICLE{Obreja2019,
       author = {{Obreja}, Aura and {Macci{\`o}}, Andrea V. and {Moster}, Benjamin and {Udrescu}, Silviu M. and {Buck}, Tobias and {Kannan}, Rahul and {Dutton}, Aaron A. and {Blank}, Marvin},
        title = "{Local photoionization feedback effects on galaxies}",
      journal = {\mnras},
         year = 2019,
        month = dec,
       volume = {490},
       number = {2},
        pages = {1518-1538},
          doi = {10.1093/mnras/stz2639},
archivePrefix = {arXiv},
       eprint = {1909.00832},
 primaryClass = {astro-ph.GA},
       adsurl = {https://ui.adsabs.harvard.edu/abs/2019MNRAS.490.1518O}
}

@ARTICLE{Taira2025,
       author = {{Taira}, Elias and {Kopenhafer}, Claire and {Oshea}, Brian W. and {Manning}, Alexis and {Fuhrman}, Evelyn and {Peeples}, Molly S. and {Tumlinson}, Jason and {Smith}, Britton D.},
        title = "{Impacts of the Metagalactic Ultraviolet Background on Circumgalactic Medium Absorption Systems}",
      journal = {arXiv e-prints},
         year = 2025,
        month = mar,
          eid = {arXiv:2503.11775},
        pages = {arXiv:2503.11775},
archivePrefix = {arXiv},
       eprint = {2503.11775},
 primaryClass = {astro-ph.GA},
       adsurl = {https://ui.adsabs.harvard.edu/abs/2025arXiv250311775T}
}

@ARTICLE{Sanchez2019,
       author = {{Sanchez}, N. Nicole and {Werk}, Jessica K. and {Tremmel}, Michael and {Pontzen}, Andrew and {Christensen}, Charlotte and {Quinn}, Thomas and {Cruz}, Akaxia},
        title = "{Not So Heavy Metals: Black Hole Feedback Enriches the Circumgalactic Medium}",
      journal = {\apj},
         year = 2019,
        month = sep,
       volume = {882},
       number = {1},
          eid = {8},
        pages = {8},
          doi = {10.3847/1538-4357/ab3045},
archivePrefix = {arXiv},
       eprint = {1810.12319},
 primaryClass = {astro-ph.GA},
       adsurl = {https://ui.adsabs.harvard.edu/abs/2019ApJ...882....8S}
}

@ARTICLE{Nelson2018,
       author = {{Nelson}, Dylan and {Kauffmann}, Guinevere and {Pillepich}, Annalisa and {Genel}, Shy and {Springel}, Volker and {Pakmor}, R{\"u}diger and {Hernquist}, Lars and {Weinberger}, Rainer and {Torrey}, Paul and {Vogelsberger}, Mark and {Marinacci}, Federico},
        title = "{The abundance, distribution, and physical nature of highly ionized oxygen O VI, O VII, and O VIII in IllustrisTNG}",
      journal = {\mnras},
         year = 2018,
        month = jun,
       volume = {477},
       number = {1},
        pages = {450-479},
          doi = {10.1093/mnras/sty656},
archivePrefix = {arXiv},
       eprint = {1712.00016},
 primaryClass = {astro-ph.GA},
       adsurl = {https://ui.adsabs.harvard.edu/abs/2018MNRAS.477..450N}
}

@ARTICLE{Gutcke2017,
       author = {{Gutcke}, Thales A. and {Stinson}, Greg S. and {Macci{\`o}}, Andrea V. and {Wang}, Liang and {Dutton}, Aaron A.},
        title = "{NIHAO - VIII. Circum-galactic medium and outflows - The puzzles of H I and O VI gas distributions}",
      journal = {\mnras},
         year = 2017,
        month = jan,
       volume = {464},
       number = {3},
        pages = {2796-2815},
          doi = {10.1093/mnras/stw2539},
archivePrefix = {arXiv},
       eprint = {1602.06956},
 primaryClass = {astro-ph.GA},
       adsurl = {https://ui.adsabs.harvard.edu/abs/2017MNRAS.464.2796G}
}

@ARTICLE{Dutta24,
       author = {{Dutta}, Sayak and {Muzahid}, Sowgat and {Schaye}, Joop and {Johnson}, Sean and {Bouch{\'e}}, Nicolas F. and {Chen}, Hsiao-Wen and {Cantalupo}, Sebastiano},
        title = "{MUSEQuBES: The column density, covering fraction, and mass of OVI-bearing gas in and around low-redshift galaxies}",
      journal = {arXiv e-prints},
         year = 2024,
        month = sep,
          eid = {arXiv:2409.15423},
        pages = {arXiv:2409.15423},
          doi = {10.48550/arXiv.2409.15423},
archivePrefix = {arXiv},
       eprint = {2409.15423},
 primaryClass = {astro-ph.GA},
       adsurl = {https://ui.adsabs.harvard.edu/abs/2024arXiv240915423D}
}

@ARTICLE{Ruan2025,
       author = {{Ruan}, Dilys and {Brooks}, Alyson M. and {Cruz}, Akaxia and {Peter}, Annika H.~G. and {Keller}, Ben and {Quinn}, Thomas and {Wadsley}, James and {Adams}, Elizabeth A.~K.},
        title = "{Predictions for Detecting a Turndown in the Baryonic Tully Fisher Relation}",
      journal = {arXiv e-prints},
         year = 2025,
        month = mar,
          eid = {arXiv:2503.16607},
        pages = {arXiv:2503.16607},
          doi = {10.48550/arXiv.2503.16607},
archivePrefix = {arXiv},
       eprint = {2503.16607},
 primaryClass = {astro-ph.GA},
       adsurl = {https://ui.adsabs.harvard.edu/abs/2025arXiv250316607R}
}

@ARTICLE{Keith2025,
       author = {{Keith}, Blake and {Munshi}, Ferah and {Brooks}, Alyson M. and {Van Nest}, Jordan and {Engelhardt}, Anna and {Cruz}, Akaxia and {Keller}, Ben and {Quinn}, Thomas and {Wadsley}, James},
        title = "{A MARVEL-ous study of how well galaxy shapes reflect Dark Matter halo shapes in Cold Dark Matter Simulations}",
      journal = {arXiv e-prints},
         year = 2025,
        month = jan,
          eid = {arXiv:2501.16317},
        pages = {arXiv:2501.16317},
          doi = {10.48550/arXiv.2501.16317},
archivePrefix = {arXiv},
       eprint = {2501.16317},
 primaryClass = {astro-ph.GA},
       adsurl = {https://ui.adsabs.harvard.edu/abs/2025arXiv250116317K}
}

@ARTICLE{Wadsley2017,
       author = {{Wadsley}, James W. and {Keller}, Benjamin W. and {Quinn}, Thomas R.},
        title = "{Gasoline2: a modern smoothed particle hydrodynamics code}",
      journal = {\mnras},
         year = 2017,
        month = oct,
       volume = {471},
       number = {2},
        pages = {2357-2369},
          doi = {10.1093/mnras/stx1643},
archivePrefix = {arXiv},
       eprint = {1707.03824},
 primaryClass = {astro-ph.IM},
       adsurl = {https://ui.adsabs.harvard.edu/abs/2017MNRAS.471.2357W}
}

@ARTICLE{McCourt2018,
       author = {{McCourt}, Michael and {Oh}, S. Peng and {O'Leary}, Ryan and {Madigan}, Ann-Marie},
        title = "{A characteristic scale for cold gas}",
      journal = {\mnras},
         year = 2018,
        month = feb,
       volume = {473},
       number = {4},
        pages = {5407-5431},
          doi = {10.1093/mnras/stx2687},
archivePrefix = {arXiv},
       eprint = {1610.01164},
 primaryClass = {astro-ph.GA},
       adsurl = {https://ui.adsabs.harvard.edu/abs/2018MNRAS.473.5407M}
}

@ARTICLE{Chevalier1974,
       author = {{Chevalier}, Roger A.},
        title = "{The Evolution of Supernova Remnants. Spherically Symmetric Models}",
      journal = {\apj},
         year = 1974,
        month = mar,
       volume = {188},
        pages = {501-516},
          doi = {10.1086/152740},
       adsurl = {https://ui.adsabs.harvard.edu/abs/1974ApJ...188..501C}
}

@ARTICLE{CowieMcKee_77,
       author = {{Cowie}, L.~L. and {McKee}, C.~F.},
        title = "{The evaporation of spherical clouds in a hot gas. I. Classical and saturated mass loss rates.}",
      journal = {\apj},
         year = 1977,
        month = jan,
       volume = {211},
        pages = {135-146},
          doi = {10.1086/154911},
       adsurl = {https://ui.adsabs.harvard.edu/abs/1977ApJ...211..135C}
}

@ARTICLE{Sanchez2021,
       author = {{Sanchez}, N. Nicole and {Tremmel}, Michael and {Werk}, Jessica K. and {Pontzen}, Andrew and {Christensen}, Charlotte and {Quinn}, Thomas and {Loebman}, Sarah and {Cruz}, Akaxia},
        title = "{One-Two Quench: A Double Minor Merger Scenario}",
      journal = {\apj},
         year = 2021,
        month = apr,
       volume = {911},
       number = {2},
          eid = {116},
        pages = {116},
          doi = {10.3847/1538-4357/abeb15},
archivePrefix = {arXiv},
       eprint = {2009.05581},
 primaryClass = {astro-ph.GA},
       adsurl = {https://ui.adsabs.harvard.edu/abs/2021ApJ...911..116S}
}

@ARTICLE{Sanchez2024,
       author = {{Sanchez}, N. Nicole and {Werk}, Jessica K. and {Christensen}, Charlotte and {Telford}, O. Grace and {Quinn}, Thomas R. and {Tremmel}, Michael and {Mead}, Jennifer and {Sharma}, Ray S. and {Brooks}, Alyson M.},
        title = "{The Scatter Matters: Circumgalactic Metal Content in the Context of the M{\textendash}{\ensuremath{\sigma}} Relation}",
      journal = {\apj},
         year = 2024,
        month = jun,
       volume = {967},
       number = {2},
          eid = {100},
        pages = {100},
          doi = {10.3847/1538-4357/ad39eb},
archivePrefix = {arXiv},
       eprint = {2305.07672},
 primaryClass = {astro-ph.GA},
       adsurl = {https://ui.adsabs.harvard.edu/abs/2024ApJ...967..100S}
}

@misc{pynbody,
  author = {{Pontzen}, A. and {Ro{\v s}kar}, R. and {Stinson}, G.~S. and {Woods},
     R. and {Reed}, D.~M. and {Coles}, J. and {Quinn}, T.~R.},
  title = "{pynbody: Astrophysics Simulation Analysis for Python}",
  note = {Astrophysics Source Code Library, ascl:1305.002},
  year = 2013
}

@ARTICLE{2020MNRAS.493.1614F,
       author = {{Faucher-Gigu{\`e}re}, Claude-Andr{\'e}},
        title = "{A cosmic UV/X-ray background model update}",
      journal = {\mnras},
         year = 2020,
        month = apr,
       volume = {493},
       number = {2},
        pages = {1614-1632},
          doi = {10.1093/mnras/staa302},
archivePrefix = {arXiv},
       eprint = {1903.08657},
 primaryClass = {astro-ph.CO},
       adsurl = {https://ui.adsabs.harvard.edu/abs/2020MNRAS.493.1614F}
}

@ARTICLE{Baumschlager2024,
       author = {{Baumschlager}, Bernhard and {Shen}, Sijing and {Wadsley}, James W.},
        title = "{Spectral reconstruction for radiation hydrodynamic simulations of galaxy evolution}",
      journal = {\aap},
         year = 2024,
        month = nov,
       volume = {691},
          eid = {A219},
        pages = {A219},
          doi = {10.1051/0004-6361/202348164},
archivePrefix = {arXiv},
       eprint = {2310.16902},
 primaryClass = {astro-ph.GA},
       adsurl = {https://ui.adsabs.harvard.edu/abs/2024A&A...691A.219B}
}

@ARTICLE{Wadsley2004,
       author = {{Wadsley}, J.~W. and {Stadel}, J. and {Quinn}, T.},
        title = "{Gasoline: a flexible, parallel implementation of TreeSPH}",
      journal = {\na},
         year = 2004,
        month = feb,
       volume = {9},
       number = {2},
        pages = {137-158},
          doi = {10.1016/j.newast.2003.08.004},
archivePrefix = {arXiv},
       eprint = {astro-ph/0303521},
 primaryClass = {astro-ph},
       adsurl = {https://ui.adsabs.harvard.edu/abs/2004NewA....9..137W}
}

@ARTICLE{2025ApJ...982L..30F,
       author = {{Faerman}, Yakov and {Zheng}, Yong and {Oppenheimer}, Benjamin D.},
        title = "{Upper Limits on the Mass of Cool Gas in the Circumgalactic Medium of Dwarf Galaxies}",
      journal = {\apjl},
         year = 2025,
        month = mar,
       volume = {982},
       number = {1},
          eid = {L30},
        pages = {L30},
          doi = {10.3847/2041-8213/adba51},
archivePrefix = {arXiv},
       eprint = {2501.02056},
 primaryClass = {astro-ph.GA},
       adsurl = {https://ui.adsabs.harvard.edu/abs/2025ApJ...982L..30F}
}
\bibliographystyle{aasjournal}

\clearpage

\appendix
\section{Individual Halo Properties in the \MM\, Sample}
\begin{longtable}{l|ccccccccccccccc}
\hline
$ID$ & $M_*$  & $M_{200c}$ & $R_{200c}$ & $sSFR_{100}$ & $M_{\mathrm{Gas}}^{\mathrm{Halo}}$ & $M_{\mathrm{HI+He}}^{\mathrm{Disk}}$ & $M_{\mathrm{Gas}}^{\mathrm{CGM}}$ & $M_{\mathrm{Cool}}^{\mathrm{CGM}}$ & $M_{\mathrm{Warm}}^{\mathrm{CGM}}$ & $M_{Z}^{\mathrm{CGM}}$ & $\rho_{\mathrm{CGM}}$ & $T_{\mathrm{CGM}}$ \\
\hline\hline

\endfirsthead
\hline
$ID$ & $M_*$  & $M_{200c}$ & $R_{200c}$ & $sSFR_{100}$ & $M_{\mathrm{Gas}}^{\mathrm{Halo}}$ & $M_{\mathrm{HI+He}}^{\mathrm{Disk}}$ & $M_{\mathrm{Gas}}^{\mathrm{CGM}}$ & $M_{\mathrm{Cool}}^{\mathrm{CGM}}$ & $M_{\mathrm{Warm}}^{\mathrm{CGM}}$ & $M_{Z}^{\mathrm{CGM}}$ & $\rho_{\mathrm{CGM}}$ & $T_{\mathrm{CGM}}$ \\
\hline\hline
\endhead
r468-1 & 9.09 & 10.85 & 87.04 & -9.54 & 9.67 & 8.86 & 9.45 & 9.27 & 8.93 & 7.39 & -3.86 & 4.41 \\
r568-1 & 9.06 & 10.82 & 84.93 & -9.46 & 9.7 & 9.26 & 9.22 & 8.69 & 9.04 & 6.91 & -4.07 & 4.69 \\
rogue-1 & 8.99 & 10.92 & 87.59 & -9.78 & 9.55 & 8.64 & 9.39 & 8.95 & 9.15 & 7.15 & -3.96 & 4.69 \\
r442-1 & 8.95 & 10.89 & 90.11 & -9.42 & 9.86 & 9.35 & 9.48 & 9.0 & 9.27 & 7.0 & -3.83 & 4.7 \\
r502-1 & 8.93 & 10.87 & 88.74 & -9.28 & 9.77 & 9.08 & 9.5 & 9.19 & 9.18 & 7.2 & -3.87 & 4.58 \\
r492-1 & 8.85 & 10.87 & 88.74 & -9.33 & 9.86 & 9.47 & 9.4 & 8.91 & 9.2 & 6.87 & -3.99 & 4.74 \\
r515-1 & 8.82 & 10.9 & 90.73 & -9.38 & 9.94 & 9.56 & 9.48 & 9.02 & 9.26 & 6.92 & -3.82 & 4.69 \\
r556-1 & 8.81 & 10.74 & 80.12 & -9.52 & 9.52 & 8.92 & 9.19 & 8.89 & 8.86 & 6.99 & -4.14 & 4.56 \\
r555-1 & 8.72 & 10.75 & 81.02 & -9.49 & 9.57 & 8.91 & 9.26 & 8.92 & 8.97 & 6.91 & -3.99 & 4.57 \\
r544-1 & 8.68 & 10.82 & 85.32 & -9.24 & 9.72 & 9.29 & 9.3 & 8.57 & 9.19 & 6.59 & -4.19 & 4.8 \\
r489-1 & 8.68 & 10.83 & 85.76 & -9.18 & 9.81 & 9.29 & 9.47 & 8.94 & 9.29 & 6.74 & -3.92 & 4.72 \\
storm-1 & 8.67 & 10.88 & 85.04 & -9.31 & 9.82 & 9.38 & 9.44 & 8.92 & 9.26 & 6.7 & -3.84 & 4.75 \\
r656-1 & 8.54 & 10.69 & 76.86 & -9.6 & 9.45 & 8.95 & 9.08 & 8.47 & 8.93 & 6.6 & -4.24 & 4.72 \\
r613-1 & 8.52 & 10.76 & 81.25 & -9.53 & 9.62 & 9.14 & 9.28 & 8.71 & 9.12 & 6.62 & -4.15 & 4.73 \\
r597-1 & 8.52 & 10.73 & 79.67 & -9.64 & 9.51 & 9.03 & 9.14 & 8.5 & 9.0 & 6.57 & -4.25 & 4.76 \\
r523-1 & 8.51 & 10.78 & 82.34 & -9.59 & 9.59 & 9.06 & 9.26 & 8.53 & 9.14 & 6.53 & -4.2 & 4.82 \\
r642-1 & 8.47 & 10.7 & 77.69 & -9.56 & 9.66 & 9.2 & 9.26 & 8.75 & 9.06 & 6.62 & -4.08 & 4.7 \\
r569-1 & 8.46 & 10.72 & 78.99 & -9.6 & 9.63 & 9.14 & 9.26 & 8.7 & 9.09 & 6.6 & -4.06 & 4.73 \\
r918-1 & 8.45 & 10.76 & 81.14 & -9.8 & 9.74 & 9.02 & 9.53 & 9.28 & 9.14 & 6.78 & -3.57 & 4.49 \\
r634-1 & 8.43 & 10.69 & 77.26 & -9.56 & 9.61 & 9.11 & 9.23 & 8.69 & 9.06 & 6.54 & -4.04 & 4.69 \\
r618-1 & 8.42 & 10.7 & 77.44 & -9.56 & 9.44 & 8.67 & 9.22 & 8.79 & 8.99 & 6.69 & -4.12 & 4.64 \\
r571-1 & 8.42 & 10.78 & 82.79 & -9.75 & 9.67 & 9.26 & 9.14 & 8.19 & 9.06 & 6.22 & -4.28 & 4.83 \\
r614-1 & 8.41 & 10.71 & 78.46 & -9.68 & 9.56 & 9.11 & 9.1 & 8.48 & 8.95 & 6.46 & -4.19 & 4.73 \\
r563-1 & 8.39 & 10.7 & 77.72 & -9.43 & 9.53 & 8.8 & 9.3 & 8.84 & 9.09 & 6.68 & -4.04 & 4.66 \\
r886-1 & 8.37 & 10.63 & 73.71 & -9.39 & 9.49 & 8.91 & 9.18 & 8.69 & 8.99 & 6.52 & -4.08 & 4.64 \\
r552-1 & 8.36 & 10.7 & 77.64 & -9.55 & 9.59 & 9.11 & 9.24 & 8.72 & 9.05 & 6.55 & -4.09 & 4.69 \\
elektra-1 & 8.17 & 10.62 & 69.59 & -9.69 & 9.35 & 8.65 & 9.11 & 8.34 & 9.02 & 6.11 & -4.21 & 4.78 \\
r718-1 & 8.02 & 10.59 & 71.27 & -10.01 & 9.31 & 8.68 & 8.93 & 8.24 & 8.81 & 5.98 & -4.32 & 4.7 \\
storm-2 & 7.98 & 10.55 & 65.73 & -9.88 & 9.55 & 8.73 & 9.38 & 9.19 & 8.93 & 6.25 & -3.59 & 4.43 \\
r761-1 & 7.91 & 10.56 & 69.84 & -9.57 & 9.48 & 9.04 & 9.02 & 8.48 & 8.85 & 5.85 & -4.16 & 4.67 \\
r916-1 & 7.85 & 10.43 & 63.29 & -9.63 & 9.02 & 8.3 & 8.78 & 8.13 & 8.65 & 6.04 & -4.4 & 4.68 \\
r552-2 & 7.84 & 10.49 & 65.9 & -10.1 & 8.94 & 8.32 & 8.62 & 7.46 & 8.58 & 5.77 & -4.67 & 4.77 \\
rogue-3 & 7.84 & 10.33 & 55.55 & -9.79 & 9.21 & 8.7 & 8.74 & 8.45 & 8.43 & 5.68 & -3.83 & 4.49 \\
r852-1 & 7.83 & 10.5 & 66.59 & -9.77 & 9.03 & 8.1 & 8.88 & 8.06 & 8.79 & 5.92 & -4.43 & 4.74 \\
r753-1 & 7.81 & 10.46 & 64.5 & -10.13 & 8.91 & 8.26 & 8.62 & 7.56 & 8.57 & 5.76 & -4.65 & 4.73 \\
r850-1 & 7.78 & 10.46 & 64.41 & -9.7 & 8.99 & 8.03 & 8.83 & 8.18 & 8.7 & 5.97 & -4.39 & 4.68 \\
r977-1 & 7.73 & 10.41 & 62.17 & -10.19 & 9.0 & 8.47 & 8.58 & 7.73 & 8.5 & 5.49 & -4.65 & 4.69 \\
rogue-8 & 7.71 & 10.12 & 47.22 & -10.12 & 8.67 & 7.93 & 8.37 & 7.8 & 8.23 & 5.47 & -4.41 & 4.67 \\
elektra-2 & 7.64 & 10.46 & 61.62 & -9.33 & 9.24 & 8.34 & 9.07 & 8.59 & 8.88 & 5.69 & -3.89 & 4.59 \\
r753-2 & 7.64 & 10.31 & 57.4 & -9.68 & 8.97 & 8.26 & 8.72 & 8.44 & 8.38 & 5.92 & -4.35 & 4.54 \\
r918-2 & 7.61 & 10.44 & 63.52 & -9.97 & 9.22 & 8.6 & 8.81 & 7.99 & 8.73 & 5.52 & -4.35 & 4.66 \\
elektra-4 & 7.56 & 10.28 & 53.66 & -10.35 & 8.51 & 7.41 & 8.38 & 7.4 & 8.33 & 5.32 & -4.64 & 4.71 \\
storm-4 & 7.56 & 10.23 & 51.38 & -10.22 & 8.72 & 7.95 & 8.42 & 7.43 & 8.37 & 5.36 & -4.57 & 4.77 \\
r1023-2 & 7.52 & 10.17 & 51.82 & -10.35 & 8.43 & 7.61 & 8.19 & 8.03 & 7.66 & 5.66 & -4.67 & 4.47 \\
cptmarvel-1 & 7.51 & 10.19 & 50.04 & -10.04 & 8.54 & 7.75 & 8.34 & 7.5 & 8.27 & 5.24 & -4.64 & 4.63 \\
storm-3 & 7.37 & 10.39 & 58.1 & -9.54 & 9.25 & 8.6 & 8.97 & 8.41 & 8.83 & 5.31 & -4.08 & 4.63 \\
rogue-7 & 7.37 & 10.17 & 49.16 & -10.15 & 8.67 & 7.84 & 8.44 & 7.83 & 8.31 & 5.33 & -4.4 & 4.63 \\
elektra-5 & 7.37 & 10.19 & 49.84 & -9.66 & 8.84 & 7.96 & 8.67 & 8.0 & 8.57 & 5.38 & -4.24 & 4.61 \\
r977-2 & 7.36 & 10.25 & 55.17 & -9.89 & 8.71 & 7.85 & 8.52 & 8.14 & 8.26 & 5.6 & -4.43 & 4.56 \\
rogue-10 & 7.14 & 10.05 & 44.96 & -10.35 & 8.2 & 7.19 & 8.04 & 6.77 & 8.01 & 4.68 & -4.77 & 4.97 \\
r544-2 & 7.07 & 10.11 & 49.37 & -9.88 & 8.57 & 7.73 & 8.38 & 7.91 & 8.19 & 5.29 & -4.58 & 4.56 \\
storm-8 & 7.02 & 9.9 & 40.04 & -10.3 & 8.21 & 7.09 & 8.09 & 7.81 & 7.77 & 5.04 & -4.54 & 4.5 \\
storm-7 & 6.99 & 9.9 & 40.15 & -10.59 & 8.14 & 6.99 & 8.04 & 7.24 & 7.96 & 4.73 & -4.7 & 4.59 \\
cptmarvel-2 & 6.95 & 9.99 & 43.04 & -9.77 & 8.79 & 8.14 & 8.38 & 8.02 & 8.12 & 4.85 & -4.17 & 4.47 \\
rogue-12 & 6.92 & 9.88 & 39.45 & -10.23 & 8.05 & 7.38 & 7.73 & 6.9 & 7.65 & 4.61 & -4.88 & 4.84 \\
storm-5 & 6.9 & 9.98 & 42.67 & -10.17 & 8.33 & 7.2 & 8.19 & 7.6 & 8.06 & 4.93 & -4.59 & 4.57 \\
cptmarvel-5 & 6.86 & 9.88 & 39.26 & -10.26 & 8.08 & 7.2 & 7.86 & 7.58 & 7.52 & 4.66 & -4.78 & 4.51 \\
cptmarvel-6 & 6.84 & 9.82 & 37.59 & -10.39 & 8.13 & 6.99 & 7.97 & 7.75 & 7.57 & 4.79 & -4.61 & 4.48 \\
cptmarvel-3 & 6.71 & 9.94 & 41.33 & -9.61 & 8.61 & 7.84 & 8.27 & 7.88 & 8.04 & 4.74 & -4.38 & 4.52 \\
rogue-11 & 6.67 & 9.91 & 40.27 & -9.83 & 8.44 & 7.69 & 8.16 & 7.43 & 8.06 & 4.57 & -4.56 & 4.58 \\
r761-4 & 6.55 & 9.71 & 36.31 & -10.09 & 8.09 & 7.43 & 7.71 & 7.69 & 6.15 & 4.74 & -4.69 & 4.27 \\
storm-6 & 6.52 & 9.93 & 40.82 & -9.24 & 8.45 & 7.4 & 8.3 & 7.86 & 8.1 & 4.83 & -4.37 & 4.52 \\
r571-2 & 6.38 & 9.84 & 40.18 & -9.79 & 8.12 & 7.11 & 7.98 & 7.92 & 7.06 & 4.86 & -4.71 & 4.42 \\
r556-3 & 6.3 & 9.7 & 36.0 & -9.72 & 8.39 & 7.81 & 8.0 & 7.94 & 7.15 & 4.55 & -4.48 & 4.4 \\

\hline
\caption{Individual Halo Properties in the \MM\, Sample. The columns, from left to right, list each galaxy's simulation volume name and halo id ($ID$), stellar mass ($M_*$), virial mass ($M_{200c}$), virial radius ($R_{200c}$) in units kpc, specific star formation over the last 100Myr ($sSFR_{100}$) in units $\rm yr^{-1}$, total gas inside $R_{200c}$ ($M_{\mathrm{Gas}}^{\mathrm{Halo}}$), HI and helium gas mass in the disk defined using \rHI\, ($M_{\mathrm{HI+He}}^{\mathrm{Disk}}$), total gas in the CGM defined as $0.15< r/R_{200c} <1$, ($M_{\mathrm{Gas}}^{\mathrm{Halo}}$), total gas mass in the cool phase ($M_{\mathrm{Cool}}^{\mathrm{CGM}}$), total gas mass in the warm phase ($M_{\mathrm{Warm}}^{\mathrm{CGM}}$), total metal mass in the CGM ($M_{\mathrm{Z}}^{\mathrm{CGM}}$), average mass weighted gas number density in the CGM ($\rho_{\mathrm{CGM}}$) in units $\rm cm^{-3}$, and average mass weighted gas temperature in the CGM $T_{\mathrm{CGM}}$ in units $\rm K$. All masses are provided in $M_{\odot}$, and all values, except $R_{200c}$, are provided as logarithms.}
\end{longtable}

\onecolumngrid
\section{Marvel vs Massive Comparison}\label{appendix:SuitebySuite}

\begin{figure}
    \centering
    \includegraphics[width=1\linewidth]{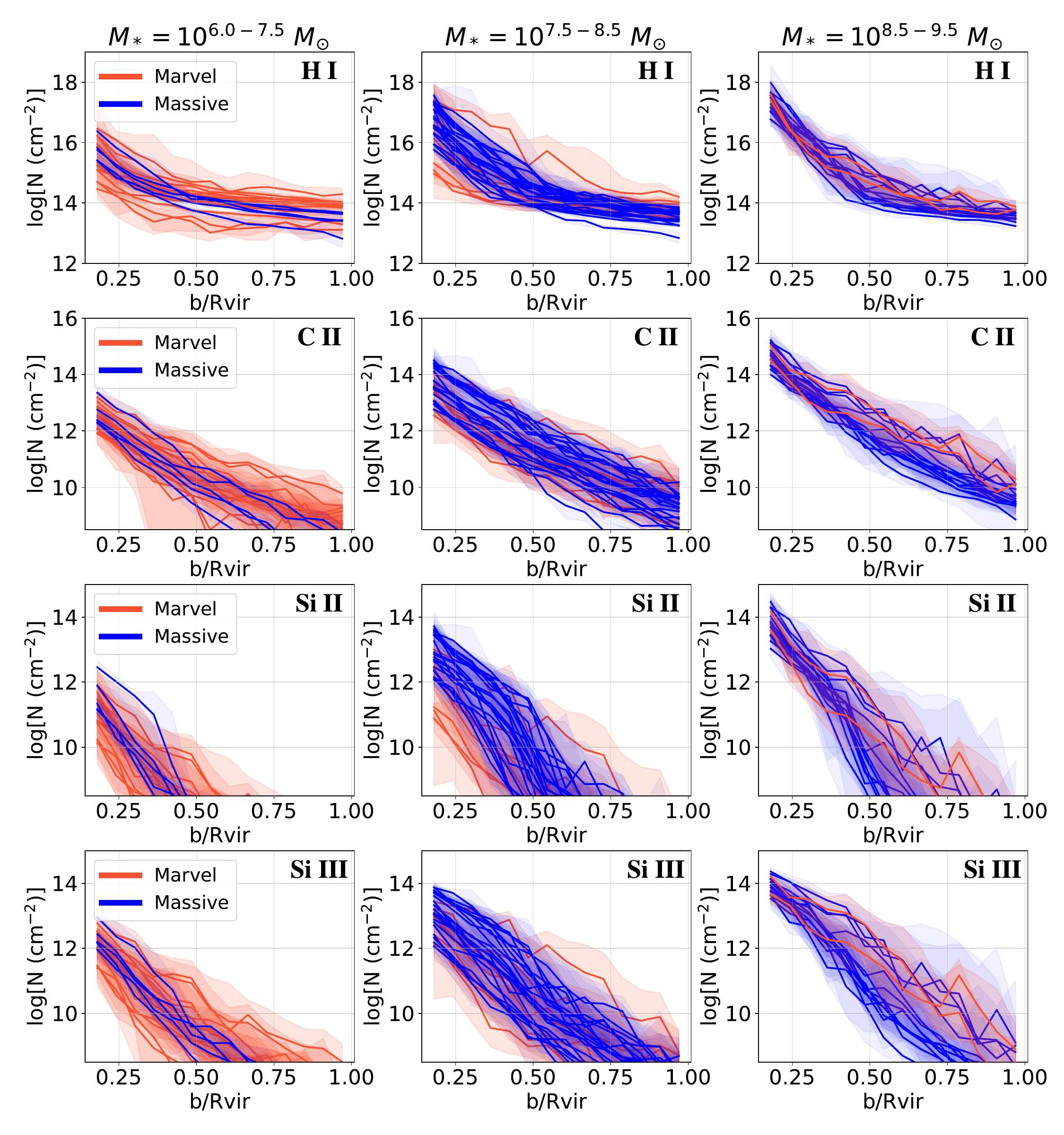}
    \caption{Column density profiles for all galaxies in the \MM\, sample. Each panel shows the median column density profiles in a given ion around a galaxy within the stellar mass bin denoted at the top of the figure. Each line represents the profile for a single galaxy and is color-coded red if the galaxy is from the Marvel suite or blue if the galaxy is from the Massive suite. Also shown is the 16th-84th percentile for a given galaxy. We find that the overall differences between the two suites are minor.
    }
    \label{fig:Appendix_N_I}
\end{figure}

\begin{figure}
    \centering
    \includegraphics[width=1\linewidth]{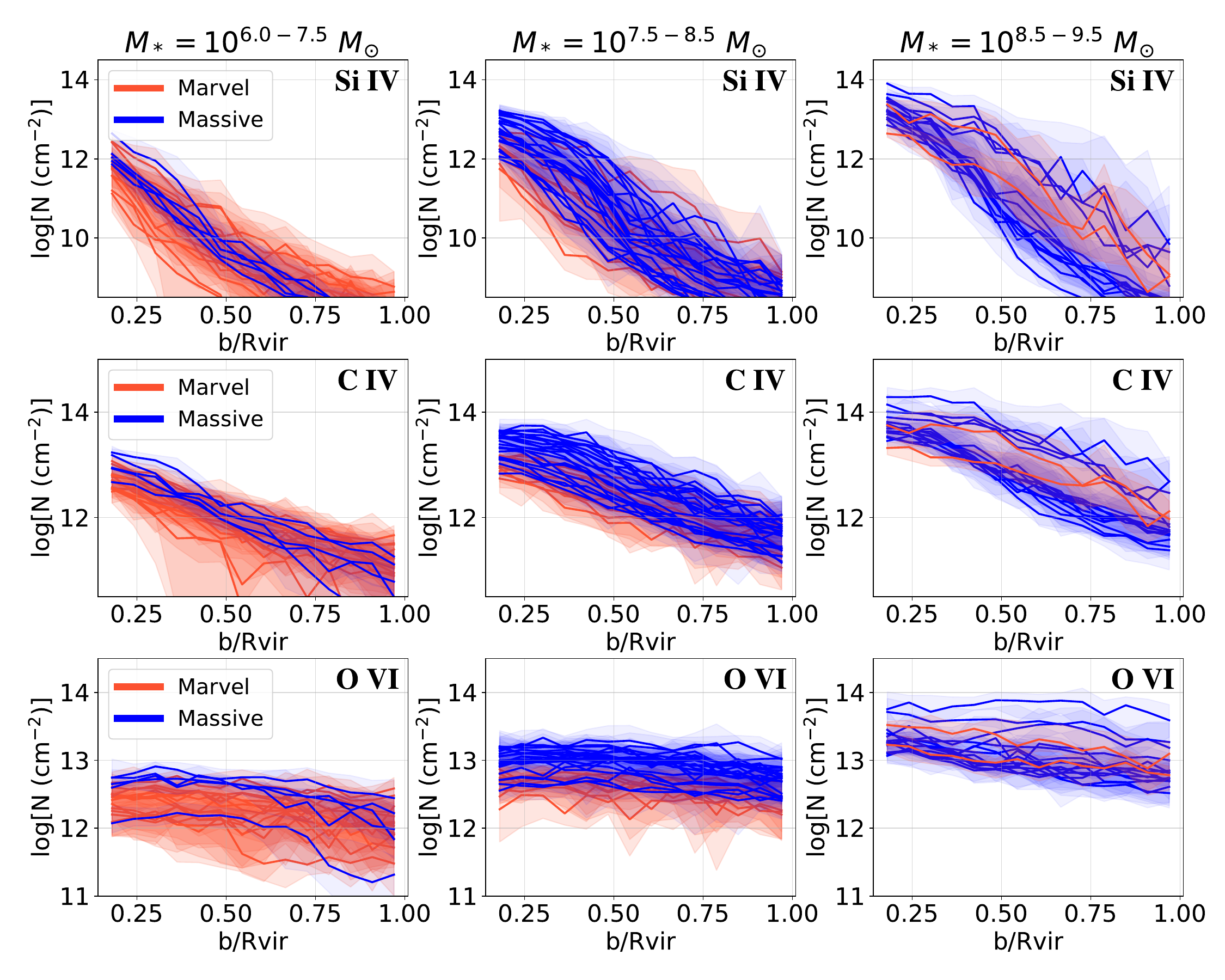}
    \caption{Same as Figure \ref{fig:Appendix_N_I} but for \SiIV, \CIV, and \OVI.}
    \label{fig:Appendix_N_II}
\end{figure}

Figures \ref{fig:Appendix_N_I} and \ref{fig:Appendix_N_II} show the column density profiles in a similar format to Figures \ref{fig:LowIonColumnDensitiesResults} and \ref{fig:HighIonColumnDensitiesResults} where each column presents data for a given stellar mass bin (denoted at the top) and each row presents the column densities in a specific ion (denoted in the top left of each panel). In these figures, we show the median column densities for each galaxy in the \MM\, sample and differentiate between suites. Specifically, red lines are profiles for galaxies in the Marvel-ous Dwarfs suite, and blue lines are profiles for galaxies in the Marvelous Massive Dwarfs suite. We find little clear variation between suites, but note that the two suites are biased toward higher or lower masses (Figure \ref{fig:MM-Summary}), and Marvel galaxies tend to span a larger range in column densities. The stellar mass range $M_*=10^{7.5-8.5}M_{\odot}$ has the most overlap and a comparable number of galaxies across the two suites, with 8 and 21 from Marvel and Massive, respectively. When comparing the median of each suite in this mass range, we find that the medians are within $1\sigma$ of one another, across impact parameters and ions---except for \NOVI.  Median \NOVI\, values tend to be within $1-1.5\sigma$, and the Massive suite produces greater column densities. The minor differences seen in column densities may be due to the buoyancy of SB vs BW \citep[see][]{Keller_2020}, or the fact that the SB model used here adopts a smaller $E_{SN}$ than the BW model.  Understanding the nuances is left to future work.  However, we conclude that the overall differences between the two runs are minor, justifying our use of both suites to span a broader mass range and increase our sample size.

\section{Consequences of $15\%R_{200c}$ CGM Definition }\label{appendix:CGMDef}
\begin{figure}[h]
    \centering
    \includegraphics[width=9cm]{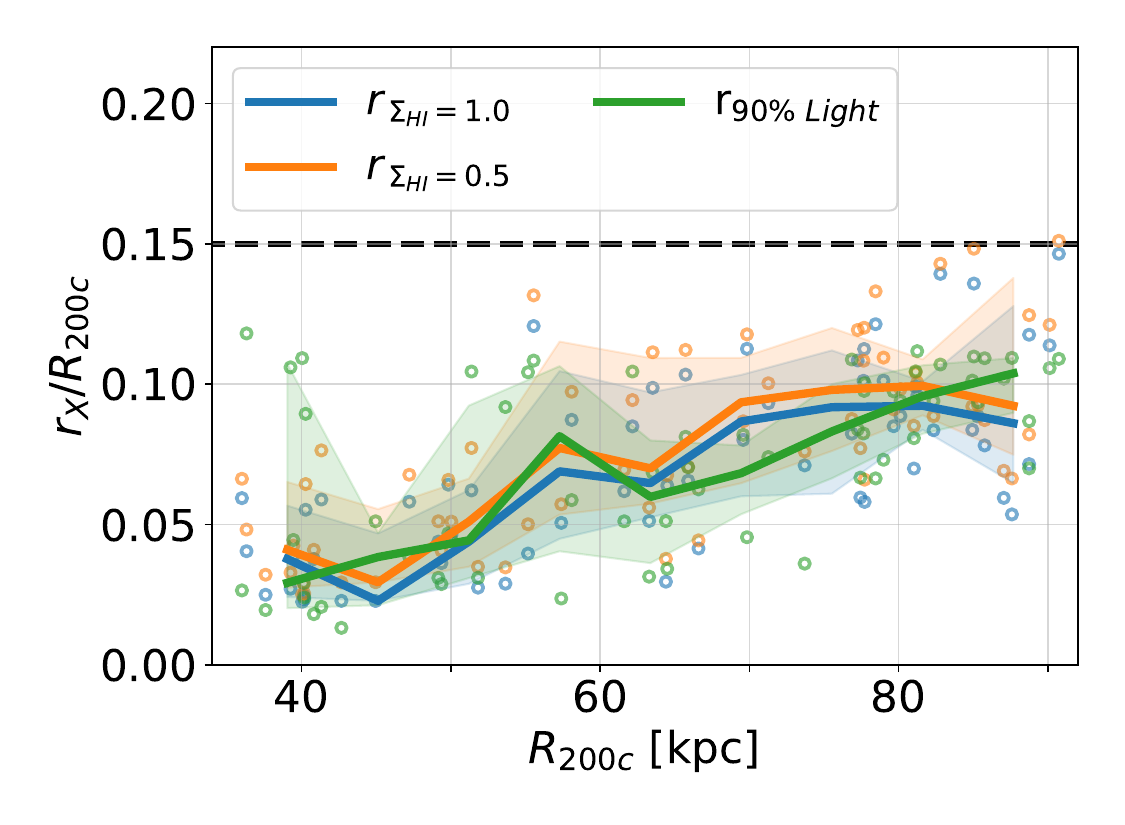}
    \caption{Fractional extent of the galaxy disk based on 3 different definitions: the radius at which the surface density of \HI\, falls below 1.0 M$_{\odot}$pc$^{-2}~(r_{\,\Sigma_{\text{HI}}=1.0}$; blue, the same definition used for \rHI); the radius at which the surface density of \HI\, falls below 0.5 M$_{\odot}$pc$^{-2}~(r_{\,\Sigma_{\text{HI}}=0.5}$; orange); and the radius within which 90\% of the starlight is contained (r$_{90\% ~Light}$; green). Solid lines show the median values, shaded regions encompass the 16th - 84th percentile, and individual halos are shown as points.}
    \label{fig:DiskSizes}
\end{figure}
While analyzing the CGM in the \MM\, sample, we opted to define the CGM as a fixed fraction of the viral radius ($0.15 R_{200c}$) to remove the disk of the galaxy and define the disk according to \rHI. We argue that the CGM definition enables better comparison to past work and observations and attempts to normalize our calculations by the gravitational influence of the galaxies. Additionally, we argue our choice of \rHI\, is more similar to how an observer would select the gaseous disk of a galaxy. However, it is a fact that, given the range in masses utilized, the physical extent of our halos differs by a factor of $\sim2\times$ from the low to high mass end. Therefore, we are selecting different physical volumes when we utilize the virial radius. In this section, we quantify the fraction of the halo that we remove when we implement these definitions. 

Figure \ref{fig:DiskSizes} shows the relation between the virial radius of a halo and the approximate size of the disk according to the radius at which the surface density of \HI\, falls below 1.0 M$_{\odot}$pc$^{-2}$ (the definition adopted for \rHI\, in the paper) and 0.5 M$_{\odot}$pc$^{-2}$ and the radius within which 90\% of the starlight is contained. Solid lines show the median values, and shaded regions show the 16th - 84th percentiles. Figure \ref{fig:DiskSizes} provides insight into the extent of the galaxy (both its ISM and stellar component) compared to the halo size, as seen by a mock observer. We find the size of the galaxy decreases relative to $R_{200c}$ with decreasing mass. For the highest-mass galaxies ($\text{log}(M_*/M_{\odot}) > 8.5$), the $r\geq0.15 R_{200c}$ CGM definition removes the ISM and stellar component plus $\sim 5\%$ of the halo. This is done to focus our study on the extended CGM. However, for the lowest mass galaxies, the fraction removed beyond the disk and from the CGM rises to $\sim 10\%$. The consequence of this is that we remove the coolest and densest material outside of the disk that may be considered CGM gas by an observer. This fact likely drives the mass dependence seen in the left column of Figure \ref{fig:CGMPhases_Profiles}, where with declining mass we see the cool, dense component disappear from the CGM region. Additionally, given the trend of increasing column densities for decreasing impact parameter in Figures \ref{fig:LowIonColumnDensitiesResults} and \ref{fig:HighIonColumnDensitiesResults}, there are likely observable metals within $0.15R_{200c}$ or 6 kpc for $\text{log}(M_*/M_{\odot}) < 7.5$ galaxies.

\end{document}